\documentclass[acmtog,authoryear, nonacm]{acmart}
\setcopyright{none}
\usepackage{graphicx} 
\usepackage{amsmath}

\usepackage{amssymb}
\usepackage{color}
\usepackage{enumerate}
\usepackage{algorithm}
\usepackage{algpseudocode}
\usepackage{natbib}

\title{Orientation in Extended Position-Based Dynamics: Application to Rigid Bodies and Cosserat Rods}
\author{Samuel Tobin}
\orcid{0009-0004-7641-7643}
\affiliation{%
  \institution{University of Tennessee-Knoxville}
  \city{Knoxville}
  \state{TN}
  \country{USA}
}
\email{stobin2@vols.utk.edu}

\author{Caleb Rucker}
\orcid{0000-0001-7181-1933}
\affiliation{%
  \institution{University of Tennessee-Knoxville}
  \city{Knoxville}
  \state{TN}
  \country{USA}
}
\email{caleb.rucker@utk.edu}



\begin{document}

\begin{teaserfigure}
  \centering
  \includegraphics[width=\textwidth]{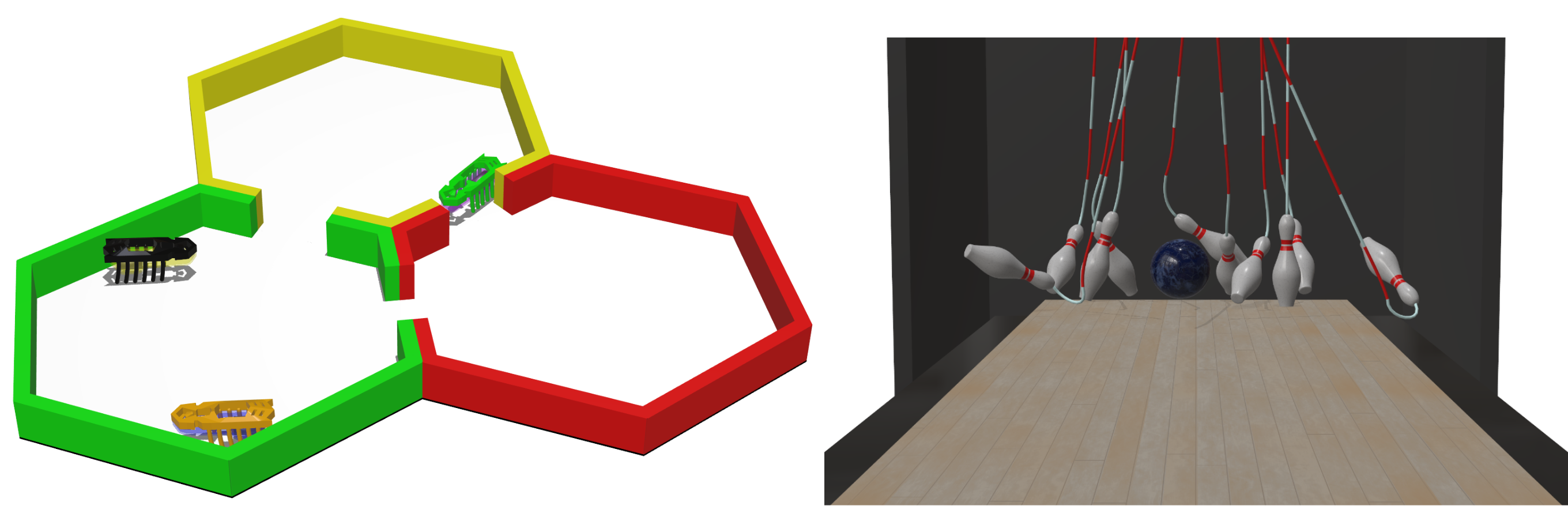}
  \caption{\textbf{Left.} A ``bristle bot'' simulated with our approach for rigid body joints and Cosserat rods. A rotating eccentric mass creates periodic vibrations that are converted to forward motion from the shape of the deformable legs and stick-slip friction. The body of the robot is rigid and the legs are modeled with only a single quadratic Cosserat rod element. Multiple robots collide with each other and their environment. \textbf{Right.} String-pin bowling. The strings are modeled as stiff Cosserat rods with only $8$ cubic elements, and each element is colored separately in the image. The strings are stable under high-velocity collisions between the ball and the pins.}
  \label{fig:teaser}
\end{teaserfigure}

\begin{abstract}
Rotational degrees of freedom in Extended Position-Based Dynamics (XPBD) require computations on the nonlinear manifold of 3D rotations.
We show that Lie theory provides a clean, unified framework for expressing rotations, constraints, interpolation, and differentiation in XPBD, enabling both improved rigid-body constraints and higher-order finite-element Cosserat rods.
We derive explicit Lie-theoretic constraint formulations and their gradients for rigid-body simulation,
improving the dynamic consistency of constrained rigid-body simulations in XPBD by a factor of over $10^4$ compared to the state-of-the-art.
Our framework naturally extends to finite-element Cosserat rods by enabling on-manifold interpolation of nodal rotations.
Linear finite elements outperform the conventional chain-of-rigid-bodies discretization, while higher-order basis functions provide even smoother solutions and faster convergence. Utility is demonstrated in a variety of examples with large deformations and contact.
\end{abstract}

\maketitle

\section{Introduction}
In computer graphics applications, Extended Position-Based Dynamics (XPBD) has emerged as a stable and efficient framework for the physics-based simulation of constrained particle systems, generalized to elastic systems. Hard constraints and elastic constraints derived from a quadratic energy potential are individually solved in a Gauss-Seidel fashion to efficiently approximate the solution to a backward Euler step in time, providing high stability under constrained time budgets. However, in its original form, XPBD (and other position-based frameworks) deals strictly with positional degrees of freedom. While this is sufficient to model cloth and elastic bodies, other objects are more naturally described by a combination of positional and rotational degrees of freedom (e.g., rigid bodies, rods). Thus, it is natural to augment each particle with independent rotational degrees of freedom, which has been done extensively in prior work for rigid bodies \cite{deul2016position, francu2017unified, muller2020detailed} and rods \cite{kugelstadt2016position, deul2018direct, angles2019viper}.

However, handling rotations in three dimensions requires some care. The set of 3D rotations is not a vector space: although rotation matrices can be added and subtracted as matrices, the result is generally not itself a valid rotation. Instead, 3D rotations have a very specific structure: together they form the \emph{Lie group} $SO(3)$, a differentiable three-dimensional manifold in the space of $3 \times 3$  matrices. 
Local coordinate charts for the manifold may be used to parameterize rotations.
The standard parameterization in graphics is with \emph{unit quaternions}, which form a four-dimensional extension of the complex numbers that encode the axis-angle representation of rotation. Unit quaternions do not have any singularities, have easily implementable formulas for how they rotate vectors and concatenate, and only require four numbers (as opposed to the nine required by storing the full rotation matrix).

In position-based frameworks, one must differentiate functions with respect to the degrees of freedom. Thus, for oriented particles, we must differentiate with respect to rotations. Because rotations lie on a manifold rather than a vector space, derivatives must respect the geometry of that manifold. Although unit quaternions provide a singularity-free representation of rotations, they are not intrinsic coordinates of the rotation manifold, so derivatives with respect to quaternion components require additional treatment to enforce the unit-norm constraint. Many prior works use a quaternion coordinate-based approach originally proposed by \cite{grassia1998practical}, for example \cite{deul2016position, deul2018direct, ferguson2021intersection}, while other prior works omit presentation of these gradients entirely \cite{muller2011solid, francu2017unified, muller2020detailed}. The coordinate-based approach of Grassia is practical for simple functions, but quickly becomes unwieldy for complex constraint functions involving multiple rotational DOF \cite{romanya2025painless}. This presents a barrier for sophisticated position-based modeling of objects with rotational DOF. Regardless of how rotations are fundamentally represented (i.e. through rotation matrices, or unit quaternions), a more natural way to describe the evolution and differentiation of rotations is through \emph{Lie theory} \cite{solà2021microlietheorystate, romanya2025painless}, which provides rigorous mathematical machinery that respects the inherent geometry of Lie groups. The use of Lie theory is commonplace in domains such as robotics \cite{murray2017mathematical}, state estimation \cite{barfoot2024state}, and computational mechanics \cite{simo1986three}, but its application remains somewhat limited in the graphics community. 

In this work, we make extensive use of Lie group theory to incorporate 3D rotations into XPBD. While our results are written assuming rotation matrix representation, the same Lie group approach can be used in the context of quaternion representation, as we point to in Section \ref{sec:quaternions}. We define joint constraints for rigid bodies using Lie theory, leading to intuitive, vector-valued expressions that linearize better than the scalar-valued constraints employed by \cite{muller2020detailed}. This results in up to $10^5$-fold reductions in primal residual for comparable computational cost. Additionally, we use Lie theoretic differentiation to compute all constraint gradients, leading to compact analytical expressions for complicated constraint functions. We show all derivations of gradient expressions in the Appendix as a blueprint for future work. Furthermore, the incorporation of Lie theory naturally enables the finite-element discretization of Cosserat rods in XPBD, with manifold rotational interpolation. This is the so-called ``geometrically exact'' Cosserat rod, pioneered by Simo and Vu Quoc \cite{simo1986three}. We show that the XPBD simulation of Cosserat rods parameterized by higher-order basis functions have greater accuracy at the same computational cost as the standard chain-of-rigid-bodies discretization commonly used in graphics.  The resulting framework illustrates how Lie-theoretic tools can be integrated into position-based simulation and may be applicable to a broader class of graphics algorithms involving rotational degrees of freedom. 

\section{Related Work}
\paragraph{Applied Lie Theory}
The explicit use of Lie theory in the graphics community has primarily appeared in applications involving rotational kinematics and interpolation \cite{park1997smooth, celledoni2015shape}, while its use in simulation frameworks remains comparatively limited. 
Outside the graphics community, Lie theory is commonly employed as a tool for rigorous treatment of 3D rotations. In the computational mechanics community, Lie theory is extensively used in the simulation of rods and shells \cite{simo1986three, simo1990stress}. In robotics, Lie theory is the standard language for describing robot kinematics \cite{murray2017mathematical, renda2020geometric, lynch2017modern, selig2005geometric}. 
In robotic state estimation, Lie theory is used to perform on-manifold geometric optimization for unknown poses \cite{barfoot2024state}. To that end, Sola et al. have developed a resource describing Lie theory specifically for applications in state estimation, with large emphasis on differentiation of geometric objects \cite{solà2021microlietheorystate}. A similar preceding resource was developed by Bloesch et al. \cite{bloesch2016primerdifferentialcalculus3d}. In these resources, they define the $\boxplus$ and $\boxminus$ operators as intuitive ways to ``add'' and ``subtract'' objects on the manifold. They then use these definitions to develop an intuitive definition of differentiation that mirrors that in Euclidean space. In this work, we adopt the same notation and methodology for differentiation of 3D rotations.

\paragraph{Position-Based Simulation with Rotational DOF}
In position-based simulation, particle positions are directly solved for or optimized over, in contrast with classical methods that solve for forces and accelerations and integrate to evolve particle positions.
Position-Based Dynamics (PBD) directly enforces geometric constraints on particle positions through iterative projection \cite{muller2007position}. While highly robust and efficient, the original formulation exhibits iteration-dependent stiffness. Extended Position-Based Dynamics (XPBD) \cite{macklin2016xpbd} resolves this limitation through a compliant constraint formulation derived from elastic energy potentials. The XPBD update equations are derived from a fixed-point iteration of the dual system of nonlinear equations of motion \cite{macklin2020primal}. 
Similarly, Position-Based Nonlinear Gauss-Seidel (PBNG) \cite{chen2023position}, developed for quasi-static problems, solves the nonlinear equilibrium equations.
An alternative viewpoint is to cast Backward Euler time-integration as an optimization problem, and optimize over the particle positions (and rotations, when applicable) to minimize the integration potential \cite{gast2015optimization, martin2011example}. This forms the basis for Projective Dynamics \cite{bouaziz2023projective}, Vertex Block Descent (VBD) \cite{chen2024vertex, giles2025augmented}, and Incremental Potential Contact (IPC) \cite{li2020incremental}.

Typically, these frameworks are developed for purely positional particles, but many works augment the particles with independent rotational degrees of freedom for the simulation of rigid bodies and rods. A key challenge is differentiating with respect to these added rotational degrees of freedom. Grassia \cite{grassia1998practical} introduced a straightforward way of computing these gradients using the chain rule: first differentiating the function with respect to quaternion coordinates as a Euclidean gradient and then differentiating through the quaternion exponential map.

\cite{deul2016position} simulated rigid bodies with oriented particles in PBD using the approach of Grassia for rotational gradients. Their treatment included a spherical joint as an explicit example of a joint constraint. More sophisticated friction modeling was subsequently incorporated by \cite{francu2017unified}, although the corresponding constraint definitions and gradients were not described in detail. \cite{muller2020detailed} provided a more detailed treatment, including update formulas for joints and motors in XPBD-based rigid body simulation. Their formulation, however, emphasizes practical update rules rather than a general constraint/Jacobian formulation. As detailed in the Supplementary Material, these update rules can be interpreted as using separate scalar constraints for the positional and rotational components of a joint, and for some joints, the resulting effective constraint definitions do not fully align with the stated update rules. In contrast, we use Lie theory to formulate the same set of joints as vector-valued constraints, which yields substantially improved accuracy in our benchmarks.

Outside the PBD family, \cite{ferguson2021intersection} extended the IPC formulation to rigid bodies for intersection-free rigid body simulations. They parameterize the rigid body rotations using exponential coordinates, and optimize over these exponential rotation vectors within each time step, and compute gradients by differentiating through the exponential map. \cite{romanya2025painless} improve upon this by instead using Lie theoretic differentiation (a la \cite{solà2021microlietheorystate}) to derive cleaner analytical gradient expressions. This enables tractable formulas for Hessians, enabling differentiable simulation. Additionally, they show that using their derivative expressions require fewer Newton iterations to converge compared to \cite{ferguson2021intersection}. The same insight applies in our XPBD setting: replacing chain-rule gradients with Lie-theoretic Jacobians yields cleaner constraint expressions and improved accuracy, while the same differentiation framework additionally supports the rotational interpolation required for Cosserat rod finite elements. 

The same differentiation challenges arise in elastic rod simulation, where prior works have taken varied approaches. The first work to simulate rods within PBD was \cite{umetani2014position}, which used ``ghost points'' with masses offset from the rod centerline to approximate torsional effects instead of particles with rotational DOF. \cite{kugelstadt2016position} instead used PBD with oriented particles with the rod strains as vector-valued constraints. The rotational gradients are derived in the ambient quaternion space, incorrectly neglecting the geometry of the rotation. \cite{deul2018direct} translate the work of Kugelstadt et al. to XPBD for inextensible rods, and correctly derive rotational gradients via the chain rule. They also introduce a direct, linear-time solver for the rod subsystem based on the approach of Baraff \cite{baraff1996linear}, mitigating the slow convergence of long constraint chains in constraint-projection-based simulation frameworks like XPBD. At the same time, \cite{soler2018cosserat} developed Cosserat rods in the Projective Dynamics framework. They derive closed-form solutions for the local projection problems associated with the rotational degrees of freedom. \cite{angles2019viper} augments the standard Cosserat rod formulation to introduce an isotropic scaling variable per vertex, capturing basic volumetric effects. They solve an optimization problem at each time step with a warm-started Gauss-Newton approach and parameterize the rotations in the rod with exponential coordinates. The required Jacobians are obtained through standard first-order linearizations of the exponential map commonly used in graphics and robotics. Recently, \cite{hsu2025stable} developed an exceptionally stable Cosserat rod method based on a decoupled quasistatic orientation optimization. This is formulated as a constrained optimization problem over unit quaternions. Our work on Cosserat rods extends the work of \cite{deul2018direct}; we move from the chain-of-rigid-bodies approach to finite elements, using Lie theoretic differentiation and interpolation of rotational degrees of freedom.

\paragraph{Cosserat Rods in Computational Mechanics}
In the computer graphics literature, rods are most commonly modeled as chains of rigid bodies, with strain energy concentrated at joints between the bodies. On the other hand, the computational mechanics community commonly employs finite-element discretizations, where position and rotational degrees of freedom are interpolated using basis functions defined on each element \cite{simo1986three}. Done correctly, a chain of rigid bodies approximation is essentially a zeroth order finite element discretization. Importantly, the finite element formalism grants freedom in the choice of basis; as increasingly higher-order basis functions are used, the accuracy per DOF increases. 

The tricky part in Cosserat rod finite elements is the rotation interpolation. The seminal work of Simo and Vu Quoc \cite{simo1986three} introduced the ``geometrically exact'' Cosserat rod formulation which parameterizes the rotational DOF of the centerline with basis functions over finite elements using a global exponential rotation vector. Additionally, they employ Lie-theoretic linearizations, effectively performing optimization on the $\mathbb{R}^3 \times SO(3)$ manifold. However, \cite{jelenic1999geometrically} show that their rotational interpolation is not frame-invariant, and instead parameterize rotation with a local rotation vector defined within each element.

\cite{saillant2024high} drew an explicit connection between XPBD and higher-order finite elements for classical volumetric FEM by defining the XPBD constraint functions at the Gauss quadrature points in each element. Thus, the sum of constraint energies within an element approximates the energy integral in that element in a Gauss integration manner. Similarly, we adapt the approach of \cite{jelenic1999geometrically} to XPBD by defining constraints at Gauss points within each element. Our incorporation of Lie theory into XPBD is necessary for the rotational interpolation and also enables compact constraint gradient expressions.

\paragraph{Positioning of Our Work}
Our work incorporates Lie-theoretic formulations into XPBD, providing a unified treatment of rotational degrees of freedom across rigid bodies and Cosserat rods. Unlike previous XPBD formulations for rigid bodies, which omit or circumvent explicit constraint definitions, we formulate common rigid body joints as explicit vector-valued constraints with analytically-derived Jacobians. Our formulation builds upon the Lie-theoretic differentiation framework of \cite{romanya2025painless} while adapting it to the constraint-based setting of XPBD.
For Cosserat rods, we connect the finite-element viewpoint from computational mechanics with position-based simulation by introducing a Lie-theoretic formulation compatible with arbitrary interpolation basis functions. This extends beyond the commonly used chain-of-rigid-bodies discretization and enables higher-order rod representations within XPBD. More broadly, our framework provides a common language for expressing rotational constraints, interpolation, and differentiation in position-based simulation.

\section{A Primer on 3D Rotations}
In the developments that follow, it is crucial to have a basic understanding of the structure and properties of 3D rotations. This section will focus only on the most relevant information for the content in the rest of the paper, but for a more in-depth introduction, see e.g. \cite{solà2021microlietheorystate,bloesch2016primerdifferentialcalculus3d}.

\subsection{The Special Orthogonal Group $SO(3)$}
\label{sec:so3}
A three-dimensional rotation is a linear transformation that preserves lengths and angles. Such transformations are represented by orthogonal matrices with determinant one:
\[
\mathbf{R}^\top \mathbf{R} = \mathbf{I}, \quad \det(\mathbf{R}) = 1.
\]
The set of all such matrices forms the \textit{special orthogonal group} $SO(3)$.
We represent vectors as column vectors and apply rotations by left multiplication, such that a rotated vector is given by $\mathbf{x}' = \mathbf{R} \mathbf{x}$.
$SO(3)$ is a \emph{Lie group}: it is both a group under matrix multiplication and a smooth manifold. 
A key distinction from vector spaces is that $SO(3)$ does \emph{not} support addition or subtraction between elements. This becomes important when differentiating or perturbing rotations. Additionally, matrix multiplication in $SO(3)$ is not commutative, so left- and right-multiplication define distinct operations.


To describe how rotations change, we work with \textit{tangent spaces}. The tangent space at $\mathbf{R}$, denoted $T_\mathbf{R}SO(3)$, consists of all instantaneous directions in which $\mathbf{R}$ can move. Each $\mathbf{R}$ has its own tangent space, which makes direct computations cumbersome.
Instead, we map tangent vectors to a common space: the tangent space at the identity, $T_\mathbf{I}SO(3)$, called the \textit{Lie algebra} $\mathfrak{so}(3)$. This process is called \textit{trivialization}. Using left- and right-multiplication, we obtain two standard ways to express $\dot{\mathbf{R}} \in T_\mathbf{R}SO(3)$ in $\mathfrak{so}(3)$:
\begin{equation}
\begin{alignedat}{2}
\text{Left-trivialization (body frame):} \quad &\boldsymbol{\Omega}_b = \mathbf{R}^\top \dot{\mathbf{R}}, \\
\text{Right-trivialization (spatial frame):} \quad &\boldsymbol{\Omega}_s = \dot{\mathbf{R}}\mathbf{R}^\top.
\end{alignedat}
\end{equation}
These correspond to expressing angular velocity in the body and spatial frames, respectively.

Differentiating the constraint $\mathbf{R}^\top \mathbf{R} = \mathbf{I}$ yields $\mathbf{R}^\top \dot{\mathbf{R}} = -\dot{\mathbf{R}} \mathbf{R}^\top$, indicating that both $\Omega_b$ and $\Omega_s$ are skew-symmetric. Therefore, the Lie algebra $\mathfrak{so}(3)$ is a vector space consisting of all $3\times3$ skew-symmetric matrices. Each element in $\mathfrak{so}(3)$ can be \textit{identified} uniquely with an element from $\mathbb{R}^3$. In other words, $\mathfrak{so}(3)$ is \textit{isomorphic} to $\mathbb{R}^3$. The isomorphism is defined via the following maps, named the \textit{hat} and \textit{vee} maps, respectively \cite{murray2017mathematical}:
\begin{align}
\label{eq:hat-map}
\begin{aligned}
    ^\wedge :\  &\mathbb{R}^3 \to \mathfrak{so}(3) \\
    &\begin{bmatrix}
        \omega_x \\
        \omega_y \\
        \omega_z
    \end{bmatrix} \mapsto \begin{bmatrix}
        0 & -\omega_z & \omega_y \\
        \omega_z & 0 & -\omega_x \\
        -\omega_y & \omega_x & 0
    \end{bmatrix}
    \end{aligned} \\
\label{eq:vee-map}
    \begin{aligned}
    ^\vee : \ &\mathfrak{so}(3) \to \mathbb{R}^3\\
    &\begin{bmatrix}
        0 & -\omega_z & \omega_y \\
        \omega_z & 0 & -\omega_x \\
        -\omega_y & \omega_x & 0
    \end{bmatrix} \mapsto \begin{bmatrix}
        \omega_x \\
        \omega_y \\
        \omega_z
    \end{bmatrix}
    \end{aligned}
\end{align}
Thus, we can write $\boldsymbol{\Omega}_b = \widehat{\boldsymbol{\omega}}_b$ and interpret $\boldsymbol{\omega}_b \in \mathbb{R}^3$ as the body-frame angular velocity (and similarly for the spatial frame).

\subsection{$SO(3)$ Exponential and Logarithmic Maps}
\begin{figure}
    \centering
    \includegraphics[width=0.65\linewidth]{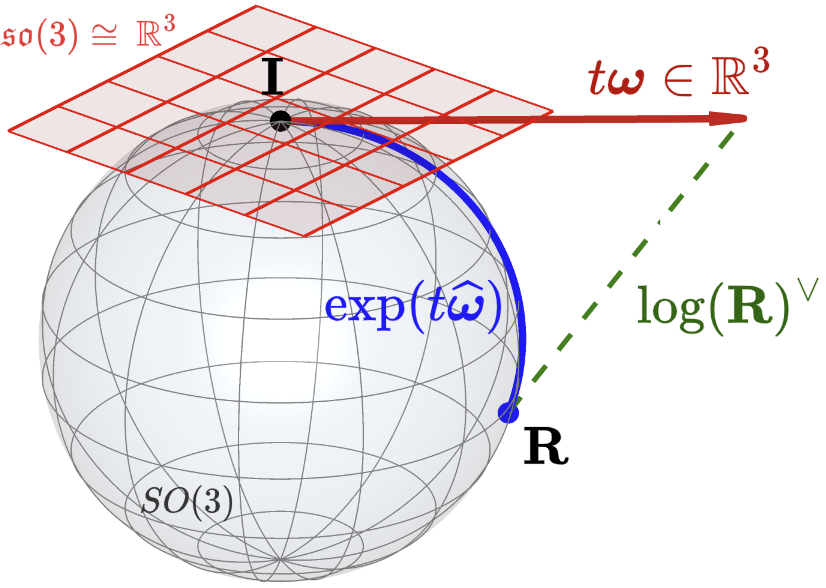}
    \caption{Graphical depiction of the exponential and logarithm maps for $SO(3)$. The sphere shown here is used as a visual representation of the smooth manifold structure of $SO(3)$.}
    \label{fig:so3-exp-log-diagram}
\end{figure}

Now, say we want to evolve a $SO(3)$ rotation $\mathbf{R}(0) = \mathbf{I}$ with some constant angular velocity $\boldsymbol{\omega}$ for a fixed amount of time $\Delta t$. The \textit{matrix exponential map} $\exp: \mathfrak{so}(3) \to SO(3)$ gives the exact change in rotation over $\Delta t$, such that:
\begin{equation}
    \mathbf{R}(\Delta t) = \exp(\Delta t\widehat{\boldsymbol{\omega}})
\end{equation}
If we start at some arbitrary rotation $\mathbf{R}(t) \neq \mathbf{I}$, increments can be applied on the left or on the right:
\begin{equation}
\label{eq:application-of-exp}
\begin{aligned}
    \mathbf{R}(t+\Delta t) &= \exp(\Delta t\widehat{\boldsymbol{\omega}}_s)\mathbf{R}(t) \\
    \mathbf{R}(t+\Delta t) &= \mathbf{R}(t) \exp(\Delta t\widehat{\boldsymbol{\omega}}_b)
\end{aligned}
\end{equation}
corresponding to spatial and body angular velocities, respectively. In this paper we use the body-frame (right-multiplication) convention since this admits the use of constant, body-frame rotational inertia matrices.


For $SO(3)$, the exponential map has the following closed-form expression, also known as Rodrigues formula:
\begin{equation}
\label{eq:so3-exp}
    \exp(\widehat{\boldsymbol{\theta}})=\mathbf{I} + \frac{\sin{||\boldsymbol{\theta}||}}{||\boldsymbol{\theta}||} \widehat{\boldsymbol{\theta}} + \frac{1 - \cos{||\boldsymbol{\theta}||}}{||\boldsymbol{\theta}||^2} \widehat{\boldsymbol{\theta}}^2
\end{equation}

The \textit{matrix logarithmic map} $\log:SO(3) \to \mathfrak{so}(3)$ is the (local) inverse of the exponential map; it returns the (shortest) rotation vector whose exponential equals $\mathbf{R}$. From \eqref{eq:application-of-exp}, we get that:
\begin{equation}
    \log (\mathbf{R}(t)^\top\mathbf{R}(t+\Delta t)) = \Delta t \widehat{\boldsymbol{\omega}}_b
\end{equation} 
A graphical depiction of the exponential and logarithm maps for $SO(3)$ can be seen in Figure \ref{fig:so3-exp-log-diagram}. The closed-form expression in $SO(3)$ is:
\begin{equation}
    \log(\mathbf{R}) = \frac{\theta}{2\sin \theta} \left(\mathbf{R} - \mathbf{R}^\top \right)
\end{equation}
where $\theta = \arccos \left( \frac{1}{2} \mathrm{tr}(\mathbf{R}) - \frac{1}{2} \right)$. Note that this formula becomes increasingly numerically imprecise as $\theta \to 0$, in which case a Taylor series expansion is typically used. Near $\theta = \pi$, the logarithm map encounters a geometric singularity because rotations by $\pi$ about the axes $\mathbf{u}$ and $-\mathbf{u}$ are indistinguishable. Since $\sin \theta \to 0$ in this limit, the standard formula becomes numerically ill-conditioned and requires special treatment; see, e.g., \cite{dellaert2012factor}.

For convenience, we define the \textit{boxplus} and \textit{boxminus} operators \cite{solà2021microlietheorystate}, which generalize vector addition and subtraction to operations on manifolds:
\begin{align}
\label{eq:so3-boxplus}
\begin{aligned}
    \boxplus:SO(3) \times \mathbb{R}^3 \to SO(3) \\
    \mathbf{R},\boldsymbol{\theta} \mapsto \mathbf{R} \exp(\widehat{\boldsymbol{\theta}})
\end{aligned} \\
\label{eq:so3-boxminus}
\begin{aligned}
    \boxminus:SO(3) \times SO(3) \to \mathbb{R}^3 \\
    \mathbf{R}_1, \mathbf{R}_2 \mapsto \log(\mathbf{R}_2^{\top} \mathbf{R}_1)^\vee
\end{aligned}
\end{align}
Intuitively, $\boxplus$ applies a local perturbation (a rotation vector) to $\mathbf{R}$, while the $\boxminus$ returns the rotation vector in $\mathbb{R}^3$ that, when added to the second argument via $\boxplus$, recovers the first:
\begin{equation}
    \mathbf{R}_2 \boxplus (\mathbf{R}_1 \boxminus \mathbf{R}_2) = \mathbf{R}_1
\end{equation}
In the definitions above, the right-multiplication convention is used, so $\boldsymbol{\theta}$ corresponds to a body-frame exponential rotation vector.


\subsection{Differentiation of 3D Rotations Using $\boxminus$}
\label{subsec:differentiation-of-3D-rotations}
Using $\boxminus$, we can define a time-derivative of $\mathbf{R}(t)$ that mirrors standard finite differences in vector calculus:
\begin{equation}
\label{eq:R-to-SO3-differential}
    \mathcal{D}_t(\mathbf{R}(t)) :=\lim_{\Delta t \to 0} \frac{\mathbf{R}(t+\Delta t) \boxminus \mathbf{R}(t)}{\Delta t} = \boldsymbol{\omega}_b
\end{equation}
As proven in Appendix \ref{sec:equivalent-definitions-angular-velocity}, $\mathcal{D}_t(\mathbf{R}(t)) = (\mathbf{R}(t)^\top \dot{\mathbf{R}}(t))^\vee = \boldsymbol{\omega}_b$. Intuitively, this makes sense: body angular velocity is the instantaneous rate of change in body-frame rotation angle, which is precisely what the definition in \eqref{eq:R-to-SO3-differential} describes.
Therefore, the resulting first-order approximation on the manifold is given by:
\begin{equation}
    \mathbf{R}(t+\Delta t) \approx \mathbf{R}(t) \boxplus \left( \Delta t \mathcal{D}_t(\mathbf{R}(t))  \right)
\end{equation}

This viewpoint extends naturally to other functions involving rotations. For a function $f: SO(3) \to \mathbb{R}$, differentiation with respect to its $SO(3)$ input argument gives us a linear map between a change in the $SO(3)$ input to a change in the $\mathbb{R}$ output. We represent a change in the $SO(3)$ input as a rotation vector $\Delta\boldsymbol{\theta} \in \mathbb{R}^3$, so we need to determine how the output changes as the input is perturbed in each of the three component directions ($\mathbf{e}_1$, $\mathbf{e}_2$, and $\mathbf{e}_3$): 
\begin{align}
\label{eq:SO3-to-R-differential}
    \mathcal{D}_\mathbf{R} (f(\mathbf{R})) &:= \lim_{\epsilon \to 0}
    \frac{1}{\epsilon}
    \begin{bmatrix}
        f(\mathbf{R} \boxplus (\mathbf{e}_1\epsilon)) - f(\mathbf{R}) \\
        f(\mathbf{R} \boxplus (\mathbf{e}_2\epsilon)) - f(\mathbf{R}) \\
        f(\mathbf{R} \boxplus (\mathbf{e}_3\epsilon)) - f(\mathbf{R}) 
    \end{bmatrix}^\top \\
\label{eq:SO3-to-R-linearization}
    f(\mathbf{R} \boxplus \Delta\boldsymbol{\theta}) &\approx f(\mathbf{R}) + \mathcal{D}_\mathbf{R}( f(\mathbf{R})) \Delta \boldsymbol{\theta}
\end{align}
Note that $\mathcal{D}_\mathbf{R} f(\mathbf{R})$ is a $1\times 3$ row vector and $\Delta \boldsymbol{\theta}$ is a $3\times 1$ column vector, so their matrix product in \eqref{eq:SO3-to-R-linearization} yields a scalar. This is straightforwardly extended to a function $\mathbf{f} = [f_1 \dots f_n]^\top: SO(3) \to \mathbb{R}^n$
\begin{align}
\label{eq:SO3-to-Rn-differential}
    \mathcal{D}_\mathbf{R} \mathbf{f}(\mathbf{R}) &:=
    \begin{bmatrix}
    \mathcal{D}_\mathbf{R}
        f_1(\mathbf{R}) \\
    \mathcal{D}_\mathbf{R}
        f_2(\mathbf{R}) \\
        \vdots \\
        \mathcal{D}_\mathbf{R}
        f_n(\mathbf{R})
    \end{bmatrix}^\top \\
\label{eq:SO3-to-Rn-linearization}
    \mathbf{f}(\mathbf{R} \boxplus \Delta\boldsymbol{\theta}) &\approx \mathbf{f}(\mathbf{R}) + \mathcal{D}_\mathbf{R}( f(\mathbf{R})) \Delta \boldsymbol{\theta}
\end{align}
For a function $g : SO(3) \to SO(3)$:
\begin{align}
\label{eq:SO3-to-SO3-differential}
    \mathcal{D}_\mathbf{R} ( g(\mathbf{R}) ) &= \lim_{\epsilon \to 0} \frac{1}{\epsilon}
    \begin{bmatrix}
        [g(\mathbf{R} \boxplus (\mathbf{e}_1\epsilon)) \boxminus g(\mathbf{R})]^\top \\
        [g(\mathbf{R} \boxplus (\mathbf{e}_2\epsilon)) \boxminus g(\mathbf{R})]^\top \\
        [g(\mathbf{R} \boxplus (\mathbf{e}_3\epsilon)) \boxminus g(\mathbf{R})]^\top
    \end{bmatrix}^\top \\
\label{eq:SO3-to-SO3-linearization}
    g(\mathbf{R} \boxplus \Delta \boldsymbol{\theta}) &\approx g(\mathbf{R}) \boxplus \left( \mathcal{D}_\mathbf{R} ( g(\mathbf{R}) ) \Delta \boldsymbol{\theta} \right)
\end{align}
These definitions express derivatives with respect to SO(3) in local tangent-space coordinates, allowing first-order variations to be computed using standard linear algebra in $\mathbb{R}^3$, which is particularly convenient for optimization and graphics applications.

\subsection{Common Jacobian Blocks}
\label{sec:jacobian-blocks}
Using the definitions above, we can derive Jacobians for common operations involving $SO(3)$ (derivations are provided in the Appendix). These serve as building blocks that can be combined via the chain rule in more complex expressions. For a point $\mathbf{p} \in \mathbb{R}^3$ and rotations $\mathbf{R}, \mathbf{R}_1, \mathbf{R}_2 \in SO(3)$:
\begin{alignat}{2}
    \label{eq:partial-of-mapping}
    &\mathcal{D}_\mathbf{R} \left( \mathbf{R} \mathbf{p} \right) &&= -\mathbf{R} \widehat{\mathbf{p}} \\
    \label{eq:partial-of-inverse}
    &\mathcal{D}_\mathbf{R} \left( \mathbf{R}^\top \right) &&= -\mathbf{R} \\
    \label{eq:partial-of-concat-1}
    &\mathcal{D}_{\mathbf{R}_1} (\mathbf{R}_1 \mathbf{R}_2) &&= \mathbf{R}_2^\top \\
    \label{eq:partial-of-concat-2}
    &\mathcal{D}_{\mathbf{R}_2} (\mathbf{R}_1 \mathbf{R}_2) &&= \mathbf{I}
\end{alignat}
These expressions capture how small perturbations in rotations propagate through rotation, inversion, and composition.

The Jacobian of the exponential map describes how a perturbation in the rotation vector affects the resulting rotation. If $\mathbf{R}=\exp({\widehat{\boldsymbol{\theta}}})$, then in left-trivialized form, 
\begin{equation}
\mathbf{R}^\top\delta \mathbf{R} = \left(\boldsymbol{\Gamma}(\boldsymbol{\theta})\delta \boldsymbol{\theta}\right)^\wedge
\end{equation}
where
\begin{equation}
\label{eq:jacobian-of-exp}
    \boldsymbol{\Gamma}(\boldsymbol{\theta})=\mathbf{I} - \frac{1 - \cos{||\boldsymbol{\theta}||}}{||\boldsymbol{\theta}||^2} \widehat{\boldsymbol{\theta}} + \frac{||\boldsymbol{\theta}|| - \sin{||\boldsymbol{\theta}||}}{||\boldsymbol{\theta}||^3} \widehat{\boldsymbol{\theta}}^2
\end{equation}
This is the \emph{right Jacobian} of $SO(3)$ (in the terminology of Sol\`a et al. \cite{solà2021microlietheorystate}), consistent with the right-multiplication (body-frame) definition of boxplus and boxminus. It computes a first-order, on-manifold approximation of the exponential map as:
\begin{equation}
\label{eq:first-order-approx-exponential-map}
    \exp \left((\boldsymbol{\theta}+\Delta \boldsymbol{\theta})^\wedge\right) \approx \exp(\widehat{\boldsymbol{\theta}}) \boxplus (\boldsymbol{\Gamma}(\boldsymbol{\theta}) \Delta\boldsymbol{\theta})
\end{equation}
The inverse of this mapping gives the differential of the logarithmic map, expressed in closed-form as:
\begin{equation}
\label{eq:jacobian-of-log}
    \boldsymbol{\Gamma}^{-1}(\boldsymbol{\theta})= \mathbf{I} + \frac{1}{2}\widehat{\boldsymbol{\theta}} + \left(\frac{1}{||\boldsymbol{\theta}||^2}-\frac{1+\cos||\boldsymbol{\theta}||}{2||\boldsymbol{\theta}||\sin||\boldsymbol{\theta}||}\right) \widehat{\boldsymbol{\theta}}^2
\end{equation}
Using these definitions, a useful Jacobian that will come up frequently is that of $\mathbf{R}_1 \boxminus \mathbf{R}_2$ (see Appendix for derivation):
\begin{equation}
    \label{eq:partial-of-box-minus}
    \begin{aligned}
        \mathcal{D}_{\mathbf{R}_1} (\mathbf{R}_1 \boxminus \mathbf{R}_2) &= \boldsymbol{\Gamma}^{-1}(\mathbf{R}_1 \boxminus \mathbf{R}_2) \\
        \mathcal{D}_{\mathbf{R}_2} (\mathbf{R}_1 \boxminus \mathbf{R}_2) &= -\boldsymbol{\Gamma}^{-1}(\mathbf{R}_1 \boxminus \mathbf{R}_2)^\top
    \end{aligned}
\end{equation}

We can chain these Jacobian blocks together according to the chain rule, which greatly simplifies the derivation of function gradients in the developments that follow. Note that the formulas \eqref{eq:partial-of-mapping}-\eqref{eq:jacobian-of-log} are specifically for the definition of differentiation given in \eqref{eq:R-to-SO3-differential} and \eqref{eq:SO3-to-R-differential} and the rotation matrix convention defined in Section \ref{sec:so3}, and may differ if different conventions are used (see e.g., \cite{solà2021microlietheorystate, bloesch2016primerdifferentialcalculus3d}).

\subsection{Quaternions}
\label{sec:quaternions}
A popular way of representing 3D rotations in graphics is with unit \emph{quaternions}, which is a 4D number system with one real part and three imaginary parts: $\mathbf{q} = a + bi + cj + dk$.  
Unit quaternions are closely related to the axis-angle rotation representation, taking the form:
\begin{equation}
\label{eq:axis-angle-quaternion}
    \mathbf{q} = \cos \left( \frac{\theta}{2} \right) + \mathbf{u} \sin\left( \frac{\theta}{2} \right)
\end{equation}
where $\mathbf{u}=iu_x + ju_y + ku_z$ is the axis of rotation and $\theta$ is the rotation angle. A 3D vector $\mathbf{x}=ix + jy + kz$ is rotated about the unit axis $\mathbf{u}$ by an angle $\theta$ with the quaternion product $\mathbf{x}'=\mathbf{q} \mathbf{x} \mathbf{q}^*$, where $\mathbf{q}^*=a-bi-cj-dk$ is the \emph{conjugate} of $\mathbf{q}$. The conjugate $\mathbf{q}^*$ is the multiplicative inverse of $\mathbf{q}$, that is, $\mathbf{q} \mathbf{q}^* = \mathbf{q}^* \mathbf{q} = 1$.

Unit quaternions belong to $S^3$, a Lie group that is the unit 3-sphere in 4D space. The group operation is multiplication, with identity element $1$ and inverse $\mathbf{q}^*$. Since $S^3$ is a Lie group, we can work with unit quaternions very similarly to how we work with rotation matrices in $SO(3)$. Like before, we work with tangent spaces $T_\mathbf{q} S^3$ to describe how quaternions change, and for convenience we wish to express everything in the tangent space at the identity (the Lie algebra $\mathfrak{s}^3$). Using left and right-multiplication, there are again two standard ways to express $\dot{\mathbf{q}} \in T_\mathbf{q}S^3$ in $\mathfrak{s}^3$:
\begin{equation}
\begin{alignedat}{2}
\text{Left-trivialization (body frame):} \quad &\boldsymbol{\phi}_b = \mathbf{q}^* \dot{\mathbf{q}}, \\
\text{Right-trivialization (spatial frame):} \quad &\boldsymbol{\phi}_s = \dot{\mathbf{q}}\mathbf{q}^*
\end{alignedat}
\end{equation}
which correspond to expressing angular velocity in the body and spatial frames, respectively.
Differentiating the constraint $\mathbf{q}^*\mathbf{q}=1$ yields $\mathbf{q}^*\dot{\mathbf{q}} = -\dot{\mathbf{q}}\mathbf{q}^*$ which indicates that both $\boldsymbol{\phi}_b$ and $\boldsymbol{\phi}_s$ have no real component (so-called \textit{pure} quaternions). Therefore, the Lie algebra $\mathfrak{s}^3$ is a vector space spanned by all pure quaternions, which is trivially isomorphic to $\mathbb{R}^3$. The exponential map $\exp : \mathfrak{s}^3 \to S^3$ can be derived as \cite{solà2021microlietheorystate}:
\begin{equation}
    \exp(\boldsymbol{\phi}) = \cos(||{\boldsymbol{\phi}}||) + \frac{{\boldsymbol{\phi}}}{||{\boldsymbol{\phi}}||} \sin(||{\boldsymbol{\phi}}||)
\end{equation}
Comparing this formula with \eqref{eq:axis-angle-quaternion} motivates the definition of
the hat and vee maps between $\mathfrak{s}^3$ and $\mathbb{R}^3$ to be:
\begin{equation}
\begin{alignedat}{2}
    ^\wedge :\  &\mathbb{R}^3 \to \mathfrak{s}^3 \quad \quad \quad \quad^\vee : \ &&\mathfrak{s}^3 \to \mathbb{R}^3\\
    &\boldsymbol{\theta} \mapsto \boldsymbol{\theta}/2 &&\boldsymbol{\phi} \mapsto 2\boldsymbol{\phi}
    \end{alignedat}
\end{equation}
The mapped vector $\boldsymbol{\theta}=\boldsymbol{\phi}^\vee$ is the corresponding exponential rotation vector, where $||\boldsymbol{\theta}||$ is the rotation angle and $\boldsymbol{\theta}/||\boldsymbol{\theta}||$ is the unit rotation axis. Thus, the axis-angle relationship given by \eqref{eq:axis-angle-quaternion} can be viewed as the exponential map with the hat map baked into it. Furthermore, the exponential rotation vector can be used to define an equivalent quaternion and rotation matrix. Specifically, using their respective exponential map definitions, the quaternion $\mathbf{q}_{\boldsymbol{\theta}}=\exp(\widehat{\boldsymbol{\theta}})$ and the rotation matrix $\mathbf{R}_{\boldsymbol{\theta}}=\exp(\widehat{\boldsymbol{\theta}})$ correspond to the same rotation: $\mathbf{q}_{\boldsymbol{\theta}} \mathbf{x} \mathbf{q}_{\boldsymbol{\theta}}^* = \mathbf{R}_{\boldsymbol{\theta}} \mathbf{x}$.

Analogously to $SO(3)$ rotation matrices, we can define the $\boxplus$ and $\boxminus$ operators using the same definition as in \eqref{eq:so3-boxplus} and \eqref{eq:so3-boxminus}. By the definitions above, and using $\mathcal{R}(\mathbf{q})$ to denote the $SO(3)$ rotation matrix corresponding to the quaternion $\mathbf{q}$:
\begin{equation}
\label{eq:box-equivalence}
    \mathbf{R} \boxplus \boldsymbol{\theta} = \mathcal{R}(\mathbf{q} \boxplus \boldsymbol{\theta}), \quad  \quad \mathbf{q}_1 \boxminus \mathbf{q}_2 = \mathcal{R}(\mathbf{q}_1) \boxminus \mathcal{R}(\mathbf{q}_2)
\end{equation}
Then, we can define a differentiation operator $\mathcal{D}$ in the same way as for $SO(3)$ rotation matrices in \eqref{eq:R-to-SO3-differential}, \eqref{eq:SO3-to-R-differential}, and \eqref{eq:SO3-to-SO3-differential}. Under the equivalence shown in \eqref{eq:box-equivalence}, for two functions $f$ and $g$ defined such that $f(\mathcal{R}(\mathbf{q})) = g(\mathbf{q})$:
\begin{equation}
    \mathcal{D}_{\mathcal{R}(\mathbf{q})}(f(\mathcal{R}(\mathbf{q})) = \mathcal{D}_\mathbf{q}( g(\mathbf{q}))
\end{equation}

Therefore, the Jacobian blocks presented in Section \ref{sec:jacobian-blocks} are also valid for the quaternion representation of 3D rotations.
In the developments that follow, we use $SO(3)$ rotation matrices to represent 3D rotations, though a unit quaternion representation may be used instead with the associated $\boxplus$ and $\boxminus$ operators.

\section{Extended Position-Based Dynamics}
This section introduces Extended Position-Based Dynamics (XPBD) as originally described by Macklin et al. in \cite{macklin2016xpbd}. XPBD is an efficient method for simulating constrained particle systems governed by a potential energy function, though its original formulation is designed to work only with particle positions, not orientations. The rest of this section will show how to augment XPBD to handle oriented particles using Lie theory to respect the manifold structure of 3D rotations.

\subsection{XPBD with Positions Only}
\label{sec:XPBD-positions-only}
As described in \cite{macklin2016xpbd}, the derivation of the XPBD equations begins with a backward-Euler discretization of Newton's second law with forces derived from an energy potential (superscript $n$ denotes the backward Euler iteration):
\begin{equation}
    \mathbf{M} \left( \frac{\mathbf{x}^\mathrm{n+1}-2\mathbf{x}^\mathrm{n}+\mathbf{x}^\mathrm{n-1}}{\Delta t^2} \right) = -\nabla U(\mathbf{x}^{n+1})^\top+ \mathbf{f}_{\mathrm{ext}}
\end{equation}
where $\mathbf{x}$ is the system state (i.e. the particle positions), $\mathbf{M}$ is a diagonal nodal mass matrix, $U$ is the potential energy of the system, and $\mathbf{f}^\mathrm{ext}$ are external applied forces. The energy potential $U$ can be expressed quadratically as a function of a "constraint" vector, $\mathbf{C}(\mathbf{x})$ such that
\begin{equation}
\label{eq:xpbd-potential}
    U(\mathbf{x}) = \frac{1}{2} \mathbf{C}(\mathbf{x})^\top \boldsymbol{\alpha}^{-1}\mathbf{C}(\mathbf{x})
\end{equation}
where $\boldsymbol{\alpha}$ is a diagonal compliance (i.e.inverse stiffness) matrix. Note that $\alpha=0$ corresponds to a hard constraint, and thus by \eqref{eq:xpbd-potential} has infinite energy associated with its violation. Lagrange multipliers $\boldsymbol{\lambda} = -\boldsymbol{\tilde\alpha}^{-1} \mathbf{C}(\mathbf{x})$ are introduced, where $\boldsymbol{\tilde\alpha}=\boldsymbol{\alpha}/\Delta t^2$, resulting in the following equations of motion:
\begin{align}
\mathbf{M} (\mathbf{x}^{n+1} - \tilde{\mathbf{x}}) - \nabla \mathbf{C}(\mathbf{x}^{n+1})^\top \boldsymbol{\lambda}^{n+1} 
&= 0, \label{eq:dynamic-equation} \\
\mathbf{C}(\mathbf{x}^{n+1}) + \boldsymbol{\tilde{\alpha}} \boldsymbol{\lambda}^{n+1} 
&= 0, \label{eq:constraint-equation}
\end{align}
where $\tilde{\mathbf{x}} = \mathbf{x}^n + \Delta t \mathbf{v}^n + \Delta t^2 \mathbf{M}^{-1} \mathbf{f}_{\mathrm{ext}}$ is the inertially predicted position in the absence of elastic or constraint forces. Linearizing Equations \eqref{eq:dynamic-equation}
and \eqref{eq:constraint-equation} about the $k^\mathrm{th}$ iteration variables $\mathbf{x}^k$ and $\boldsymbol{\lambda}^k$, and neglecting the constraint Hessian and residual of \eqref{eq:dynamic-equation} results in the following Newton fixed-point iteration that will solve for $\mathbf{x}^{n+1}$ and $\boldsymbol{\lambda}^{n+1}$:
\begin{align}
    &\left[ \nabla \mathbf{C}(\mathbf{x}^k) \mathbf{M}^{-1} 
    \nabla \mathbf{C}(\mathbf{x}^k)^\top + \boldsymbol{\tilde{\alpha}} \right] 
    \Delta \boldsymbol{\lambda}^k = -\mathbf{C}(\mathbf{x}^k) 
    - \boldsymbol{\tilde{\alpha}} \boldsymbol{\lambda}^k
\label{eq:lambda-update}
\end{align}
\begin{equation}
\Delta\mathbf{x}^k=\mathbf{M}^{-1}\nabla\mathbf{C}(\mathbf{x}^k)^\top\Delta\boldsymbol{\lambda}^k
\label{eq:position-update}
\end{equation}
with initial iterates $\mathbf{x}_0=\tilde{\mathbf{x}}$ and $\boldsymbol{\lambda}_0=\mathbf{0}$. Rather than solve the above equations globally, Macklin et al. \cite{macklin2016xpbd} employ a projected Gauss-Seidel update strategy which considers only a single constraint with index \(j\):
\begin{equation}
    \Delta\lambda_j=\frac{-C_j(\mathbf{x}^k) - \tilde\alpha_j\lambda_{j}^k}{\nabla C_j(\mathbf{x}^k)\mathbf{M}^{-1}\nabla C_j(\mathbf{x}^k)^\top + \tilde\alpha_j}
\label{eq:single-constraint-lambda-update}
\end{equation}
After each computation of $\Delta \lambda_j$, the corresponding $\Delta\mathbf{x}$ is computed and both $\mathbf{x}$ and $\boldsymbol{\lambda}$ are updated. One Gauss-Seidel iteration consists of iterating through all constraints.

\subsection{XPBD with Positions and Orientation}
\label{sec:XPBD-with-orientation}
We can extend the XPBD algorithm described in the previous section to accommodate particles with both position and orientation. The new system state $\boldsymbol{x} = (\mathbf{p}_1, \mathbf{R}_1, \dots , \mathbf{p}_N, \mathbf{R}_N)$ is the collection of global particle poitions $\mathbf{p}_i$ and particle orientations $\mathbf{R}_i$, where $N$ is the number of particles in the system.

Following the derivation steps in the previous section, we begin by discretizing a Newtonian potential system in time. As in standard XPBD, the positional degrees of freedom are governed by Newton's second law with a backward Euler discretization.  For particle $i$ this is:
\begin{equation}
    m_i \mathbf{I}_{3\times3} \left( \frac{\mathbf{p}_i^{n+1} - 2\mathbf{p}_i^n + \mathbf{p}_i^{n-1}}{\Delta t^2} \right) = -\frac{\partial}{\partial \mathbf{p}_i} U(\boldsymbol{x}^{n+1})^\top + \mathbf{f}_i^{\mathrm{ext}}
\end{equation}
The term $-\frac{\partial}{\partial \mathbf{p}_i} U^\top$ is the force on particle $i$ resulting from the potential energy function $U$. This can be confirmed through a virtual work argument. Similarly, it can be shown (see Appendix \ref{sec:foces-torques-derived-from-energy}) that the body torque on particle $i$ resulting from the potential energy function is $-\mathcal{D}_{\mathbf{R}_i} U^\top$.
Combined with Euler's rigid body rotation equations, the rotational degrees of freedom for particle $i$ are governed by:
\begin{equation}
\label{eq:rotational-dof}
    \boldsymbol{\mathcal{I}}_i \left( \frac{\boldsymbol{\omega}_i^{n+1} - \boldsymbol{\omega}_i^{n}}{\Delta t}\right) = -\mathcal{D}_{\mathbf{R}_i} U(\boldsymbol{x}^{n+1})^\top + \boldsymbol{\tau}_i^{\mathrm{ext}} - \boldsymbol{\omega}_i^n \times \boldsymbol{\mathcal{I}_i} \boldsymbol{\omega}_i^n
\end{equation}
where $\boldsymbol{\omega}_i$ is the body-frame angular velocity of particle $i$, $\boldsymbol{\mathcal{I}}_i$ is the body-frame rotational inertia tensor for particle $i$, and $\boldsymbol{\tau}_i^{\mathrm{ext}}$ is the external applied torque on particle $i$, expressed in the body frame. Here, a semi-implicit Euler time-discretization (the gyroscopic cross-product term is evaluated at the old time step) is used, allowing us to lump the gyroscopic cross-product into the inertial update.

Note that Euler's rotation equations are defined at the angular velocity level, but our system state involves the rotation matrices themselves. From a finite-difference approximation of \eqref{eq:R-to-SO3-differential}, the body angular velocity
$\boldsymbol{\omega}_i^{n+1}$ can be written in terms of $\mathbf{R}_i^{n+1}$ and $\mathbf{R}_i^{n}$ as:
\begin{equation}
\label{eq:omega-n+1}
    \boldsymbol{\omega}_i^{n+1} \approx \frac{\mathbf{R}_i^{n+1} \boxminus \mathbf{R}_i^n}{\Delta t} 
\end{equation}
As shown in Appendix \ref{sec:derivation-of-rotational-inertial-update}, under the assumption that $\Delta t$ (and therefore also the change in rotation over the time step) is sufficiently small, we can substitute in \eqref{eq:omega-n+1} and lump known quantities together \eqref{eq:rotational-dof} to obtain:
\begin{equation}
\label{eq:particle-i-rotation-eom}
    \boldsymbol{\mathcal{I}}_i \left( \mathbf{R}_i^{n+1} \boxminus \tilde{\mathbf{R}}_i^{n} \right) = -\Delta t^2 \mathcal{D}_{\mathbf{R}_i} U(\boldsymbol{x}^{n+1})^\top
\end{equation}
where
\begin{equation}
\label{eq:rotation-inertial-update}
    \tilde{\mathbf{R}}_i^n = \mathbf{R}_i^n \boxplus \left( \Delta t \boldsymbol{\omega}_i^n + \Delta t^2 \boldsymbol{\mathcal{I}}_i^{-1} (\boldsymbol{\tau}_i^{\mathrm{ext}} - \boldsymbol{\omega}_i^n \times \boldsymbol{\mathcal{I}}_i \boldsymbol{\omega}_i^n ) \right)
\end{equation}
is the inertially predicted orientation for particle $i$ in the absence of constraint forces. 
Observe that this has the exact same form as the positional equations of motion (which are the same as in Section \ref{sec:XPBD-positions-only}):
\begin{equation}
\label{eq:particle-i-position-eom}
    m_i \mathbf{I}_{3\times3} \left( \mathbf{p}_i^{n+1} - \tilde{\mathbf{p}}_i^n \right) = -\Delta t^2 \frac{\partial}{\partial \mathbf{p}_i} U(\boldsymbol{x}^{n+1})^\top
\end{equation}
where
\begin{equation}
\label{eq:position-inertial-update}
    \tilde{\mathbf{p}}_i^n = \mathbf{p}_i^n +\Delta t \mathbf{v}_i^n + \Delta t^2 \frac{1}{m_i}\mathbf{f}_i^{\mathrm{ext}}
\end{equation}
is the inertially predicted position for particle $i$ in the absence of constraint forces.

If we define $\Delta \tilde{\boldsymbol{x}}^{n+1}$ as the following:
\begin{equation}
\label{eq:dx-tilde}
\Delta \tilde{\boldsymbol{x}}^{n+1} = \begin{bmatrix}
    \vdots \\
    \mathbf{p}_i^{n+1} - \tilde{\mathbf{p}}_i^n \\
    \mathbf{R}_i^{n+1} \boxminus \tilde{\mathbf{R}}_i^n \\
    \vdots
\end{bmatrix}
\end{equation}
then we can combine equations \eqref{eq:particle-i-rotation-eom} and \eqref{eq:particle-i-position-eom} into one nonlinear system for all particles:
\begin{equation}
\label{eq:xpbd-with-orientation-eom}
\boldsymbol{\mathcal{M}} \Delta \tilde{\boldsymbol{x}}^{n+1} = -\Delta t^2 \nabla U(\boldsymbol{x}^{n+1})^\top
\end{equation}
where $\boldsymbol{\mathcal{M}}=\mathrm{diag}(m_1 \mathbf{I}_{3\times3},\  \boldsymbol{\mathcal{I}}_1, \dots , m_N \mathbf{I}_{3\times3},\  \boldsymbol{\mathcal{I}}_N)$ is the block-diagonal system inertia matrix. 
Since all the rotational inertia tensors $\boldsymbol{\mathcal{I}}_i$ are defined in the body frame of their respective particles, $\boldsymbol{\mathcal{M}}$ is constant throughout the simulation (and diagonal if for each body the body-attached coordinate system aligns with the principal axes). Note that the $\nabla$ symbol is used to represent a generalized derivative that contains both $\partial_{\mathbf{p}_i}$ and $\mathcal{D}_{\mathbf{R}_i}$ sub-blocks, see Section \ref{sec:constraint-gradients}.

As before, we introduce Lagrange multipliers $\boldsymbol{\lambda} = -\tilde{\boldsymbol{\alpha}}^{-1} \mathbf{C}(\boldsymbol{x})$ such that \eqref{eq:xpbd-with-orientation-eom} becomes:
\begin{align}
\label{eq:position-orientation-primary-eq}
\boldsymbol{\mathcal{M}} \Delta \tilde{\boldsymbol{x}}^{n+1} - \nabla \mathbf{C}(\boldsymbol{x}^{n+1})^\top \boldsymbol{\lambda}^{n+1} &= 0 \\
\label{eq:position-orientation-constraint-eq}
\mathbf{C}(\boldsymbol{x}^{n+1}) + \tilde{\boldsymbol{\alpha}} \boldsymbol{\lambda}^{n+1} &= 0
\end{align}
Then, linearizing about the $k^{\mathrm{th}}$ iteration state $\boldsymbol{x}^k$ and Lagrange multipliers $\boldsymbol{\lambda}^k$, approximating the partial of \eqref{eq:position-orientation-primary-eq} with respect to $\boldsymbol{x}^k$ as $\boldsymbol{\mathcal{M}}$ and neglecting the residual of \eqref{eq:position-orientation-primary-eq} results in the following Newton subproblem:
\begin{equation}
\label{eq:position-orientation-newton-subproblem}
    \begin{bmatrix}
        \boldsymbol{\mathcal{M}} & -\nabla \mathbf{C}(\boldsymbol{x}^k)^\top \\
        \nabla \mathbf{C}(\boldsymbol{x}^k)^\top & \tilde{\boldsymbol{\alpha}}
    \end{bmatrix}
    \begin{bmatrix}
        \Delta \boldsymbol{x}^k \\
        \Delta \boldsymbol{\lambda}^k
    \end{bmatrix}
    =
    \begin{bmatrix}
        \boldsymbol{0} \\
        -\mathbf{C}(\boldsymbol{x}^k) - \tilde{\boldsymbol{\alpha}}\boldsymbol{\lambda}_i
    \end{bmatrix}
\end{equation}
where 
\begin{equation}
\Delta \boldsymbol{x}^k = \begin{bmatrix}
    \vdots \\
    \mathbf{p}_i^{k+1}-\mathbf{p}_i^k \\
    \mathbf{R}_i^{k+1} \boxminus \mathbf{R}_i^k \\
    \vdots
\end{bmatrix}
\end{equation}
with initial iterates $\boldsymbol{x}^{(0)}= (\tilde{\mathbf{p}}_1, \ \tilde{\mathbf{R}}_1, \dots)$ and $\boldsymbol{\lambda}^{(0)}=\mathbf{0}$. We can solve for $\Delta \boldsymbol{\lambda}$ as before:
\begin{equation}
\label{eq:position-orientation-lambda-update}
\left[ \nabla \mathbf{C}(\boldsymbol{x}^k) \boldsymbol{\mathcal{M}}^{-1} 
    \nabla \mathbf{C}(\boldsymbol{x}^k)^\top + \boldsymbol{\tilde{\alpha}} \right] 
    \Delta \boldsymbol{\lambda}^k = -\mathbf{C}(\boldsymbol{x}^k) 
    - \boldsymbol{\tilde{\alpha}} \boldsymbol{\lambda}^k
\end{equation}
and use an iterative Gauss-Seidel solution strategy of the same form as \eqref{eq:single-constraint-lambda-update} to solve for each $\Delta \lambda_j^k$ individually on a constraint-by-constraint basis:
\begin{equation}
    \label{eq:position-orientation-single-lambda-update}
    \Delta\lambda_j=\frac{-C_j(\boldsymbol{x}^k) - \tilde\alpha_j\lambda_{j}^k}{\nabla C_j(\boldsymbol{x}^k)\boldsymbol{\mathcal{M}}^{-1}\nabla C_j(\boldsymbol{x}^k)^\top + \tilde\alpha_j}
\end{equation}
The position and orientation update can be computed analogously to before as:
\begin{equation}
    \label{eq:position-orientation-position-update}
    \Delta \boldsymbol{x}^k = \boldsymbol{\mathcal{M}}^{-1} \nabla \mathbf{C}(\boldsymbol{x}^k)^\top \Delta \boldsymbol{\lambda}^k
\end{equation}
Then, we can apply these updates to compute the latest estimates of $\mathbf{p}_i$ and $\mathbf{R}_i$ with
\begin{equation}
\label{eq:position-orientation-application-of-update}
    \begin{bmatrix}
        \vdots \\
        \mathbf{p}_i^{k+1} \\
        \mathbf{R}_i^{k+1} \\
        \vdots
    \end{bmatrix} = \begin{bmatrix}
        \vdots \\
        \mathbf{p}_i^{k} + \Delta \mathbf{p}_i^k \\
        \mathbf{R}_i^{k} \boxplus \Delta \mathbf{R}_i^k\\
        \vdots
    \end{bmatrix}
\end{equation}


We emphasize that extending XPBD to handle particles with rotational DOF requires very few modifications to the original algorithm. The update formulas for XPBD with orientation \eqref{eq:position-orientation-single-lambda-update}, \eqref{eq:position-orientation-position-update} and the update formulas for normal XPBD \eqref{eq:single-constraint-lambda-update}, \eqref{eq:position-update} are identical. The key differences lie in the computation of the inertially predicted positions \eqref{eq:rotation-inertial-update}, \eqref{eq:position-inertial-update}, and the application of position updates \eqref{eq:position-orientation-application-of-update}, which require the use of the $\boxplus$ operator for rotational DOF to stay on the $SO(3)$ manifold.

\subsection{Constraint Gradients with Rotational DOF}
\label{sec:constraint-gradients}
In the previous section, the constraint gradient $\nabla \mathbf{C}$ is required to calculate both the Lagrange multiplier updates \eqref{eq:position-orientation-single-lambda-update} and the position and orientation updates \eqref{eq:position-orientation-position-update}. If we perturb each particle with a positional perturbation $\Delta \mathbf{p}_i$ and a body-frame rotational perturbation $\Delta \boldsymbol{\theta}$, the constraint gradient maps the total perturbation $\Delta\boldsymbol{x}=[\Delta\mathbf{p}_1^\top \ \Delta\boldsymbol{\theta}_1^\top \  \cdots \  \Delta\mathbf{p}_N^\top \  \Delta\boldsymbol{\theta}_N^\top]^\top$ to a change in the constraint vector $\mathbf{C}$. The constraint gradient can therefore be broken into the following sub-blocks:
\begin{equation}
    \nabla \mathbf{C}(\boldsymbol{x}) =\begin{bmatrix}
        \frac{\partial C_1}{\partial \mathbf{p}_1} & \mathcal{D}_{\mathbf{R}_1}C_1 & \cdots & \frac{\partial C_1}{\partial \mathbf{p}_N} & \mathcal{D}_{\mathbf{R}_N}C_1  \\
        \vdots & \vdots & & \vdots & \vdots \\
        \frac{\partial C_M}{\partial \mathbf{p}_1} & \mathcal{D}_{\mathbf{R}_1}C_M  & \cdots & \frac{\partial C_M}{\partial \mathbf{p}_N} & \mathcal{D}_{\mathbf{R}_N}C_M
    \end{bmatrix}
\end{equation}
Each block of the constraint gradient $\frac{\partial C_j}{\partial \mathbf{p}_i}$ and $\mathcal{D}_{\mathbf{R}_i} C_j$ are $1\times 3$ row vectors (see \eqref{eq:SO3-to-R-differential}) and are implicitly functions of the current state $\boldsymbol{x}$. The entire constraint gradient is therefore a matrix of size $M \times 6N$, where $M$ is the number of constraints and $N$ is the number of particles. 

\subsection{Velocity-Level Damping Solve}
\label{sec:velocity-damping-solve}
In the original XPBD paper, damping forces are simultaneously combined with elastic forces at the position level \cite{macklin2016xpbd}. While elegant, this greatly decreases the convergence and stability of XPBD \cite{tobin2025efficient}. Instead, we follow \cite{muller2020detailed} and apply damping at the velocity level, after the position updates for the time step have already been computed. We start with the Rayleigh dissipation potential \cite{macklin2016xpbd}:
\begin{equation}
\label{eq:dissipation-potential}
\begin{aligned}
    U_d &= \frac{1}{2} \dot{\mathbf{C}}(\mathbf{x})^\top \boldsymbol{\beta} \dot{\mathbf{C}}(\mathbf{x}) \\
    &= \frac{1}{2} \boldsymbol{v}^\top \nabla \mathbf{C}^\top \boldsymbol{\beta} \nabla \mathbf{C} \boldsymbol{v}
\end{aligned}
\end{equation}
where $\boldsymbol{\beta}$ is a diagonal damping matrix and $\boldsymbol{v}=[\mathbf{v}_1^\top \ \boldsymbol{\omega}_1^\top \ \cdots \ \mathbf{v}_N^\top \ \boldsymbol{\omega}_N^\top]^\top$ is a concatenated velocity vector for all particles. We define the damping in terms of the constraint compliances $\boldsymbol{\beta} = b \boldsymbol{\alpha}^{-1}$, where $b$ is a single scalar, so that all constraints have the same damping ratio regardless of stiffness. The force derived from the dissipation potential in \eqref{eq:dissipation-potential} is given by the partial derivative with respect to velocity:
\begin{equation}
    \mathbf{f}_d = -\frac{\partial}{\partial \boldsymbol{v}} U_d = -\nabla \mathbf{C}^\top \boldsymbol{\beta} \nabla \mathbf{C} \boldsymbol{v}
\end{equation}
Thus, from a backwards Euler discretization we have that:
\begin{equation}
    \boldsymbol{\mathcal{M}} (\boldsymbol{v}^{n+1} - \boldsymbol{v}^n) = -\Delta t \nabla \mathbf{C}^\top \boldsymbol{\beta} \nabla \mathbf{C} \boldsymbol{v}^{n+1}
\end{equation}
Following the same procedure as before, we define velocity-level Lagrange multipliers $\boldsymbol{\mu} = -\tilde{\boldsymbol{\beta}} \nabla \mathbf{C} \boldsymbol{v}$ where $\tilde{\boldsymbol{\beta}} = \Delta t \boldsymbol{\beta}$, leading to the following nonlinear system:
\begin{align}
    \boldsymbol{\mathcal{M}} (\boldsymbol{v}^{n+1} - \boldsymbol{v}^n) - \nabla \mathbf{C}^\top \boldsymbol{\mu}^{n+1} &= 0 \\
    \nabla \mathbf{C}^\top \boldsymbol{v}^{n+1} + \tilde{\boldsymbol{\beta}}^{-1} \boldsymbol{\mu}^{n+1} &= 0
\end{align}
This has exactly the same structure as \eqref{eq:position-orientation-primary-eq} and \eqref{eq:position-orientation-constraint-eq}, so we follow the same XPBD linearization procedure as before to obtain:
\begin{align}
\label{eq:velocity-constraint-solve}
    \left[\nabla \mathbf{C} \boldsymbol{\mathcal{M}} \nabla \mathbf{C}^\top + \tilde{\boldsymbol{\beta}}^{-1} \right] \Delta \boldsymbol{\mu}^k &= -\nabla \mathbf{C} \boldsymbol{v}^{k} - \tilde{\boldsymbol{\beta}}^{-1} \boldsymbol{\mu}^k \\
    \Delta \boldsymbol{v}^k &= \boldsymbol{\mathcal{M}}^{-1} \nabla \mathbf{C}^\top \Delta \boldsymbol{\mu}^k
\end{align}
The system for $\Delta \boldsymbol{\mu}^k$ in \eqref{eq:velocity-constraint-solve} can be solved constraint-by-constraint analogous to \eqref{eq:single-constraint-lambda-update} and \eqref{eq:position-orientation-single-lambda-update}. Note that since the constraints do not depend on velocity, the constraint gradients only need to be computed once for the entire velocity solve. Additionally, constraints that are not damped are skipped, as there is nothing to be done. Furthermore, hard constraints ($\alpha=0$) are also skipped, because $\dot{C}=0$ always for constraints that are strictly enforced.

\section{XPBD for Rigid Bodies}
Each ``particle'' in the constrained particle system in \eqref{eq:position-orientation-primary-eq} and \eqref{eq:position-orientation-constraint-eq} can be considered as a rigid body, so if we are able to define the appropriate constraint functions and their gradients, we can efficiently model and simulate constrained rigid body dynamics using XPBD. Gradients of the constraints presented in this section can be found in the Appendix.

\subsection{Scalar vs. Vector Constraints}
In XPBD, there is some freedom in defining the constraint functions: an infinite number of constraint functions can represent the same physical constraint. For example, take the simple case where we want to constrain two particles to have the same position. Letting $\mathbf{p}_1$ and $\mathbf{p}_2$ denote the particle positions, we can think of the constraint as enforcing the distance between the particles to be 0:
\begin{equation}
    C_\mathrm{dist} = || \mathbf{p}_1 - \mathbf{p}_2 ||
\end{equation}
Alternatively, we can think of the constraint as enforcing the difference in each of the coordinates to be 0, leading to the following vector of constraints:
\begin{equation}
    \mathbf{C}_\mathrm{coord} = \mathbf{p}_1 - \mathbf{p}_2
\end{equation}

\begin{figure}
    \centering
    \includegraphics[width=1\linewidth]{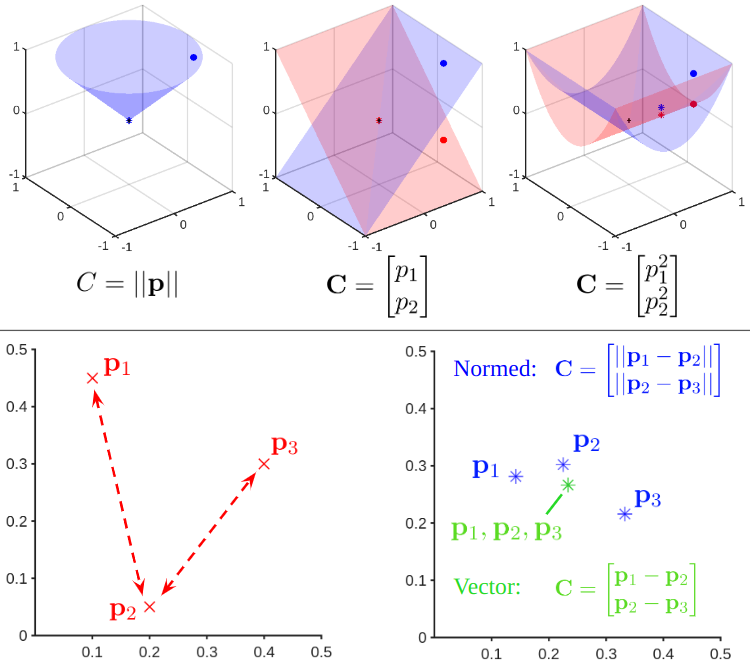}
    \caption{Constraint definitions affect XPBD solver accuracy. Solid circles indicated initial positions, and asterisks indicate positions after a single XPBD update using a global solve of \eqref{eq:position-orientation-lambda-update}. All constraints have $\alpha=0$. \textbf{Top.} Constraining a 2D point to $\mathbf{0}$. Different constraint definitions are used, indicated below their respective figures: normed constraint ({Top Left}), vector constraint ({Top Middle}), and nonlinear vector constraint ({Top Right}). The nonlinear vector constraint fails to move the point to (0,0) in a single XPBD update as a result of its nonlinearity. The normed vector constraint succeeds (despite being nonlinear) due to its linearity in the radial direction. \textbf{Bottom.} Simple 3-particle system in 2D. Constraints are defined so that the distance between $\mathbf{p}_1$ and $\mathbf{p}_2$ is $\mathbf{0}$, and the distance between $\mathbf{p}_2$ and $\mathbf{p}_3$ is $\mathbf{0}$. After a single XPBD update, the particles are moved to the locations on the bottom right. The blue asterisks are the resulting positions using normed constraints, and the green asterisks are the resulting positions using vector constraints. Due to the linearity of the vector constraints, a single XPBD update results in exact alignment, which is not the case for the normed constraints.}
    \label{fig:constraint-linearization}
\end{figure}
Both constraints are mathematically valid, so which should we choose?
On one hand, the scalar constraint requires only a single constraint update, compared to the three that are required by the vector constraint. From a computational standpoint, the scalar constraint is then three times faster than the vector constraint. However, the vector constraint is \emph{linear} in the positions, whereas the scalar constraint introduces nonlinearity through the Euclidean norm. This is relevant because when we are solving for the Lagrange multiplier updates $\Delta \boldsymbol{\lambda}$, we are solving the \emph{linearized} Newton subproblem in \eqref{eq:position-orientation-newton-subproblem}. 
Therefore, the more ``linear'' the constraints are, the better the linearization about our current estimates will be, and the more accurate our solution to the nonlinear system will be. 

The benefit of constraint linearity is demonstrated in Figure \ref{fig:constraint-linearization}. In the top plots, a point $\mathbf{p}=[0.6,\ -0.4]^\top$ is constrained to $\mathbf{0}$ using three equivalent constraint definitions: a scalar, distance constraint using the Euclidean norm; a vector-valued constraint that is linear in $\mathbf{p}$; and a vector-valued constraint that is nonlinear in $\mathbf{p}$. The constraint surfaces are shown for each constraint, and the plotted circles indicated the initial constraint violation. Then, a single XPBD update is performed involving a global solve of \eqref{eq:lambda-update} for $\Delta \boldsymbol{\lambda}$ and consequently a position update \eqref{eq:position-update}. The new position on the constraint surfaces is indicated with plotted asterisks. We use $\boldsymbol{\alpha}=\mathbf{0}$, so ideally after a single XPBD update, $\mathbf{p}=\mathbf{0}$ and the constraints are exactly satisfied.

A single XPBD update using the nonlinear vector-valued constraints does not bring $\mathbf{p}$ to $\mathbf{0}$, due to the nonlinearity of the constraints themselves. The linearization does not hold globally, and thus the step is not globally exact. But, when the constraints are linear, the step is indeed globally exact, and brings $\mathbf{p}$ to $\mathbf{0}$ exactly, as shown in the top middle plot of Figure \ref{fig:constraint-linearization}. Despite being nonlinear, the scalar distance constraint exactly brings $\mathbf{p}$ to $\mathbf{0}$ after a single XPBD update. This is because the scalar constraint happens to be linear in the position update direction (i.e. it is radially symmetric), so the step happens to be exact. 

This symmetry is not true in general, as evidenced by Figure \ref{fig:constraint-linearization}, bottom, where three particles are constrained such that the distance between them is $\mathbf{0}$. This is achieved with two constraints indicated with the dotted arrows: between $\mathbf{p}_1$ and $\mathbf{p}_2$ and between $\mathbf{p}_2$ and $\mathbf{p}_3$. Using scalar constraints (bottom right, blue), a single XPBD update (with a global solve) does not successfully join the three points, due to the nonlinearity in the constraints. On the other hand, the vector-valued, linear constraints (bottom right, green), succeed in exactly joining the three points after a single XPBD update. 

This experiment demonstrates the downside of unnecessarily nonlinear constraints; the linearization is only valid locally, so in general more iterations are required for convergence. This results in higher residuals, as shown in Section \ref{sec:rigid-body-constraints}. It should be noted that constraint nonlinearity is often natural and unavoidable; the main takeaway here is that we should strive for as ``linear'' of constraints as possible.


In \cite{muller2020detailed}, Muller et al. derive the XPBD update formulas for rigid bodies using impulse-based dynamics, circumventing explicit constraint definitions and constraint gradient computations. Their derivations are consistent with scalar (i.e. normed) versions of the rigid-body joint constraints presented in the following sections, thereby introducing unnecessary nonlinearity and decreasing solver convergence. Additionally, their lack of explicit constraint definitions (1) prevents alternate constraint formulations, such as the vector constraints presented in the following sections and (2) leads to incorrect gradient expressions. An explicit side-by-side comparison of the approach and equations presented by Muller et al. in \cite{muller2020detailed} and the approach detailed below is presented in the Supplemental Material.

\subsection{Fixed Joint}
\begin{figure}
    \centering
    \includegraphics[width=0.9\linewidth]{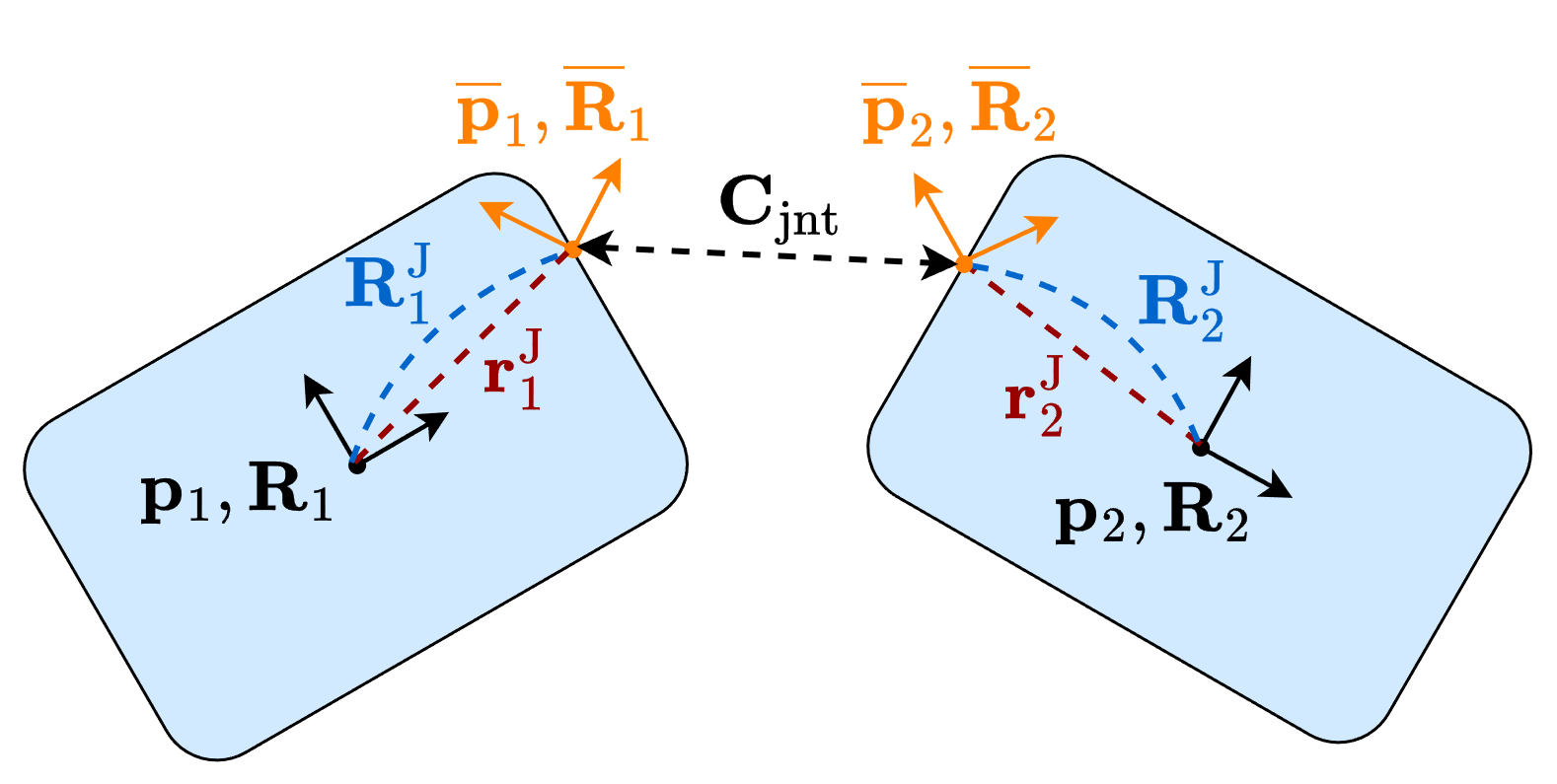}
    \caption{Diagram of joint frame definitions for rigid body constraints.}
    \label{fig:rigid-body-joint-defs}
\end{figure}
\subsubsection{General Joint Frame Definitions}
To define a joint between rigid bodies $1$ and $2$, we need to first define the joint frame on each rigid body that we would like to align. For rigid body $1$, we define a a fixed, body-frame translation vector $\mathbf{r}_1^\mathrm{J}$ from the body-attached rigid body frame to the origin of the joint frame, and a fixed, body-frame rotation $\mathbf{R}_1^\mathrm{J}$ from the body-attached rigid body frame to the orientation of the joint frame. As an equation:
\begin{align}
\label{eq:joint-frame-position}
    \overline{\mathbf{p}}_1 &=\mathbf{p}_1+\mathbf{R}_1\mathbf{r}_1^\mathrm{J} \\
\label{eq:joint-frame-orientation}
    \overline{\mathbf{R}}_1 &=\mathbf{R}_1 \mathbf{R}_1^\mathrm{J}
\end{align} 
where $\overline{\mathbf{p}}_1$ and $\overline{\mathbf{R}}_1$ are the global position and orientation of the joint frame on rigid body $1$, with similar definitions for rigid body $2$. These definitions are illustrated in Figure \ref{fig:rigid-body-joint-defs} Typical rigid body joint constraints are hard constraints, so $\boldsymbol{\alpha}=\mathbf{0}$ is used unless otherwise stated. 

Additionally, we note that ``one-sided'' joints (e.g., joints between a rigid body and the ground) can be accommodated by treating $\mathbf{p}_1$ and $\mathbf{R}_1$ as completely fixed. The resulting joint constraints are only functions of $\mathbf{p}_2$ and $\mathbf{R}_2$, and thus only the gradients with respect to these quantities should be used.

\subsubsection{Vector Fixed Joint Constraint}
For a fixed joint, both the position and orientation of the joint frames on bodies $1$ and $2$ are exactly the same. Thus, we can simply enforce that the positional difference $\overline{\mathbf{p}}_1 - \overline{\mathbf{p}}_2$ and the angular difference $\overline{\mathbf{R}}_1 \boxminus \overline{\mathbf{R}}_2$ is equal to 0:
\begin{equation}
    \label{eq:rigid-body-fixed-constraint}
    \mathbf{C}_\mathrm{fix}=\begin{bmatrix}
        \overline{\mathbf{p}}_1 - \overline{\mathbf{p}}_2 \\
        \overline{\mathbf{R}}_1 \boxminus \overline{\mathbf{R}}_2
    \end{bmatrix}
\end{equation}

\subsection{Revolute Joint}
A revolute joint has a single rotational DOF, allowing body~$2$ to rotate freely about an axis of the joint frame on body $1$. This amounts to an axis in the body $1$ joint frame being aligned with an axis in the body $2$ joint frame. The joint axis is arbitrarily chosen to be $\mathbf{e}_3$ expressed in the joint frame of each body.

\subsubsection{Vector Revolute Joint Constraint}
The positional requirement is satisfied as for a fixed joint, with $\overline{\mathbf{p}}_1 - \overline{\mathbf{p}}_2$. Aligning the two joint axes can be achieved by driving two components of the rotational difference $\overline{\mathbf{R}}_1 \boxminus \overline{\mathbf{R}}_2$ to $\mathbf{0}$ while allowing relative rotation about the joint axis. Thus, the revolute joint constraint is simply:
\begin{equation}
    \mathbf{C}_\mathrm{rev} = \begin{bmatrix}
        \overline{\mathbf{p}}_1 - \overline{\mathbf{p}}_2 \\
        \left[ \overline{\mathbf{R}}_1 \boxminus \overline{\mathbf{R}}_2 \right]_{1-2}
    \end{bmatrix}
\end{equation}
where $[\cdot]_{1-2}$ corresponds to the first two rows of a vector or matrix.
\subsubsection{Revolute Joint Limits}
Since the first two components of $\overline{\mathbf{R}}_1~\boxminus~\overline{\mathbf{R}}_2$ are driven to $\mathbf{0}$, the joint angle $\theta_{\mathrm{rev}}$ is simply the third component:
\begin{equation}
\label{eq:boxminus-theta-rev}
    \theta_\mathrm{rev} = \left[ \overline{\mathbf{R}}_1 \boxminus \overline{\mathbf{R}}_2 \right]_3
\end{equation}
where $[\cdot]_3$ corresponds to the third row of a vector or matrix.

We can impose angle limits on the revolute joint $\theta_\mathrm{rev,min}~\leq~\theta_\mathrm{rev} \leq \theta_\mathrm{rev,max}$ by imposing inequality constraints on the joint angle:
\begin{equation}
\label{eq:revolute-joint-angle-limit-constraints}
\begin{aligned}
    C_\mathrm{rev,min} = \theta_\mathrm{rev}-\theta_\mathrm{rev,min} \geq 0 \\
    C_\mathrm{rev,max} = \theta_\mathrm{rev,max} - \theta_\mathrm{rev} \geq 0
\end{aligned}
\end{equation}
Being inequality constraints, these constraints are only enforced when they are violated, i.e. when $C<0$. 

\subsection{Spherical Joint}
A spherical joint only constrains the positions of the joint frames to be co-located, allowing complete rotational freedom of the two bodies. We define rotation of body $2$ about the z-axis of the joint frame on body $1$ to be ``twist'', and the combined rotation about the x- and y-axes to be ``swing''.

\subsubsection{Vector Spherical Joint Constraint}
As the rotational DOF are not constrained, the appropriate vector constraint in this case is just the positional constraint:
\begin{equation}
    \mathbf{C}_\mathrm{sph} = \overline{\mathbf{p}}_1 - \overline{\mathbf{p}}_2
\end{equation}

\subsubsection{Joint Limits}
Often times, the joint limits for swing are different than the joint limit for twist, such as for a classical ball-in-socket joint. This can easily be enforced by decomposing the rotation between the joint frames, $\mathbf{R}_\Delta=\overline{\mathbf{R}}_1^\top \overline{\mathbf{R}}_2$, into a swing rotation and a twist rotation:
\begin{equation}
\begin{aligned}
    \mathbf{R}_\Delta &= \mathbf{R}_\mathrm{swing} \mathbf{R}_\mathrm{twist}
    \\ &=\mathrm{Exp}(\theta_\mathrm{swing} \widehat{\mathbf{a}}_\mathrm{swing}) \mathbf{R}_z(\theta_{\mathrm{twist}})
\end{aligned}
\end{equation}
where $\theta_\mathrm{swing}$ and $\theta_\mathrm{twist}$ are the swing and twist angles, respectively, $\widehat{\mathbf{a}}_\mathrm{swing}$ is the axis of rotation (in the xy plane) for the swing rotation, and $\mathbf{R}_z(\cdot)$ denotes a rotation about the z-axis.
Because the twist rotation is strictly a rotation about the z-axis, it leaves the z-axis unchanged. Therefore, any change in the z-axis is solely due to the swing rotation. So, the swing angle can be determined by finding the angle between the original z-axis $\mathbf{e}_3$ and the transformed z-axis $\mathbf{R}_\Delta \mathbf{e}_3$:
\begin{equation}
    \theta_\mathrm{swing} = \arccos(\mathbf{e}_3^\top \mathbf{R}_\Delta \mathbf{e}_3)
\end{equation}
By a similar argument, it can be shown that the twist angle can be determined directly from the rotation matrix $\mathbf{R}_\Delta$ as:
\begin{equation}
    \theta_\mathrm{twist} = \mathrm{atan2}(\mathbf{R}_\Delta(2,1)-\mathbf{R}_\Delta(1,2), \mathbf{R}_\Delta(1,1) + \mathbf{R}_\Delta(2,2))
\end{equation}
Swing and twist angle limits can be imposed using inequality constraints as was done with the revolute joint angle limits in \eqref{eq:revolute-joint-angle-limit-constraints}. 

\subsection{Prismatic Joint}
A prismatic joint allows a single sliding DOF along the joint axis. We choose the joint axis to be the z-axis of the body $1$ joint frame.

\subsubsection{Vector Prismatic Joint Constraint}
If the body $2$ joint frame origin is on the joint axis, its position expressed in the body $1$ joint frame should have its first two components equal to 0. And, since we are aligning the orientations of the two joint frames, the full 5D constraint is:
\begin{equation}
\label{eq:rigid-body-prismatic-constraint}
    \mathbf{C}_{\mathrm{pr}} = \begin{bmatrix}
        [\overline{\mathbf{R}}_1^\top (\overline{\mathbf{p}}_2 - \overline{\mathbf{p}}_1)]_{1-2} \\
        \overline{\mathbf{R}}_1 \boxminus \overline{\mathbf{R}}_2
    \end{bmatrix}
\end{equation}

\subsubsection{Joint Limits}
The translation of body $2$ along the joint axis is found by projecting the vector between the joint frame origins onto the joint axis. This amounts to expressing the position difference in the body $1$ joint frame and extracting the z-component:
\begin{equation}
\label{eq:prismatic-joint-translation}
d_\mathrm{pr} = \left[ \overline{\mathbf{R}}_1^\top (\overline{\mathbf{p}}_2 - \overline{\mathbf{p}}_1) \right]_3
\end{equation}
As with the other joints, we can form inequality constraints to enforce translation limits of the same form as \eqref{eq:revolute-joint-angle-limit-constraints}. 

\subsection{Joints with Target Joint Variables}
Some joints are driven to a target joint position via internal springs inside the joint. This can be achieved using a modified fixed joint constraint from \eqref{eq:rigid-body-fixed-constraint}:
\begin{equation}
\label{eq:rigid-body-target-joint-constraint}
\mathbf{C}_\mathrm{target} = \begin{bmatrix}
        \overline{\mathbf{R}}_1^\top(\overline{\mathbf{p}}_2 - \overline{\mathbf{p}}_1) \\
        \overline{\mathbf{R}}_1 \boxminus \overline{\mathbf{R}}_2
    \end{bmatrix} - \begin{bmatrix}
    \mathbf{p}_\mathrm{target} \\
    \boldsymbol{\theta}_\mathrm{target}
\end{bmatrix}
\end{equation}
where $\mathbf{p}_\mathrm{target}$ is the target position offset and $\boldsymbol{\theta}_\mathrm{target}$ is the target rotational offset, expressed as a body-frame rotation vector. The positional part is expressed in the body $1$ joint frame to accommodate prismatic joints. Crucially, some of the constraint compliances are not 0, allowing for some relative motion depending on the corresponding DOF. Specific examples for the joint types described earlier are given below.

\subsubsection{Revolute Joint Constraint}
The target position offset is 0, and $\boldsymbol{\theta}_\mathrm{target}=\left[0\ \ 0\ \ \theta_\mathrm{rev,target} \right]^\top$ where $\theta_\mathrm{rev,target}$ is the target joint angle. The constraint compliances are then $\boldsymbol{\alpha}=\left[0\ \ 0\ \ 0\ \ 0\ \ 0\ \ k^{-1} \right]^\top$ where $k$ is the torsional spring stiffness.

\subsubsection{Spherical Joint Constraint}
The target position offset is 0, and $\boldsymbol{\theta}_\mathrm{target} \neq \mathbf{0}$. The constraint compliances are then $$\boldsymbol{\alpha}=\left[0\ \ 0\ \ 0\ \ k_1^{-1}\ \ k_2^{-1}\ \ k_3^{-1} \right]^\top$$ for spring stiffnesses $k_i$.

\subsubsection{Prismatic Joint Constraint}
Now, the target rotational offset is 0, and $\mathbf{p}_\mathrm{target} = \left[0 \ \ 0 \ \ d_\mathrm{target} \right]^\top$ where $d_\mathrm{target}$ is the target joint translation. The constraint compliances are then $\boldsymbol{\alpha}=\left[0\ \ 0\ \ k^{-1}\ \ 0\ \ 0\ \ 0 \right]^\top$ where $k$ is the spring stiffness.

\section {XPBD for Cosserat Rods}
Another application of XPBD with oriented particles is modeling Cosserat rods, as done in \cite{deul2018direct,angles2019viper}. We start by introducing the theory of continuous Cosserat rods, and then present a rigid-body discretization that follows \cite{deul2018direct, gaston20263d}. Using our $SO(3)$ framework, we can develop a finite-element discretization of a Cosserat rod with higher-order Lagrangian basis functions. 

\subsection{Continuous Cosserat Rod}
\label{sec:continuous-cosserat-rod}
\begin{figure}
    \centering
    \includegraphics[width=0.6\linewidth]{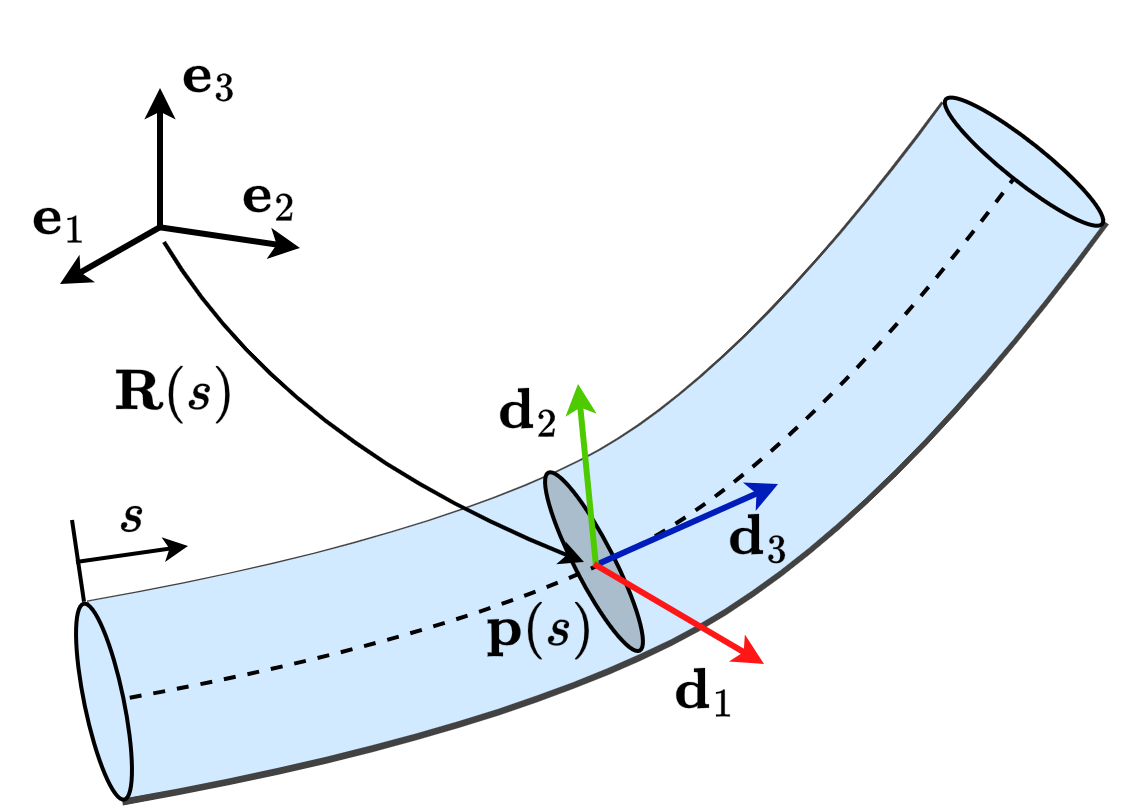}
    \caption{A continuous Cosserat rod.}
    \label{fig:continuous-cosserat}
\end{figure}

A Cosserat rod is modeled as a smooth curve in $\mathbb{R}^3$, $\mathbf{p}(s) :[0, L] \to \mathbb{R}^3$ which defines the centerline of the rod. At each location $s$ along the rod's centerline a coordinate frame $\{\mathbf{d}_1, \mathbf{d}_2, \mathbf{d}_3\}$ is defined, with $\mathbf{d}_3$ initially in the direction tangent to $\mathbf{p}(s)$ at $s$. The orientation $\mathbf{R}(s) \in SO(3)$ relates the local, director-defined frame at $s$ to the global, fixed frame $\{ \mathbf{e}_1, \mathbf{e}_2, \mathbf{e}_3 \}$. A graphical depiction of these quantities can be seen in Figure \ref{fig:continuous-cosserat}.

According to Cosserat theory, the local shear and extension strains along the rod $\mathbf{v}(s)$, and the curvature along the rod $\mathbf{u}(s)$ are defined as:
\begin{align}
\label{eq:continuous-shear-stretch}
\mathbf{v}(s) &= \mathbf{R}(s)^\top \mathbf{p}'(s) \\
\label{eq:continusous-bending}
\mathbf{u}(s) &= \mathbf{R}(s)^\top \mathbf{R}'(s) = \mathcal{D}_s (\mathbf{R}(s))
\end{align}
with $(\cdot)'$ used to denote differentiation with respect to $s$.
Recall that the notation $\mathcal{D}_s (\mathbf{R}(s))$ corresponds to the definition given in \eqref{eq:R-to-SO3-differential}, that is:
\begin{equation}
    \mathbf{u}(s) = \lim_{\Delta s \to 0} \frac{\mathbf{R}(s+\Delta s) \boxminus \mathbf{R}(s)}{\Delta s}
\end{equation}

We can interpret $\mathbf{v}_1$ and $\mathbf{v}_2$ as the shear strains in the local $x$ and $y$ directions, respectively, and $\mathbf{v}_3-1$ as the axial strain along the local $z$ direction (i.e. along the centerline). Similarly, $\mathbf{u}_1$ and $\mathbf{u}_2$ are angular rates of change about the local $x$ and $y$ directions, with respect to $s$, representing bending, and $\mathbf{u}_3$ is the angular rate about the local $z$, representing torsion. 
While these have units of curvature (angle/length), not strain (unitless), they are commonly referred to as bending and torsional
strain variables because they are used to variables are used to define the total strain energy in the rod, which is the sum of the energy associated with shear strains $\mathbf{v}$ and the energy associated with bending strains $\mathbf{u}$ (the dependence on $s$ is implied):
\begin{align}
\label{eq:continuous-shear-stretch-energy}
U_\mathrm{shear} &= \frac{1}{2} \int_{0}^{L} (\mathbf{v} - \mathbf{e}_3)^\top \mathbf{K}^\mathrm{v} (\mathbf{v} - \mathbf{e}_3) ds \\
\label{eq:continuous-bending-torsion-energy}
U_\mathrm{bend} &= \frac{1}{2} \int_{0}^{L} (\mathbf{u} - \mathbf{u}^*)^\top \mathbf{K}^\mathrm{u} (\mathbf{u} - \mathbf{u}^*) ds 
\end{align}
where $\mathbf{u}^*$ is the precurvature in the rod and the local shear and bending stiffness matrices $\mathbf{K}^\mathrm{v}$ and $\mathbf{K}^\mathrm{u}$ are defined as
\begin{align}
\label{eq:shear-stiffness}
\mathbf{K}^\mathrm{v} &= \begin{bmatrix}
    GA & 0 & 0 \\
    0 & GA & 0 \\
    0 & 0 & EA
\end{bmatrix} \\
\label{eq:bending-stiffness}
\mathbf{K}^\mathrm{u} &= \begin{bmatrix}
    EI_1 & 0 & 0 \\
    0 & EI_2 & 0 \\
    0 & 0 & GJ
\end{bmatrix}
\end{align}
where $G$ is the shear modulus, $E$ is the Young's modulus, $A$ is the cross-sectional area, $I_1$ and $I_2$ are the second moments of area of the cross section (in the local $x$ and $y$ directions, respectively), and $J = I_1 + I_2$ is the second polar moment of area of the cross section.

\subsection{Rigid-Body Discretization of Cosserat Rod}
\label{sec:rigid-body-discretization-of-cosserat-rod}
\begin{figure}
    \centering
    \includegraphics[width=1\linewidth]{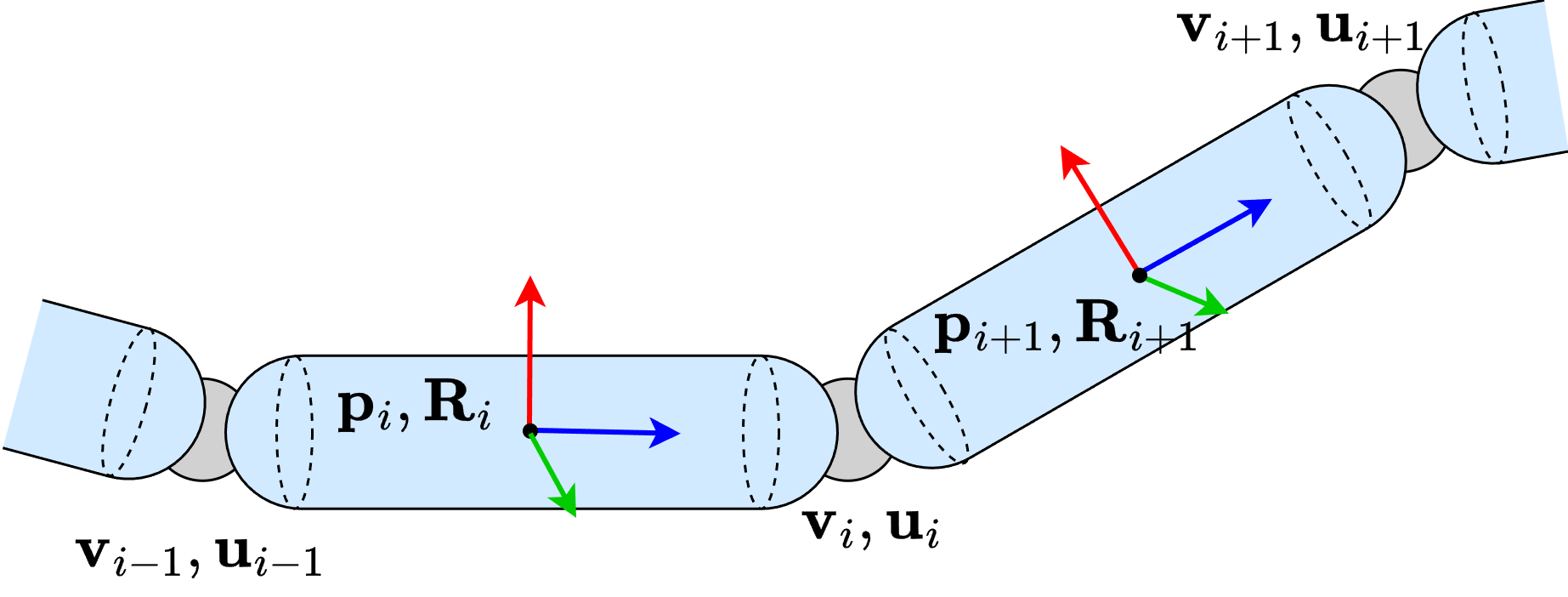}
    \caption{Rigid-body discretization of a Cosserat rod. Positions and orientations along the rod are defined at rigid body centers of mass. Strains are defined at joints between the rigid bodies.}
    \label{fig:rigid-body-discretization}
\end{figure}
To enable computation, the continuous Cosserat rod must be discretized in some way. One common approach is the chain-of-rigid-bodies approximation, modeling the rod as a chain of $N$ rigid bodies connected by elastic joints that encode the strain energy of the rod. The two end rigid bodies are each assigned half the length of the interior rigid bodies, with their DOF located at the outermost end, corresponding to either the tip or base of the rod. This enables simple enforcement of boundary conditions and joints involving the base or tip of the rod. 

With this convention, each interior rigid body has length $l=L/(N-1)$, translational mass $m=l\rho A$, and rotational inertia $\boldsymbol{\mathcal{I}}=l\rho \mathbf{J}$. The end rigid bodies have length $l_\mathrm{end}=L/2(N-1)$, translational mass $m=l_\mathrm{end} \rho A$ and rotational inertia $\boldsymbol{\mathcal{I}}_{\mathrm{end}} = l_\mathrm{end} \rho \mathbf{J} + m_\mathrm{end} \mathrm{diag}(0,l_\mathrm{end}^2,0)$. Here, $\rho$ is the density of the rod, $A$ is the cross-sectional area, $L$ is the total length of the rod, and $\mathbf{J}$ is a diagonal matrix of the second moments of area of the cross section. 

The shear and bending strains are defined at each joint in terms of the DOF of the adjacent rigid bodies. That is, at the joint between rigid bodies $j$ and $j+1$: 
\begin{align}
\label{eq:discretized-stretch-shear}
    \mathbf{v}_j &= \frac{1}{2l} \left( \mathbf{R}_{j}^\top + \mathbf{R}_{j+1}^\top \right) \left( \mathbf{p}_{j+1} - \mathbf{p}_{j}\right) \\
\label{eq:discretized-bending-torsion}
    \mathbf{u}_j &= \frac{1}{l} (\mathbf{R}_{j+1} \boxminus \mathbf{R}_j)
\end{align}
where $l=L/(N-1)$ is the rest length between adjacent rigid body positions in the discretization.
The shear strain definition arises from a symmetric discretization \cite{gaston20263d}, where the discrete shear strains evaluated at rigid bodies $j$ and $j+1$ are averaged. The bending strain is the effective curvature of the between bodies $j$ and $j+1$, which is the angle change divided by the rest length between them. The strains are assumed to be constant between any two rigid bodies, resulting in a piecewise-constant approximation of strain in the rod. Thus, the discretized elastic potentials become:
\begin{align}
\label{eq:discretized-shear-potential}
    {U}_{\mathrm{shear}} &= \frac{1}{2}\sum_{j=1}^{N-1} l (\mathbf{v}_j - \mathbf{e}_3)^\top \mathbf{K}_j^\mathrm{v} (\mathbf{v}_j - \mathbf{e}_3) \\
\label{eq:discretized-bending-potential}
    U_{\mathrm{bend}} &= \frac{1}{2}\sum_{j=1}^{N-1} l (\mathbf{u}_j - \mathbf{u}_j^*)^\top \mathbf{K}_j^\mathrm{u} (\mathbf{u}_j - \mathbf{u}_j^*)
\end{align}
where $\mathbf{u}_j^*$ is the precurvature in the rod between nodes $j$ and $j+1$.
By inspection, we can associate quantities in each of the discretized energy expressions above with the definition of XPBD constraints and compliances $U_\mathrm{XPBD}=\frac{1}{2} \mathbf{C}^\top \boldsymbol{\alpha}^{-1} \mathbf{C}$, such that a 6D XPBD constraint is defined at each of the $N-1$ joints between rigid bodies: 
\begin{equation}
\label{eq:cosserat-constraints}
\begin{aligned}
\mathbf{C}_j(\mathbf{p}_{j}, \mathbf{R}_j, \mathbf{p}_{j+1}, \mathbf{R}_{j+1}) =
    \begin{bmatrix}
        \mathbf{C}_j^\mathrm{v} \\
        \mathbf{C}_j^\mathrm{u}
    \end{bmatrix}
    =
    \begin{bmatrix}
        \mathbf{v}_j-\mathbf{e}_3 \\
        \mathbf{u}_j - \mathbf{u}_j^*
    \end{bmatrix} \\
    =
    \begin{bmatrix}
        \frac{1}{2l} (\mathbf{R}_{j}^\top + \mathbf{R}_{j+1}^\top ) \left( \mathbf{p}_{j+1} - \mathbf{p}_{j}\right) - \mathbf{e}_3\\
        \frac{1}{l} \left( \mathbf{R}_{j+1} \boxminus \mathbf{R}_j \right)  - \mathbf{u}^*_j 
    \end{bmatrix}
\end{aligned}
\end{equation}
with constraint stiffnesses
\begin{equation}
\label{eq:cosserat-constraint-stiffnesses}
    \boldsymbol{\alpha}_j^{-1}= l\begin{bmatrix}
        \mathbf{K}_j^\mathrm{v} & \\
        & \mathbf{K}_j^\mathrm{u}
    \end{bmatrix}
\end{equation}
The gradients of these constraints can be found in the Appendix.

\subsection{Finite-Element Discretization of Cosserat Rod}
\subsubsection{Discretization and Basis Functions}
Alternatively, we may adopt a finite-element discretization of the continuous Cosserat rod, where the continuous rod is discretized into $N_{el}$ elements, each with $N_{en}$ nodes. We label the DOF of the $i$th element $(\mathbf{p}^i_j, \mathbf{R}^i_j)$ for $j=1,...,N_{en}$. Under this labeling, $(\mathbf{p}^i_{N_{en}}, \mathbf{R}^i_{N_{en}})$ and $(\mathbf{p}^{i+1}_1, \mathbf{R}^{i+1}_1)$ are the same.

\begin{figure*}
    \centering
    \includegraphics[width=0.8\textwidth]{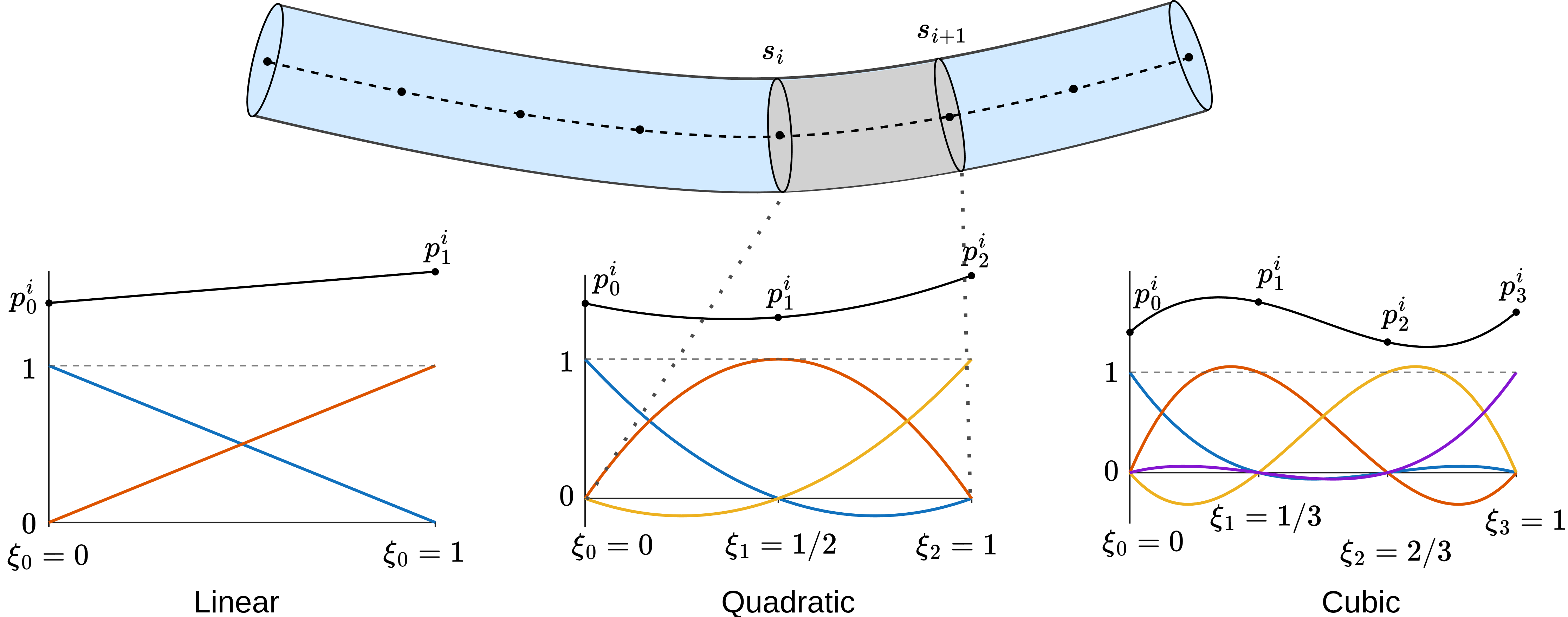}
    \caption{Basis function parameterization of a Cosserat rod finite element. Using \eqref{eq:reference-coordinate-affine-map}, element $i$ is mapped from its physical interval $[s_i, s_{i+1}]$ to the reference interval $[0, 1]$, on which the basis functions are defined. Linear, quadratic, and cubic Lagrange polynomials are shown.}
    \label{fig:basis-functions}
\end{figure*}

The domain of the $i$th element is spanned by arc-length values $[s_i, s_{i+1}]$.
This physical domain is mapped to a reference domain $[0,1]$ using the affine transformation
\begin{equation}
\label{eq:reference-coordinate-affine-map}
	\xi(s) = \frac{1}{l_i}(s - s_i)
\end{equation}
where $l_i$ is the rest length of element $i$.
Node $1$ in the element corresponds to $\xi=0$, and node $N_{en}$ in the element corresponds to $\xi=1$. The location of node $j$ in the element in terms of $\xi$ is denoted $\xi_j$.
We interpolate the position and orientation in an element using $N_{en}$ local orthonormal basis functions $\phi_j(\xi)$, defined over the reference domain of each element. Common choices of basis include linear basis functions, for which $N_{en}=2$ and $\xi_j=[0,1]$:
\begin{equation}
\label{eq:linear-basis-functions}
\phi_1(\xi) = 1-\xi \ \ \ \ \ \ \ \phi_2(\xi) = \xi
\end{equation}
and quadratic Lagrange basis functions, for which $N_{en} = 3$ and $\xi_j=[0, 0.5, 1]$:
\begin{equation}
\label{eq:quadratic-basis-functions}
\begin{aligned}
\phi_1(\xi) &= 2\xi^2 - 3\xi + 1 \\ \phi_2(\xi) &= -4\xi^2 + 4\xi \\ \phi_3(\xi) &= 2\xi^2 - \xi
\end{aligned}
\end{equation}
We require that the basis functions satisfy the Kronecker delta property $\phi_j(\xi_k) = \delta_{jk}$, that is $\phi_j(\xi_k)=1$ for $j=k$ and $0$ otherwise. This property is depicted in Figure \ref{fig:basis-functions}.
\subsubsection{Interpolation of Position and Orientation}
\begin{figure}
    \centering
    \includegraphics[width=0.55\linewidth]{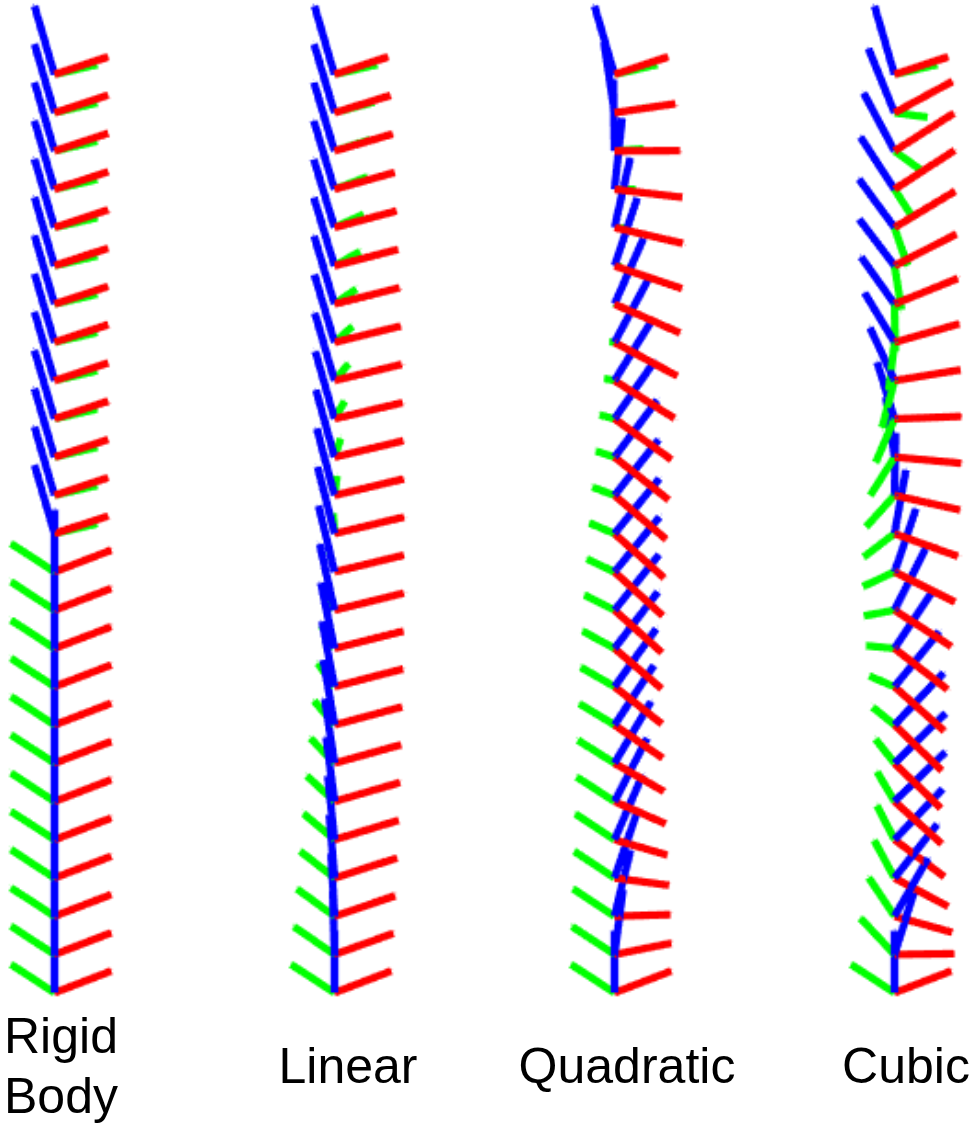}
    \caption{Rotational interpolation using \eqref{eq:fem-rotational-dof-parameterization} and different basis functions.}
    \label{fig:rotation-interpolation}
\end{figure}

The set of local basis functions is used to interpolate the nodal values of position and orientation. The positional degrees of freedom on the $i$th element are interpolated with basis functions as:
\begin{equation}
\label{eq:fem-positional-dof-parameterization}
\mathbf{p}(s) = \sum_{j=1}^{N_{en}} \phi_j(\xi(s)) \mathbf{p}_j^i, \ \ \ \ \  s \in [s_i, s_{i+1}]
\end{equation}
The dependence of $\xi$ on $s$ is explicitly noted here but will be dropped in future equations. 
Because of the nonlinear group structure of $SO(3)$, the interpolation of the rotational degrees of freedom requires more care than the positional degrees of freedom. Originally proposed by Jeleni\`c and Crisfield \cite{jelenic1999geometrically}, we parameterize the \emph{local} exponential rotation vector with the chosen basis. Thus, for element $i$, we start at $\mathbf{R}_1^i$ and perturb it by a local change in rotation:  
\begin{equation}
\label{eq:fem-rotational-dof-parameterization}
\mathbf{R}(s) = \mathbf{R}_1^i \boxplus \boldsymbol{\theta}(\xi)
\end{equation}
where
\begin{equation}
    \boldsymbol{\theta}(\xi) = \sum_{j=2}^{N_{en}} \phi_j(\xi)(\mathbf{R}_j^i \boxminus \mathbf{R}_1^i) 
\end{equation}
This not only aligns well with our body-frame definitions of exponential rotation vectors, but also maintains the strain-objectivity of the finite element \cite{jelenic1999geometrically}. Additionally, by the Kronecker delta property, plugging in $\xi_j$ into \eqref{eq:fem-rotational-dof-parameterization} yields $\mathbf{R}_j^i$, confirming that it is a valid interpolation of rotation. The resulting interpolation is illustrated in Figure \ref{fig:rotation-interpolation}. We note that with linear basis functions, the interpolation above corresponds to spherical linear interpolation (SLERP).

\subsubsection{Element Strains}
With expressions for $\mathbf{p}(s)$ and $\mathbf{R}(s)$ defined on an element, we can use \eqref{eq:continuous-shear-stretch} and \eqref{eq:continusous-bending} to define the strains over an element. Starting with the shear strains on element $i$, and using $(\cdot)'$ to denote differentiation with respect to the arc length parameter $s$: 
\begin{equation}
\label{eq:fem-shear-streatch}
\begin{aligned}
\mathbf{v}(s) &= \mathbf{R}(s)^\top \mathbf{p}'(s) \\
&= \mathbf{R}(s)^\top \sum_{j=1}^{N_{en}} \phi_j'(\xi) \mathbf{p}^i_j
\end{aligned}
\end{equation}
It is important to invoke the chain rule when evaluating $\phi'$, so on element $i$:
\begin{equation}
\phi_j'(s) = \frac{\partial \phi_j}{\partial \xi} \frac{d \xi}{d s} = \frac{1}{l_i} \frac{\partial \phi_j}{\partial \xi},
\end{equation}
which shows how the factor $1/l_j$ arises naturally in the expressions for rod strains from the parameterization of position and orientation with local basis functions.
The bending and torsional strains on element $i$ are:
\begin{equation}
\label{eq:fem-bending-torsion}
\begin{aligned}
\mathbf{u}(s) &=  \mathcal{D}_s(\mathbf{R}(s)) \\
&= \boldsymbol{\Gamma}(\boldsymbol{\theta}(s)) \boldsymbol{\theta}'(s)
\end{aligned}
\end{equation}
where
\begin{equation}
\label{eq:fem-theta-prime}
\boldsymbol{\theta}'(s) = \sum_{j=2}^{N_{en}} \phi_j'(\xi) (\mathbf{R}_j \boxminus \mathbf{R}_1)
\end{equation}
The full derivation can be found in the Appendix. It is worth noting that with linear basis functions, \eqref{eq:fem-shear-streatch} and \eqref{eq:fem-bending-torsion} reduce to:
\begin{equation}
\begin{aligned}
	\mathbf{v}(s) &= \frac{1}{l_i} [\mathbf{R}^i_1 \boxplus s(\mathbf{R}^i_2 \boxminus \mathbf{R}^i_1)]^\top (\mathbf{p}^i_2 - \mathbf{p}^i_1) \\
	\mathbf{u}(s) &= \frac{1}{l_i} (\mathbf{R}^i_2 \boxminus \mathbf{R}^i_1)
\end{aligned}
\end{equation}
where the identity $\boldsymbol{\Gamma}(s\boldsymbol{\theta}) \boldsymbol{\theta} = \boldsymbol{\theta}$ was used to simplify $\mathbf{u}(s)$.

\subsubsection{XPBD Constraints}
With expressions for the strains $\mathbf{v}(s)$ and $\mathbf{u}(s)$ defined along the rod length, we can return to the continuous Cosserat rod energy expressions in \eqref{eq:continuous-shear-stretch-energy} and \eqref{eq:continuous-bending-torsion-energy}:
\begin{equation}
\label{eq:fem-continuous-cosserat-energy}
U_\mathrm{rod} = \frac{1}{2} \int_{0}^{L} \left[ (\mathbf{v} - \mathbf{e}_3)^\top \mathbf{K}^\mathrm{v} (\mathbf{v} - \mathbf{e}_3) + (\mathbf{u} - \mathbf{u}^*)^\top \mathbf{K}^\mathrm{u} (\mathbf{u} - \mathbf{u}^*) \right] ds 
\end{equation}
To use the XPBD machinery developed earlier, we need to define a finite vector of constraints $\mathbf{C}$ and associated compliances $\boldsymbol{\alpha}$ such that the product $$U_{\mathrm{XPBD}} = \frac{1}{2} \mathbf{C}^\top \boldsymbol{\alpha}^{-1} \mathbf{C}$$ approximates the continuous energy expression above. Following \cite{saillant2024high}, the energy expression is evaluated at discrete points using a local Gaussian quadrature on each element:
\begin{equation}
\label{eq:fem-discretized-cosserat-energy}
U_\mathrm{rod} \approx \frac{1}{2} \sum_{i=1}^N \sum_{k=1}^m l_i w_k \begin{bmatrix}
	\mathbf{v}(\tilde{\xi}_k) - \mathbf{e}_3 \\
	\mathbf{u}(\tilde{\xi}_k)
\end{bmatrix}^\top \begin{bmatrix}
	\mathbf{K}^{\mathrm{v}} & \\ & \mathbf{K}^{\mathrm{u}}
\end{bmatrix} \begin{bmatrix}
\mathbf{v}(\tilde{\xi}_k) - \mathbf{e}_3 \\
\mathbf{u}(\tilde{\xi}_k) - \mathbf{u}^*(\tilde{\xi}_k)
\end{bmatrix}
\end{equation}
where $m$ is the number of Gauss quadrature points used per element, and $\tilde{\xi}_k$ corresponds to the location of the $k$th Gauss point within the reference domain $[0,1]$. For example, a one-point Gauss quadrature evaluates the integral over an element at a single point $\tilde{\xi}_0=1/2$ with weight $w_0=1$, while a two-point Gauss quadrature evaluates the integral over an element at two points $\tilde{\xi}_0, \tilde{\xi}_1 = 1/2 \pm \sqrt{3}/2$ with weights $w_0,w_1=1/2$. By inspection, we can define $m$ constraints on each element:
\begin{equation}
\mathbf{C}_i^k = \begin{bmatrix}
	\mathbf{v}(\tilde{\xi}_k) - \mathbf{e}_3 \\
	\mathbf{u}(\tilde{\xi}_k) - \mathbf{u}^*(\tilde{\xi}_k)
\end{bmatrix}
\end{equation}
with associated compliances defined by:
\begin{equation}
(\boldsymbol{\alpha}_i^k)^{-1} = l_i w_k \begin{bmatrix}
	\mathbf{K}^{\mathrm{v}} & \\ & \mathbf{K}^{\mathrm{u}}
\end{bmatrix}
\end{equation}
The gradients of $\mathbf{C}_i^j$ are simply the gradients of $\mathbf{v}(s)$ and $\mathbf{u}(s)$, which can be found in the Appendix. We can see that with linear basis functions and a one-point Gauss quadrature, there is a single constraint for each element:
\begin{equation}
\label{eq:fem-linear-basis-constraints}
\mathbf{C}_i = \frac{1}{l_i} \begin{bmatrix}
	 [\mathbf{R}_1^i \boxplus \frac{1}{2}(\mathbf{R}_2^i \boxminus \mathbf{R}_1^i)]^\top (\mathbf{p}_2^i - \mathbf{p}_1^i) - \mathbf{e}_3 \\
	 (\mathbf{R}_2^i \boxminus \mathbf{R}_1^i) - l\mathbf{u}^*_i
\end{bmatrix}
\end{equation}
with constraint stiffness:
\begin{equation}
\alpha_i^{-1} = l_i \begin{bmatrix}
	\mathbf{K}^{\mathrm{v}} & \\ & \mathbf{K}^{\mathrm{u}}
\end{bmatrix}
\end{equation}
This is nearly identical to the rigid-body rod constraints and constraint stiffnesses given in \eqref{eq:cosserat-constraints} and \eqref{eq:cosserat-constraint-stiffnesses}. The only difference is the interpolation method used to obtain the rotation at the midpoint between two nodes for the shear strains: \eqref{eq:cosserat-constraints} simply averages the two rotation matrices (which does not guarantee the result remains in $SO(3)$), while \eqref{eq:fem-linear-basis-constraints} uses a genuine $SO(3)$ interpolation.

\begin{figure}
    \centering
    \includegraphics[width=1\linewidth]{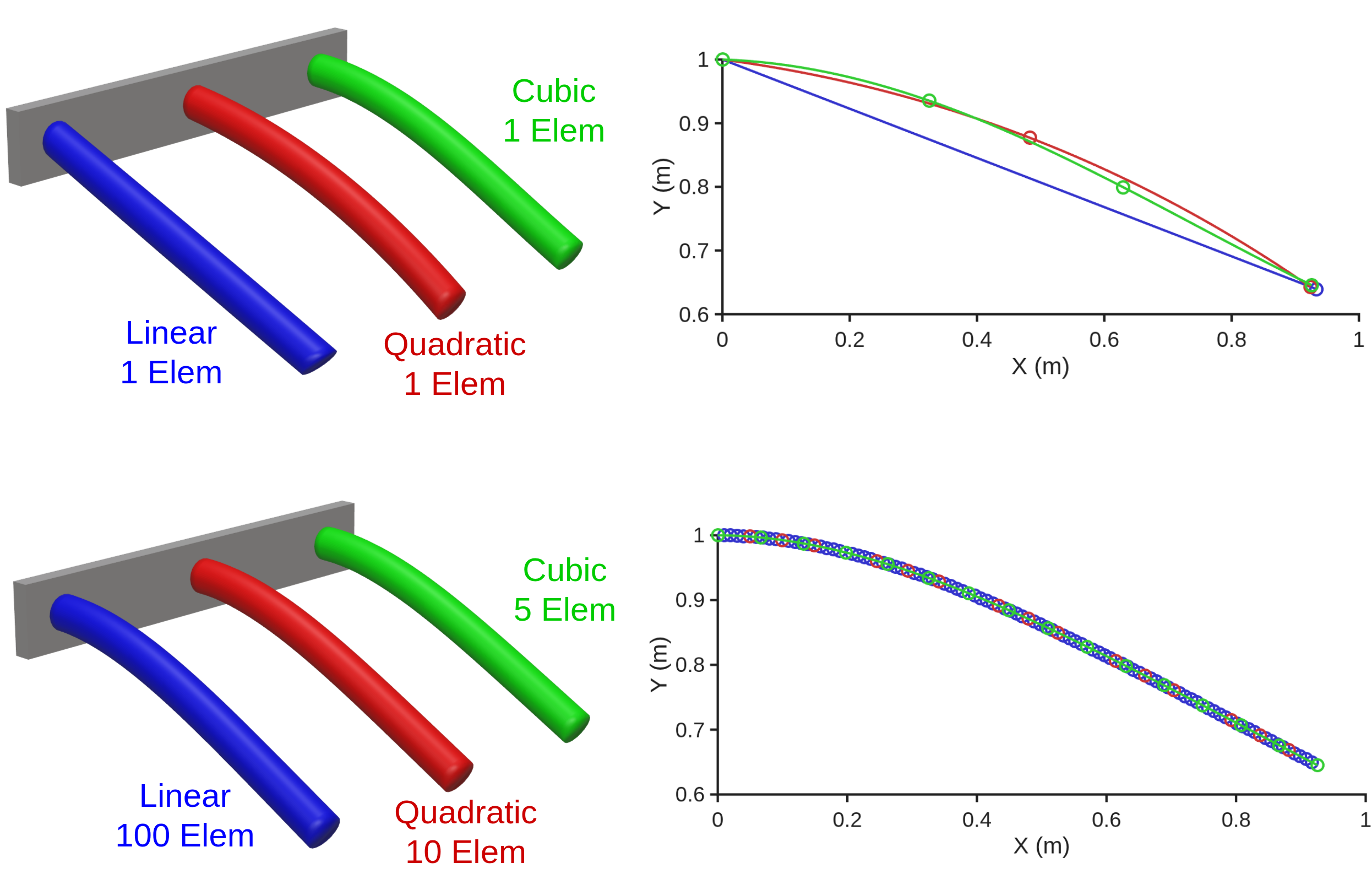}
    \caption{Demonstration of different basis functions for a large-deflection cantilever rod. For only a single element (\textbf{top}), the degree of basis function used is obvious. As more elements are used (\textbf{bottom}), the rods converge to the same shape.}
    \label{fig:single-vs-many}
\end{figure}

\subsubsection{Shear Locking and Number of Gauss Points}
We can choose how many Gauss points to use when approximating the energy integrals. At first glance, this appears to be a straightforward accuracy–speed tradeoff: using more Gauss points can improve accuracy, but requires projecting a larger number of constraints.
However, a classic issue that plagues beam and rod finite elements is so-called \textit{shear locking}. As a rod becomes more slender (the ratio $L/r$ increases), the shear stiffness increases. Under certain combinations of basis functions and integral approximations, artificial constraints are introduced which couple the bending behavior in the rod to the shear stiffness, resulting in a rod that is artificially stiff in bending \cite{prathap1982reduced}. In beams and rods, this is typically avoided using ``reduced'' integration \cite{prathap1982reduced, simo1986three}, where a lower order approximation is used to avoid these artificial constraints. In this work, we use ``uniformly reduced'' integration \cite{simo1986three}, where rods with linear basis functions use a one-point Gauss quadrature, rods with quadratic basis functions use a two-point Gauss quadrature, and rods with cubic basis functions use a three-point Gauss quadrature. As demonstrated in Figure \ref{fig:shear-locking}, uniformly reduced integration successfully avoids shear locking at high slenderness ratios. 

\subsubsection{Assignment of Node Inertias}
Part of the efficiency of XPBD relies on a diagonal mass matrix $\boldsymbol{\mathcal{M}}$, so that its inverse is also diagonal. Physically, this corresponds to negligible inertial coupling between particles. If $\boldsymbol{\mathcal{M}}^{-1}$ is not diagonal, the update formulas in \eqref{eq:position-orientation-lambda-update} and \eqref{eq:position-orientation-position-update} are still correct but involve significantly more coupling, since the matrix product $\boldsymbol{\mathcal{M}}^{-1} \nabla \mathbf{C}^\top$ may be dense.

In the finite element method in one dimension, the mass matrix is banded, leading to a dense inverse mass matrix \cite{zienkiewicz2005finite}. Furthermore, in the geometrically-exact Cosserat rod formulation, the mass matrix is configuration-dependent for large rotations \cite{simo1986three}. For simplicity, the mass matrix is diagonalized in the initial configuration using row-sum mass lumping \cite{zienkiewicz2005finite} and held constant throughout the simulation. Each element has a total translational inertia of $\rho A l_i$ and a rotational inertia described by the tensor $\rho \mathbf{J}$ where $\mathbf{J}= \mathrm{diag}( I_x, I_y, I_z)$. For linear elements, row-sum lumping corresponds to each node in the element receiving half of the total element inertia. For quadratic elements, this corresponds to a split of $1/6$ to the element endpoint nodes and $2/3$ to the midpoint node, and for cubic elements, the element endpoints nodes receive $1/8$ the total mass and the two element internal nodes receive $3/8$ the total mass.

\section{Contact and Friction}
\subsection{Collision Detection}

Broad-phase collision detection is performed using spatial hashing \cite{teschner2003optimized}. Potentially colliding pairs are identified, and narrow-phase collision detection is performed to determine if the pair of objects is truly colliding. If there is penetration, pairs of contact points defining the contact manifold are generated, along with the collision normal. 


\subsubsection{Rigid-Rod}
\label{sec:collision-rigid-rod}
Collisions between a rigid body and a rod element require finding the point(s) on the rod element that are closest to the rigid body. For linear rod elements and simple convex shapes, cheap analytical formulas may be used. For higher-order rod elements and arbitrary rigid bodies, closed-form expressions do not exist. We follow the approach of \cite{macklin2020local} and represent the rigid body by a Signed Distance Function (SDF),  $\psi : \mathbb{R}^3 \to \mathbb{R}$. If a point $\mathbf{x}$ is inside a rigid body, $\psi(\mathbf{x}) < 0$, with $\psi(\mathbf{x})=0$ representing the object surface.
Thus, for a single rod element, we seek all $\xi$ that are local minima of:
\begin{equation}
\label{eq:rigid-rod-collision-minimization}
\min_{\xi \in [0,1]} \psi(\mathbf{p}(\xi))
\end{equation}
where $\mathbf{p}(\xi)$ is the centerline of the rod element, given by \eqref{eq:fem-positional-dof-parameterization}. If $\psi(\mathbf{p}(\xi)) < r$ with $r$ being the radius of the rod, then there is a collision at that $\xi$, and the collision normal is $\mathbf{n} = \nabla \psi(\mathbf{p}(\xi))$. The contact point (expressed in the global frame) on the rigid body is the corresponding point on the SDF surface $\mathbf{p}(\xi) -\psi(\mathbf{p}(\xi)) \nabla \psi(\mathbf{p}(\xi))$ and the contact point (expressed in the global frame) on the rod is $\mathbf{p}(\xi) - r \nabla \psi(\mathbf{p}(\xi))$.

Note that in general \eqref{eq:rigid-rod-collision-minimization} is a non-smooth and non-convex scalar minimization problem, because $\psi$ may not be smooth. We make the assumption that a single minimum exists for the element and initially test multiple seed points. The deepest penetrating of seed point is then refined with golden section search, illustrated in Figure \ref{fig:collision-optimization}, Left.

\subsubsection{Rod-Rod}
\label{sec:collision-rod-rod}
Collisions between rod elements are determined by finding the point(s) on each rod element that minimize the distance between them. When both rod elements are linear, this simply amounts to finding the closest points between two line segments. But, for two rod elements of higher order, clean analytical expressions do not exist. Instead, we formulate the collision detection problem as a minimization:
\begin{equation}
\label{eq:rod-rod-collision-minimization}
\min_{\xi_1, \xi_2 \in [0,1]} || \mathbf{p}_1 (\xi_1) - \mathbf{p}_2(\xi_2)||^2
\end{equation}
where $\mathbf{p}_i$ and $\xi_i$ are respectively the centerline position and interpolation parameter for rod $i$. For some $(\xi_1, \xi_2)$ that correspond to a local minima of \eqref{eq:rod-rod-collision-minimization}, a collision is registered if $|| \mathbf{p}_1(\xi_1) - \mathbf{p}_2(\xi_2) || < r_1 + r_2$, with collision normal $\mathbf{n} = \mathbf{p}_1(\xi_1) - \mathbf{p}_2(\xi_2)$. The contact points (expressed in the global frame) on rod 1 and rod 2 are then $\mathbf{p}(\xi_1) - r_1 \mathbf{n}$ and $\mathbf{p}_2(\xi_2) + r_2 \mathbf{n}$.

\begin{figure}
    \centering
    \includegraphics[width=1\linewidth]{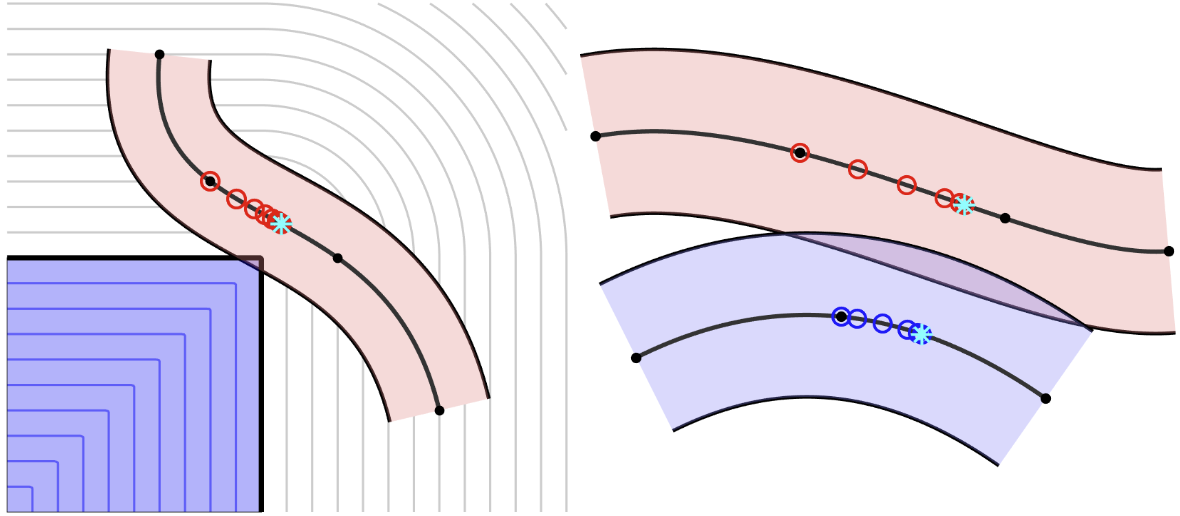}
    \caption{Optimization over a rod element for collision detection. Hollow circles denote the closest point(s) at each iteration, and cyan asterisks denote the final iterates. \textbf{Left.} Collision detection between cubic rod element and SDF. \textbf{Right.} Iterative optimization for collision detection between a cubic rod element (red) and a quadratic rod element (blue), for a specific pair of seed points.}
    \label{fig:collision-optimization}
\end{figure}

The minimization problem in \eqref{eq:rod-rod-collision-minimization} can be solved using a damped Gauss-Newton approach:
\begin{equation}
\label{eq:rod-rod-collision-gauss-newton}
(\mathbf{J}^\top \mathbf{J} + \lambda \mathbf{I}) \begin{bmatrix}
    \Delta \xi_1 \\
    \Delta \xi_2
\end{bmatrix} = -\mathbf{J}^\top (\mathbf{p}_1(\xi_1) - \mathbf{p}_2(\xi_2))
\end{equation}
where $\mathbf{J}= [ \frac{\partial \mathbf{p}_1}{\partial \xi_1}, \  -\frac{\partial \mathbf{p}_2}{\partial \xi_2}]$ is the Jacobian of the residual. The damping factor $\lambda$ is chosen adaptively depending on the trace of $\mathbf{J}^\top \mathbf{J}$. A trust region is used to prevent overly large steps from occurring. To find all local minima, we initialize the algorithm from all possible pairs of $\xi_1$ and $\xi_2$ locations corresponding to nodes in the elements. The iterative optimization for a specific pair of seed points is illustrated in Figure \ref{fig:collision-optimization}, Right.

\subsection{Contact Constraints}

For each pair of contact points, an inequality contact constraint is defined:
\begin{equation}
\label{eq:contact-constraint}
C_c = \mathbf{n}^\top (\mathbf{a}_1 - \mathbf{a}_2) \geq 0
\end{equation}
where $\mathbf{n}$ is the collision normal (pointing from body 2 to body 1) expressed in the global frame, and $\mathbf{a}_1$ and $\mathbf{a}_2$ are the contact points on bodies 1 and 2, expressed in the global frame. The contact points are themselves functions of the underlying state. For rigid bodies, this is simply:
\begin{equation}
\label{eq:contact-points-rigid-bodies}
\mathbf{a}_i = \mathbf{p}_i + \mathbf{R}_i \mathbf{a}^\mathrm{loc}_i
\end{equation}
where $\mathbf{p}_i$ and $\mathbf{R}_i$ are the position and orientation of the center of mass of the rigid body, and $\mathbf{a}^\mathrm{loc}_i$ is the vector from the center of mass to the contact point, expressed in the local frame. For rods, the contact point is defined in the interpolated frame:
\begin{equation}
\label{eq:contact-points-rods}
\mathbf{a}_i = \mathbf{p}(s_i) + \mathbf{R}(s_i) \mathbf{a}^\mathrm{loc}_i
\end{equation}
where $s_i$ is the position along the rod of the interpolated frame, which is obtained directly from the optimizations in \eqref{eq:rigid-rod-collision-minimization} and \eqref{eq:rod-rod-collision-minimization}. 

\subsection{Friction}
\label{sec:friction}
After all other constraints have been solved for, Coulombic friction is applied for active contact constraints (contact constraints with associated $\lambda > 0$) through a frictional position correction applied in the tangential direction of relative motion between the two contact points. The relative motion of the two contact points in the tangential direction can be found as:
\begin{equation}
\label{eq:friction-relative-tangential-direction}
\begin{aligned}
\Delta \mathbf{a}_{\mathrm{rel}} &= (\mathbf{a}_1 - \mathbf{a}_{1,\mathrm{prev}}) - (\mathbf{a}_2 - \mathbf{a}_{2, \mathrm{prev}}) \\
\Delta \mathbf{a}_{\mathrm{tan}} &= \Delta \mathbf{a}_{\mathrm{rel}} - (\Delta \mathbf{a}_{\mathrm{rel}}^\top \mathbf{n}) \mathbf{n}
\end{aligned}
\end{equation}
where $\mathbf{a}_{1,\mathrm{prev}}$ is the position of the contact point $\mathbf{a}_1$ at the end of the last time step, with a similar definition for $\mathbf{a}_{2,\mathrm{prev}}$. 
The maximum frictional impulse we can apply is one that undoes all relative motion between the two contact points. In this case, we essentially have a constraint $C_f = || \Delta \mathbf{a}_\mathrm{tan} ||$. If we project this constraint, we will get a $\lambda_f$ corresponding to this maximum frictional impulse. However, this $\lambda_f$ may be outside the bounds of physically possible frictional impulses. So, we must clamp $\lambda_f$ to the interval $[-\mu \lambda_n, \mu \lambda_n]$, where $\mu$ is the coefficient of friction (static or dynamic) and $\lambda_n$ is the normal impulse (the Lagrange multiplier for the associated contact constraint). Then, we apply the positional correction with \eqref{eq:position-orientation-position-update}:
\begin{equation*}
\Delta \boldsymbol{x}_f = \boldsymbol{\mathcal{M}}^{-1} \nabla C_f^\top \lambda_f
\end{equation*}

To determine whether the static or dynamic friction coefficient should be used, we must look at the relative velocities of the contact points in the tangential direction:
\begin{equation}
\begin{aligned}
    \dot{\mathbf{a}}_\mathrm{rel} &= \dot{\mathbf{a}}_1 - \dot{\mathbf{a}}_2 \\
    \dot{\mathbf{a}}_{\mathrm{tan}} &=  \dot{\mathbf{a}}_{\mathrm{rel}} - ( \dot{\mathbf{a}}_{\mathrm{rel}}^\top \mathbf{n}) \mathbf{n}
\end{aligned}
\end{equation}
If $|| \dot{\mathbf{a}}_{\mathrm{tan}}|| > \epsilon $ for some small $\epsilon > 0$, then the contact points were moving relative to each other last time step in the tangential direction, so the dynamic friction coefficient $\mu_d$ should be used. Otherwise, the static friction coefficient $\mu_s$ should be used.

In order to calculate $\dot{\mathbf{a}}_{\mathrm{tan}}$, the time derivative of each contact point $\dot{\mathbf{a}}_i$ must be calculated. For contact points on rigid bodies, this is simply:
\begin{equation}
\label{eq:rigid-body-contact-point-velocity}
\dot{\mathbf{a}}_i = \frac{d}{dt} (\mathbf{p} + \mathbf{R} \mathbf{a}_i^\mathrm{loc}) = \mathbf{v} + \mathbf{R} \widehat{\boldsymbol{\omega}} \mathbf{a}_i^{\mathrm{loc}}
\end{equation}
For contact points on rods, we can use the chain rule to find that:
\begin{equation}
    \dot{\mathbf{a}}_i = \sum_{j=1}^{N_{en}} \frac{\partial \mathbf{p}(s_i)}{\partial \mathbf{p}_j} \mathbf{v}_j + \sum_{j=1}^{N_{en}} \mathcal{D}_{\mathbf{R}_j} (\mathbf{R}(s_i)) \widehat{\boldsymbol{\omega}}_j \mathbf{a}_i^\mathrm{loc}
\end{equation}
An expression for $\mathcal{D}_{\mathbf{R}_j} (\mathbf{R}(s_i))$ can be found in the Appendix.

\subsection{Restitution}
\label{sec:restitution}
Following \cite{muller2020detailed}, we incorporate restitution at the velocity level. When a contact constraint is active ($\lambda > 0$), we reflect the previous relative normal velocity between contact points about the contact plane, scaled by the coefficient of restitution. The relative normal velocity between the contact points is:
\begin{equation}
    \dot{\mathbf{a}}_\mathrm{norm} = (\dot{\mathbf{a}}_\mathrm{rel}^\top \mathbf{n}) \mathbf{n}
\end{equation}
with a similar definition for the previous relative normal velocity. Then, we define a constraint at the velocity level:
\begin{equation}
    C_\mathrm{res} = \dot{\mathbf{a}}_{\mathrm{norm}} + e \dot{\mathbf{a}}_{\mathrm{norm,prev}}
\end{equation}
to drive $\dot{\mathbf{a}}_{\mathrm{norm}}$ to $-e \dot{\mathbf{a}}_{\mathrm{norm,prev}}$, the reflected normal velocity. We project the constraint using the particle velocities as the primary variables; the constraint gradients are computed with respect to the velocities and the velocities are updated directly.

\section{Solving the XPBD Linear Subsystem}
\label{sec:solving-xpbd-linear-subsystem}
The core of the XPBD algorithm involves solving the linear system for the Lagrange multipliers in \eqref{eq:position-orientation-lambda-update}, rewritten below:
\begin{equation*}
\left[ \nabla \mathbf{C}(\boldsymbol{x}^k) \boldsymbol{\mathcal{M}}^{-1} 
    \nabla \mathbf{C}(\boldsymbol{x}^k)^\top + \boldsymbol{\tilde{\alpha}} \right] 
    \Delta \boldsymbol{\lambda}^k = -\mathbf{C}(\boldsymbol{x}^k) 
    - \boldsymbol{\tilde{\alpha}} \boldsymbol{\lambda}^k
\end{equation*}
Traditionally, this system is solved with a Gauss-Seidel-style approach, where only a single constraint is considered at a time, allowing for a simple scalar computation of the Lagrange multiplier update, as in \eqref{eq:position-orientation-single-lambda-update} rewritten below:
\begin{equation*}
    \Delta\lambda_j=\frac{-C_j(\mathbf{x}^k) - \tilde\alpha_j\lambda_{j}^k}{\nabla C_j(\mathbf{x}^k)\mathbf{M}^{-1}\nabla C_j(\mathbf{x}^k)^\top + \tilde\alpha_j}
\end{equation*}
In the theme of Gauss-Seidel, the positions and orientations are then immediately updated after each constraint solve, so \eqref{eq:position-orientation-single-lambda-update} always uses the most recent information, in other words, our current best guesses for the new state $\boldsymbol{x}^{k+1}$. 

\subsection{Block Constraint Solve}
\label{sec:block-constraint-solve}
The single-constraint update \eqref{eq:position-orientation-single-lambda-update} is not truly a Gauss-Seidel update as it does not use a full row of the system matrix, but instead ignores cross-coupling terms between constraints that affect the same particles.
This means that the individual constraint solves have no knowledge of other constraints, and therefore may violate other constraints in order to satisfy the constraint that is currently being solved. This may undo some of the previous progress made in earlier constraint solves, leading to higher residuals.

Similar to the approach introduced in \cite{ton2023parallel} for Neo-Hookean constraints, we can simultaneously consider multiple constraints that affect the same particles and solve the resulting linear subsystem directly. By considering a group of constraints simultaneously, the XPBD algorithm will compute a position update that moves towards satisfying all the constraints in the group at the same time. There is a tradeoff, however; a larger group of constraints solved simultaneously means a larger linear system to solve, which in general scales with the number of constraints cubed. Thus, we desire some middle ground that incorporates some constraint coupling effects while maintaining computational efficiency.


For rigid bodies, a natural choice for constraint groupings is the set of constraints that make up the vector-valued joint constraints. These groups of constraints not only affect exactly the same particles, but also have computational formulas described by matrices, naturally leading to simultaneous computation.
For example, considering the general $N \times 1$ joint constraint $\mathbf{C}_\mathrm{jnt}$, we would solve the $N\times N$ linear subsystem:
\begin{equation}
\begin{aligned}
    \left[ \nabla \mathbf{C}_\mathrm{jnt}(\boldsymbol{x}^k) \boldsymbol{\mathcal{M}}^{-1} 
    \nabla \mathbf{C}_\mathrm{jnt}(\boldsymbol{x}^k)^\top + \boldsymbol{\tilde{\alpha}}_\mathrm{jnt} \right] 
    \Delta \boldsymbol{\lambda}^k_\mathrm{jnt} = \\ -\mathbf{C}_\mathrm{jnt}(\boldsymbol{x}^k) 
    - \boldsymbol{\tilde{\alpha}}_\mathrm{jnt} \boldsymbol{\lambda}^k_\mathrm{jnt}
\end{aligned}
\end{equation}
where $\boldsymbol{\lambda}^k_\mathrm{jnt}$ and $\boldsymbol{\tilde{\alpha}}_\mathrm{jnt}$ are respectively the Lagrange multipliers and compliances associated with the fixed joint constraint. The above system is SPD and never exceeds $6 \times 6$ (for joint constraints) so computing its solution is quite efficient.

\subsection{Block-Banded Solver for Rods}
\label{sec:block-banded-solver-for-rods}
\begin{figure}
    \centering
    \includegraphics[width=1\linewidth]{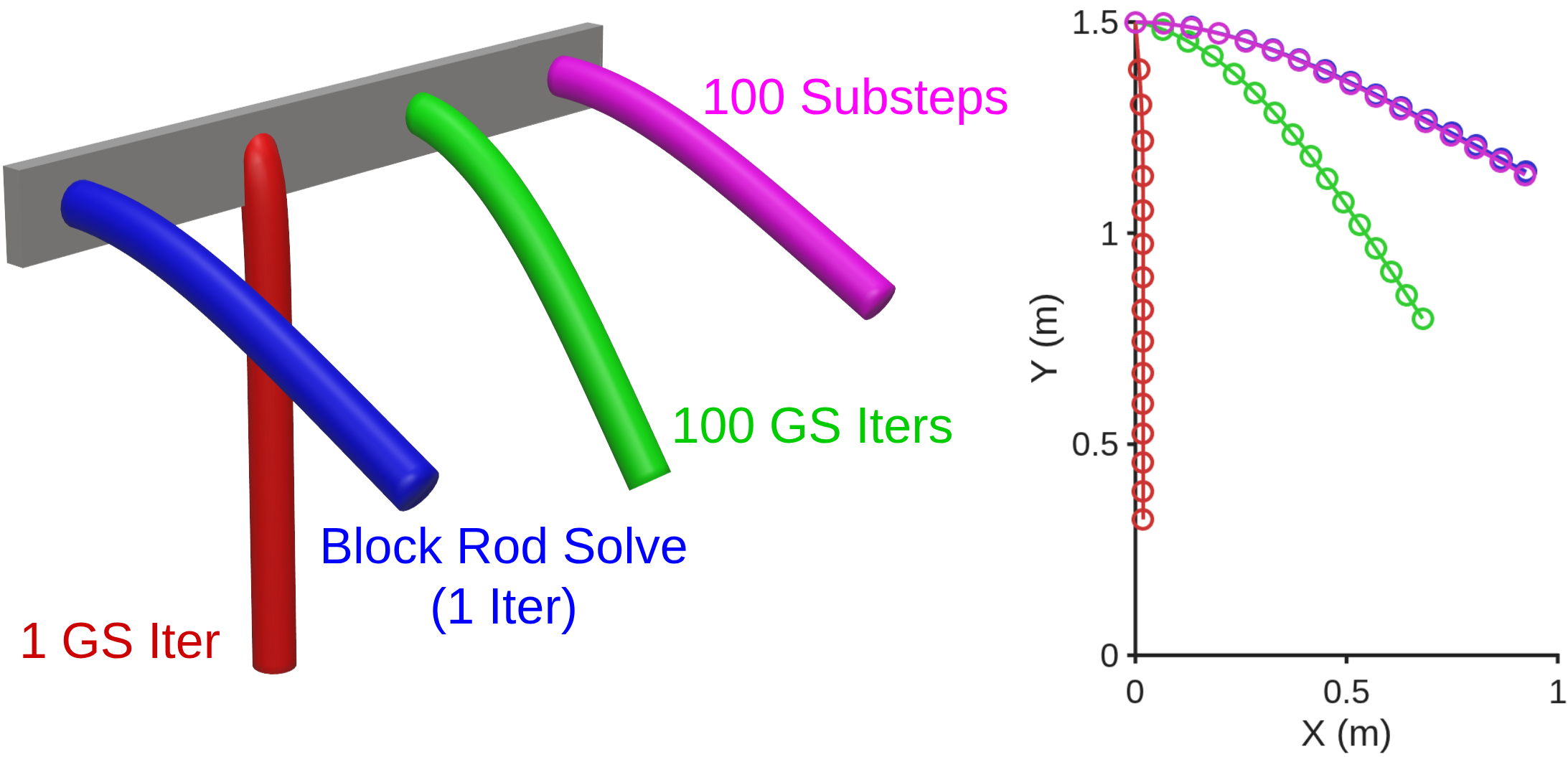}
    \caption{Different methods for solving the linear system for $\Delta \boldsymbol{\lambda}$ for a cantilevered rod with 5 cubic elements. A time step of $1\times 10^{-2}$ seconds is used. The $O(n)$ block rod solve (\textbf{blue}) maintains the correct stiffness in the rod in one iteration, while a single Gauss-Seidel iteration of individual solves (\textbf{red}) is completely flaccid, indicating non-convergence. 100 Gauss Seidel iterations per time step (\textbf{green}) is stiffer, but still not fully converged. 100 substeps (\textbf{pink}) are needed to approach the stiffness of a single global solve.}
    \label{fig:rod-global-solve}
\end{figure}

Another manifestation of the local nature of the single-constraint update in \eqref{eq:position-orientation-lambda-update} is poor convergence for long chains of constraints, simply because information takes many iterations to propagate throughout the structure. For example, using the single-constraint solver, a position correction at the base of a rod will take at a minimum $N$ iterations to affect the tip, where $N$ is the number of nodes in the rod. In practice, it takes many more iterations to converge \cite{kugelstadt2016position, hsu2025stable}.
Poor convergence often manifests as nonphysical "floppiness" and/or other simulation artifacts. 
Instead of the single-constraint solve, we follow \cite{deul2018direct} and perform a block solve on the rod elastic constraints, excluding contact constraints. That is, we isolate only the elastic constraints associated with the rod and solve the linear subsystem of equations for $\Delta \boldsymbol{\lambda}$ using a direct solver. 

In general, the computation time for the solution to a dense linear system of $n$ equations scales with $n^3$. As the number of equations $n$ grows, it quickly becomes intractable to solve the system within the time constraints imposed by real-time applications. However, for the isolated rod subsystem, the system matrix $\mathbf{A} = (\nabla \mathbf{C} \boldsymbol{\mathcal{M}}^{-1} \nabla \mathbf{C}^\top + \tilde{\boldsymbol{\alpha}})$ is SPD and \textit{block-banded}. The specialized structure of the system matrix permits the use of a specialized, block-banded solver \cite{quarteroni2000-numerical-mathematics}, which solves the linear system in $O(np^3)$ time, where $p$ is the block size ($p=6$ for the rod constraints).
The block bandwidth (number of nonzero block off-diagonals) is equal to $2m-1$, where $m$ is the number of Gauss quadrature points used per element. For example, using a single Gauss quadrature point per rod element (i.e. for rods with linear basis functions) leads to a block-tridiagonal system (block bandwidth = 1).  

Solving the system this way results in vastly superior convergence to the individual constraint solves (and at the same time complexity). This is demonstrated in Figure \ref{fig:rod-global-solve}, where nonconvergence manifests as extreme floppiness in the rod, and 100 substeps are required to even approach the behavior of a single global solve.
Additionally, we note that constraints fixing the base or tip node of the rod can be included in the rod subsystem without changing the block bandwidth of the system. These constraints are arguably the most important to include in the block rod solve solve, so that the computed XPBD updates for the rod can react appropriately to its base and/or tip being fixed.


\section{Experiments and Results}
We have implemented our method in C++ for execution on a single CPU core. Pseudocode for a single simulation time step can be found in Algorithm \ref{alg:time-step}. The simulations were performed on a AMD Ryzen 9 7900X CPU, and timings and parameters for experiments with Cosserat rods are detailed in Table \ref{tab:experiment-timings}.

\begin{algorithm}
\caption{Simulation time step}
\label{alg:time-step}
\begin{algorithmic}[1]
\State collision detection using $\boldsymbol{x}$ \Comment{Sec. \ref{sec:collision-rigid-rod}, \ref{sec:collision-rod-rod}}
\State $\boldsymbol{x} \gets$ inertial update \Comment{Eqs. \eqref{eq:position-inertial-update},\eqref{eq:rotation-inertial-update}}
\Statex \footnotesize{\textit{Position-level solve}}
\For{$n_{iter}$ iterations}
    \State \Call{BlockRodSolve}{each rod} \Comment{Sec. \ref{sec:block-banded-solver-for-rods}}
    \State \Call{BlockJointSolve}{each joint} \Comment{Sec. \ref{sec:block-constraint-solve}}
    \State \Call{SolveConstraint}{each remaining constraint} \Comment{Eq. \eqref{eq:position-orientation-single-lambda-update}}
\EndFor
\State \Call{ApplyFriction}{each active col. constraint} \Comment{Sec. \ref{sec:friction}}
\State $\boldsymbol{v}^{n+1} \gets \frac{1}{\Delta t}( \boldsymbol{x}^{n+1} \boxminus \boldsymbol{x}^n)$ \Comment{velocity update}
\Statex \footnotesize{\textit{Velocity-level solve}}
\State \normalsize\Call{BlockRodVelSolve}{each rod} \Comment{Secs. \ref{sec:velocity-damping-solve}, \ref{sec:block-banded-solver-for-rods}}
\State \Call{BlockJointVelSolve}{each joint} \Comment{Secs. \ref{sec:velocity-damping-solve}, \ref{sec:block-constraint-solve}}
\State \Call{ConstraintVelSolve}{each remaining constraint} \Comment{Sec. \ref{sec:velocity-damping-solve}}
\State \Call{ApplyRestitution}{each active col. constraint} \Comment{Sec. \ref{sec:restitution}}
\end{algorithmic}
\end{algorithm}

\subsection{Rigid Body Constraints}
\label{sec:rigid-body-constraints}

\begin{figure*}
    \centering
    \includegraphics[width=1\textwidth]{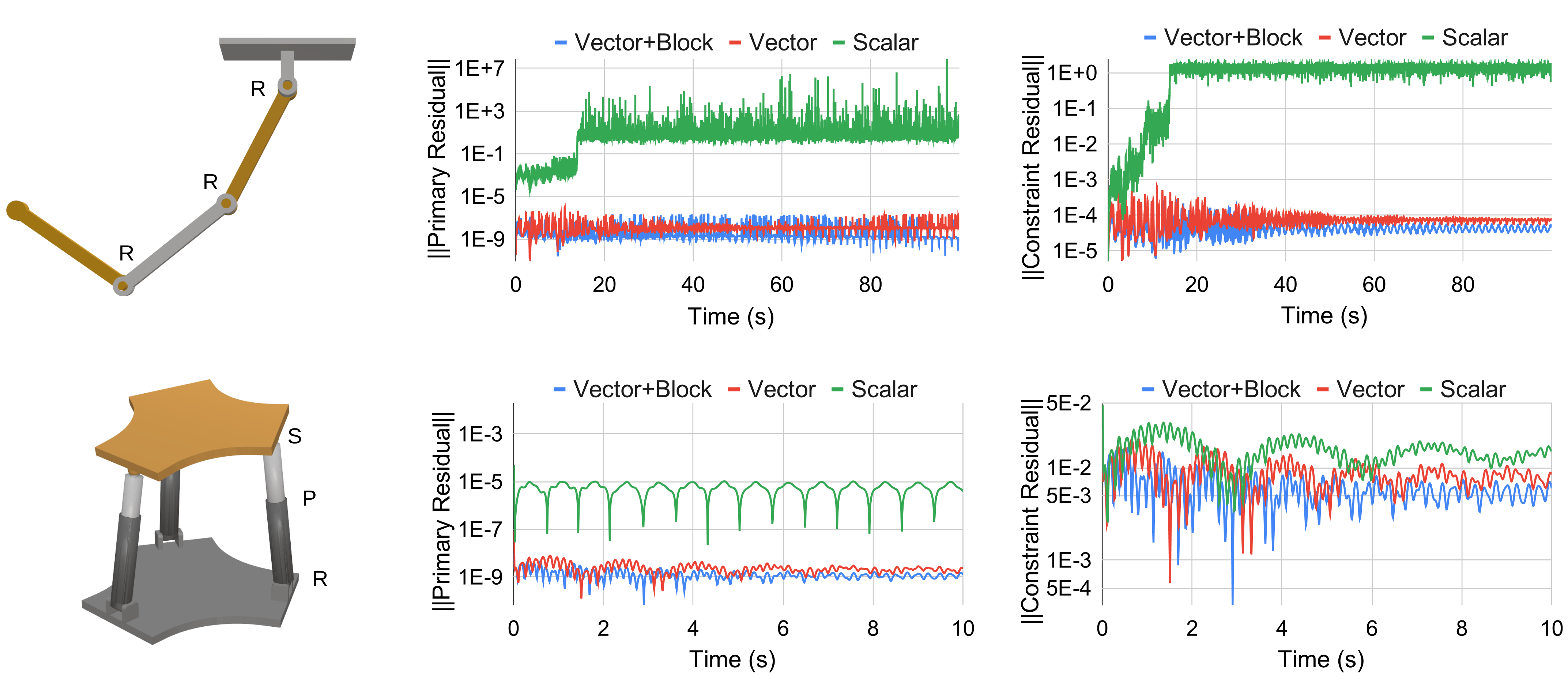}
    \caption{Plots of residuals for rigid body scenarios with different combinations of constraint formulations and solve types (vector constraints + block solves, vector constraints + individual solves, scalar constraints + individual solves). \textbf{Left.} A triple pendulum released from near horizontal. The scalar constraint formulation goes unstable at $t=15$ seconds. The vector constraint formulation has $10^5\times$ lower primal residual from $0$ to $15$ seconds. \textbf{Right.} Inverse dynamics of a 3-RPS parallel robot. The vector constraint formulation with block constraint solves has on average a $10^4\times$ lower primal residual than the scalar constraint formulation.}
    \label{fig:rigid-body-experiments}
\end{figure*}

\begin{table}[]
    \centering
    \caption{Timings per substep (in $\mu$s) for the rigid-body scenarios performed with different constraint and solver combinations.}
    \label{tab:pendulum-timings}
    \begin{tabular}{|c||c|c|c|}
    \hline
        & Vector + Block & Vector & Scalar \\
    \hline
       Triple Pendulum  & 1.29 & 1.39 & 1.24  \\
     \hline
      3-RPS Robot & 3.41 & 5.36 & 3.36 \\
      \hline
    \end{tabular}
\end{table}

For the following rigid-body experiments, we compared three different solver and constraint combinations: individual constraint solves with scalar constraints (i.e. the approach used in \cite{muller2020detailed}), individual constraint solves with vector-valued constraints, and block constraint solves with vector-valued constraints. The approaches are compared in terms of the primary and constraint residuals, which are the left hand side of \eqref{eq:position-orientation-primary-eq} and \eqref{eq:position-orientation-constraint-eq}, respectively. The primal residual quantifies the violation of the discrete equations of motion. It measures how well the update from the inertial prediction $\tilde{\mathbf{x}}^n$ to $\boldsymbol{x}^{n+1}$ is explained by the applied constraint forces. Smaller residuals indicate greater dynamic consistency and closer convergence to the fixed point. The constraint residual measures the violation of the imposed constraints. Smaller residuals indicate greater kinematic consistency and more accurate constraint satisfaction.

\paragraph{Triple Pendulum} In the first experiment, a triple pendulum was released from an initial angle of $80$ degrees from the downward vertical, and swings for $100$ seconds. The residuals are plotted in Figure \ref{fig:rigid-body-experiments}, left. Upon initial release, the vector constraint formulation has over $10^5$ times better primal residual and over $10$ times better constraint residual compared to the scalar constraint formulation. As the simulation progresses, the scalar constraints become unstable, and the simulation blows up at roughly $t=15$ seconds, as indicated by the spike in residuals. As the simulation progresses, the block solve with vector constraints maintains a $10\times$ lower primal residual and a $2\times$ lower constraint residual than the individual constraint solves.

\paragraph{3-RPS Robot} In the second experiment, the inverse dynamics of a 3-RPS parallel robot was simulated for $10$ seconds. The robot consists of three parallel revolute-prismatic-spherical joint chains, resulting in three degrees of freedom of the platform. A constraint dictating the height and normal of the top platform evolved in time, causing the platform to move and the joints to respond accordingly. The vector constraint formulation with individual constraint solves has nearly $10^4$ times lower primal residual compared to using scalar constraints, and the vector constraints with block constraint solves has two times lower primal residual than that. The difference in constraint residual is less pronounced, but again the block solve with vector constraints is the best approach. 

\paragraph {Performance} The average time taken per time step (in microseconds) is given in Table \ref{tab:pendulum-timings}. Of particular note is the nearly negligible difference in execution time between the scalar constraints and vector constraints with the block solve. This makes sense, as the only real difference in computation between them is a small SPD system solve, which is a nearly negligible amount of operations. The vector constraints with individual solves suffers from having to recompute parts of the constraints multiple times, leading to slower overall execution times. 


\subsection{Cosserat Rod Convergence}
\begin{figure}
    \centering
    \includegraphics[width=1\linewidth]{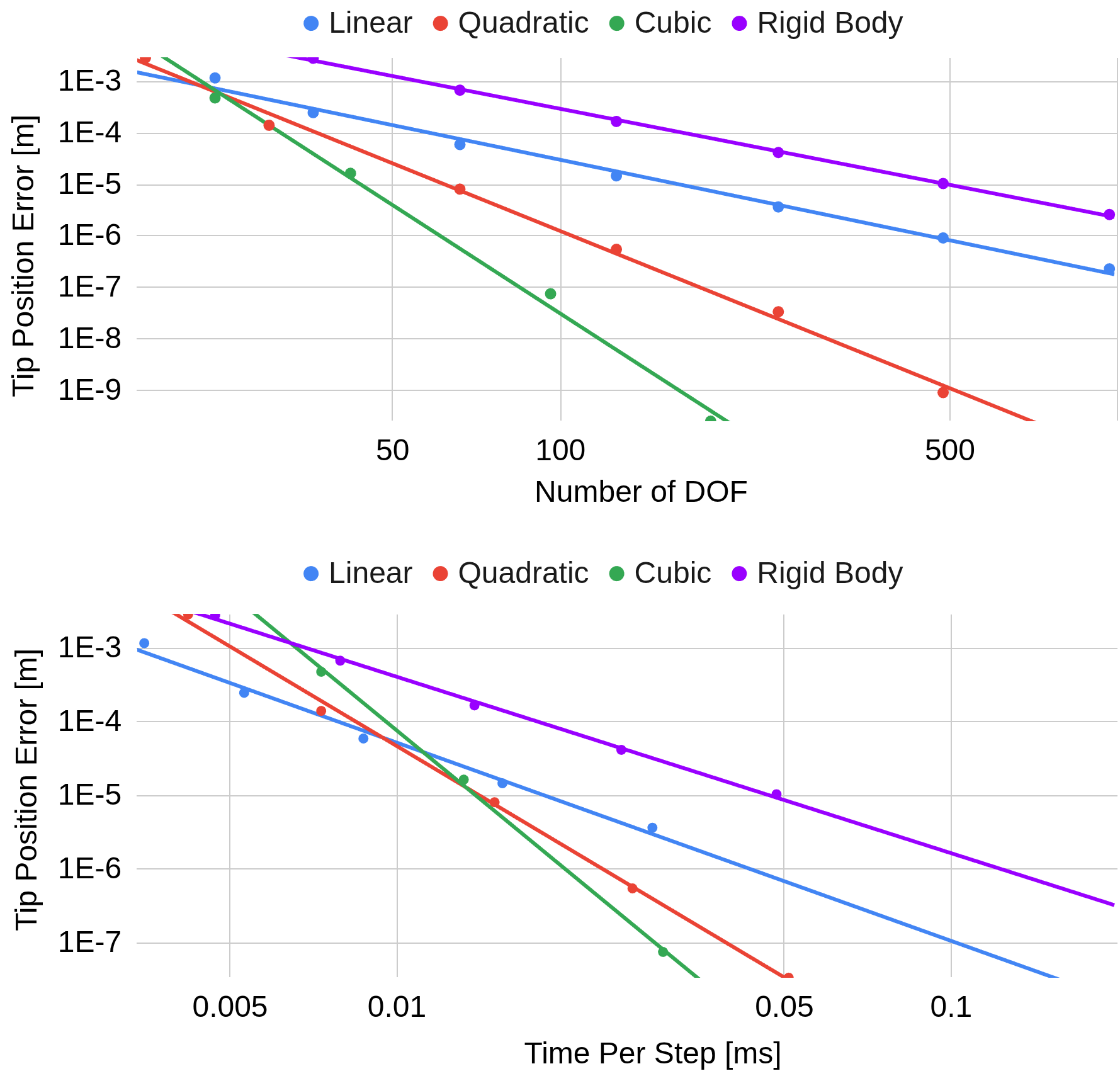}
    \caption{Convergence study for a large-deflection cantilever rod subjected to gravitational loading. The error metric was tip position error, relative to a converged simulation of a cubic rod with $500$ elements. \textbf{Top.} Convergence of different discretizations in terms of total degrees of freedom (6 times number of nodes). The higher-order basis functions exhibit higher accuracy and the expected accelerated convergence rates towards the reference solution. Despite having the same convergence rate, the finite-element approach with linear basis functions is more accurate than the chain-of-rigid-bodies approach due to the proper handling of $SO(3)$ rotations. \textbf{Bottom.} Efficiency of different Cosserat rod discretizations. Execution time per time step is on the X-axis, which accounts for the increased computation required for higher-order basis functions. The linear and quadratic finite-element parameterizations always have greater accuracy for the same time budget than the rigid-body discretization. }
    \label{fig:convergence}
\end{figure}

The steady-state accuracy of different Cosserat rod discretizations was compared at different levels of refinement. The discretizations compared were a rigid-body discretization (similar to \cite{deul2018direct}) and finite-element discretizations with linear, quadratic, and cubic Lagrange basis functions. The tested scenario was a large-deflection, cantilevered Cosserat rod under gravitational loading, with a Young's modulus of 5 MPa, a diameter of $0.1$ meters, a length of $1$ meter, and a density of $1000$ kg/m$^3$. For each rod, a steady-state result was achieved by allowing a dynamic simulation with a time step of $5\times10^{-3}$ seconds to come to rest from the inherent numerical damping in backward Euler integration. A converged reference solution was generated with a finite-element rod with $500$ cubic elements. The results of the convergence study are shown in Figure \ref{fig:convergence}.

In the top plot, the tip error is plotted against the number of DOF in the rod. Firstly, the linear finite elements are more accurate than the rigid-body discretization for the same number of DOF, though they have the same $O(h^2)$ convergence rate. As shown in \eqref{eq:fem-linear-basis-constraints}, the only difference between the rigid-body and linear finite-element XPBD constraints is the rotational interpolation. Thus, the increase in accuracy can be directly attributed to the proper handling of $SO(3)$ rotation interpolation in the finite-element formulation. As expected, the higher-order elements have superior convergence ($O(h^4)$ for quadratic elements and $O(h^6)$ for cubic elements) and accuracy for the same number of DOF, quickly beating the linear finite-element discretization.

In the bottom plot, the tip error is plotted against execution time per time step (in ms). This way, we normalize against the increased computation required for higher-order basis functions. Due to the linear time complexity of all discretizations, the convergence rates with time on the x-axis remains the same. Despite the increased complexity of linear and quadratic basis functions compared to the rigid-body discretization, they always have greater accuracy for the same time budget than the rigid-body discretization. Furthermore, three quadratic elements gives a more accurate result than ten linear elements at roughly the same computation time, and three cubic elements gives a more accurate result than five quadratic elements at roughly the same computation time. Thus, higher-order parameterizations should be preferred for high-fidelity simulations where accuracy is a concern.

\subsection{Shear Locking}
\label{sec:shear-locking}
\begin{figure}
    \centering
    \includegraphics[width=1\linewidth]{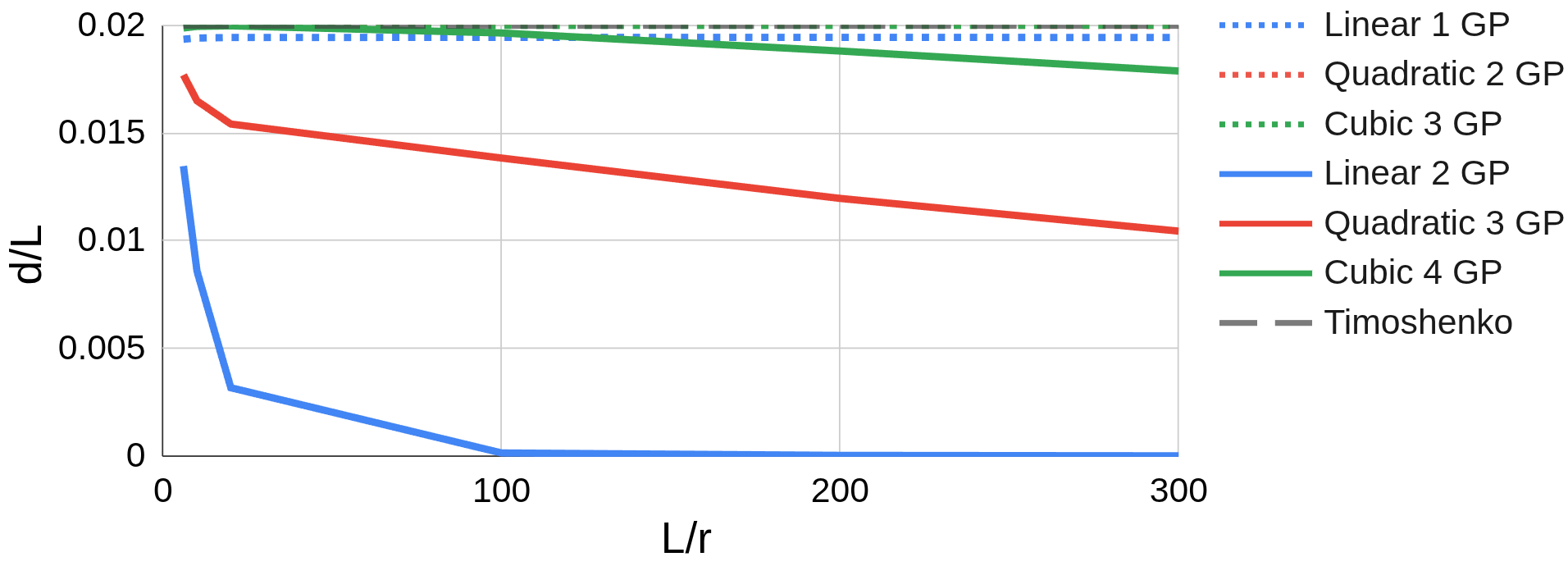}
    \caption{Shear locking of a small-deflection cantilever rod subject to a tip force. A 3-element linear rod, 1-element quadratic rod, and 1-element cubic rod are examined. The length of the rod $L$ is varied while the radius $r$ is kept fixed. The Young's modulus $E$ is adjusted accordingly to maintain an analytical deflection to length ratio ($d/L$) of $0.02$. Under-integration (dotted lines) maintain a fixed error as slenderness ($L/r$) is varied, indicating a lack of shear locking. Full integration (solid lines) exhibit significant shear locking at high slenderness ratios.}
    \label{fig:shear-locking}
\end{figure}

To examine the effect of shear locking in our implementation, we performed a sweep over slenderness ratio $L/r$ for a small-deflection, cantilevered rod subjected to a downward tip force. The analytical solution the tip deflection of a cantilevered, linear Timoshenko beam \cite{timoshenko1930strength} under a prescribed tip load $P$ is:
\begin{equation}
    d = \frac{PL^3}{3EI} + \frac{PL}{GA^*}
\end{equation}
where $A^* = A\kappa_s$ is the shear-corrected cross-sectional area which accounts for the shear stress is not uniformly distributed over the cross section. For a circular cross-section \cite{cowper1966shear}:
\begin{equation}
    \kappa_s = \frac{6(1+\nu)}{7+6\nu}
\end{equation}
where $\nu$ is the Poisson ratio. Using $P=25$ N, $r=0.05$ m, and $\nu=0.4$, we varied the length of the rod and computed the theoretical $E$ required to maintain a $d/L$ ratio of $0.02$. Then, we ran cantilevered beam experiments to steady-state convergence with $h=5\times10^{-3}$ s for a 3-element rod with linear basis functions, a 1-element rod with quadratic basis functions, and a 1-element rod with cubic basis functions. For each type of basis, we tested uniformly reduced integration (one Gauss point, two Gauss points, and three Gauss points for linear, quadratic, and cubic basis functions, respectively) and ``full'' integration (two Gauss points, three Gauss points, and four Gauss points). The results are shown in Figure \ref{fig:shear-locking}.

Clearly, the full integration showcases significant shear locking for higher slenderness ratios as evidenced by the reduction in displacement and the deviation from the analytical solution. On the other hand, reduced integration matches the analytical solution regardless of the slenderness ratio. In fact, the 1-element quadratic and cubic cantilevered rods matched the analytical solution almost exactly with only a $0.1\%$ error.

\subsection{Helical Spring Buckling}
\begin{figure}
    \centering
    \includegraphics[width=1\linewidth]{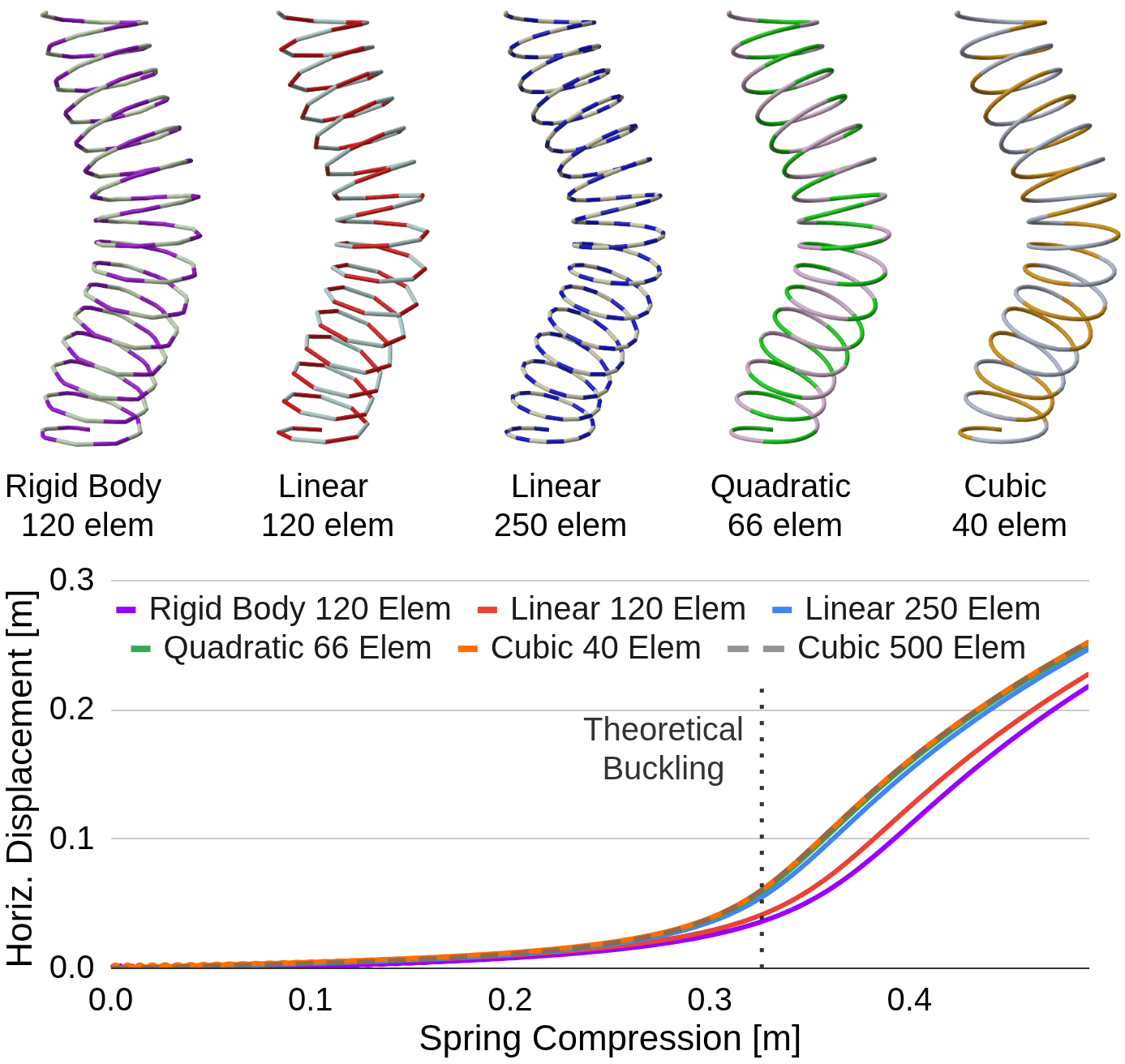}
    \caption{Helical compression springs simulated with different finite-element discretizations are compressed past the point of buckling. \textbf{Top.} Screenshots from the simulation at a spring compression of 0.4 m, well past the buckling point. The rigid-body and linear discretizations are visually jagged compared to the higher-order discretizations. \textbf{Bottom.} Total lateral displacement of the middle rod node against the prescribed compression displacement. Theoretical buckling compression was extracted from \cite{becker1992buckling}, and a ground truth simulation was performed using cubic basis functions and $500$ elements.}
    \label{fig:spring-buckling}
\end{figure}

Helical compression springs can be modeled in terms of the pre-curved elastic wire that they are made of. We tested the bending and torsional coupling of our Cosserat implementation through comparison of 
buckling point to classical mechanics. All springs are made of steel ($E=60$ GPa, $\nu=0.3$), and tests are performed without gravitational effects.


A helical spring can be treated as a vertical column with low shear that buckles under a critical compressive displacement or load.
Theoretical critical displacements can be obtained by solving the differential equations governing the behavior of the elastic wire which makes up the spring. From \cite{becker1992buckling}, a steel spring with a rest length of $1.4$ m, a coil radius of $0.1$ m, and $15$ coils has a critical buckling displacement of about $0.32$ m. We simulated quasistatic compression for different discretizations and measured the total lateral distance that the middle rod node moved throughout the compression. A ground truth simulation was performed using cubic basis functions and $500$ elements. The results are shown in Figure \ref{fig:spring-buckling}. 

All tested discretizations agree reasonably well with the theoretical buckling compression and show qualitatively similar behavior. The rigid-body discretization with 120 elements (purple), linear basis functions with $120$ elements (red), the quadratic basis functions with $66$ elements (green), and the cubic basis functions with $40$ elements (orange) have roughly the same number of DOF, but have different degrees of accuracy and visual quality. The rigid-body discretization and the linear basis functions with $120$ elements are visually jagged and have by far the worst accuracy, with its displacement versus compression curve being noticeably far from the other discretizations. Between the two of them, the displacement-compression curve for the rigid-body discretization is farther from the ground truth. Additionally, the rigid-body discretization has its nodes at the midpoint of the rigid body elements, so the spring appears to have a wider diameter than the other discretizations. The quadratic basis discretization with $66$ elements is smooth and is quite close to the ground truth, while the cubic basis discretization with $40$ elements agrees almost exactly with the ground truth.

The linear basis functions with $250$ elements (blue), the quadratic basis functions with $66$ elements (green), and the cubic basis functions with $40$ elements (orange) all take roughly the same amount of time to complete the simulation. The linear discretization still shows some visual jaggedness, and while it is much closer to the ground truth than the linear discretization with $120$ elements, it is still slightly worse than the quadratic and cubic discretizations.

\subsection{Plectoneme}
\begin{figure}
    \centering
    \includegraphics[width=1\linewidth]{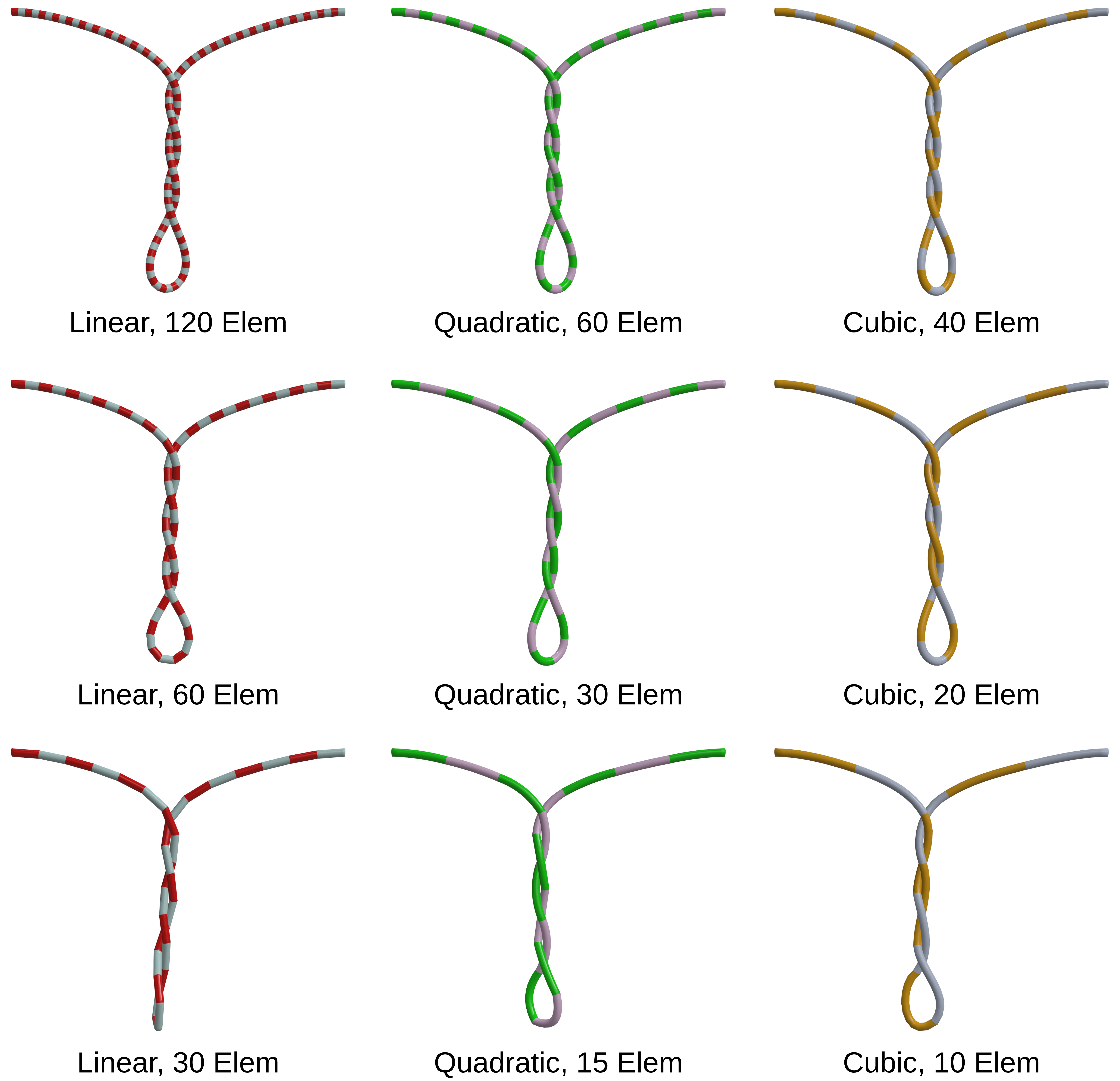}
    \caption{A rod with slack is twisted, forming a plectoneme. Different basis functions and numbers of elements are tested. Each row has approximately the same number of nodes in the discretization. For the same number of nodes, higher-order basis functions produce smoother results, but at a higher computational cost.}
    \label{fig:plectoneme}
\end{figure}

A common benchmark in the rod graphics literature is the formation of plectonemes, where a rod with slack is twisted with one end fixed, causing torsional windup in the rod. We test the formation of a plectoneme for the same rod and same applied boundary conditions with different discretizations, shown in Figure \ref{fig:plectoneme}. Each row in the figure has approximately the same number of nodes in the discretization, with different basis functions (linear, quadratic, cubic). Regardless of the discretization, the behavior of the plectoneme is roughly the same. At coarser discretizations (Figure \ref{fig:plectoneme}, bottom row), the use of quadratic and cubic basis functions results in much smoother solutions than with linear basis functions. Furthermore, coarser approximations with higher-order basis functions result in qualitatively similar behavior to more refined approximations, demonstrating their improved convergence. In the bottom row, the quadratic and cubic basis functions yield a result more similar to the converged solution in the top row. Similarly, in the middle row of Figure \ref{fig:plectoneme}, the quadratic and cubic basis functions produce a result nearly identical to the row above it, while the loop of the linear solution has not turned all the way.

\subsection{Bowling}
The string-pin bowling scene shown in Figure \ref{fig:teaser}, Right consists of ten rigid pins, each attached to a long, stiff rope ($E=1$ GPa) simulated with only $8$ cubic elements. The other end of the stiff rope is attached to an elastic prismatic joint to simulate slack in the rope. A bowling ball is rolled towards the pins at high velocity ($10$ m/s), creating high-impact collisions between pins and strings. After the throw, the pins are raised and untangled before set down again for the next throw. Despite only consisting of $8$ elements, the rope exhibits smooth deformation and high stability. With a time step of $h=2\times 10^{-3}$ seconds, this scenario can be simulated faster than real-time.

\subsection{Bristlebot}

\begin{figure}
    \centering
    \includegraphics[width=1\linewidth]{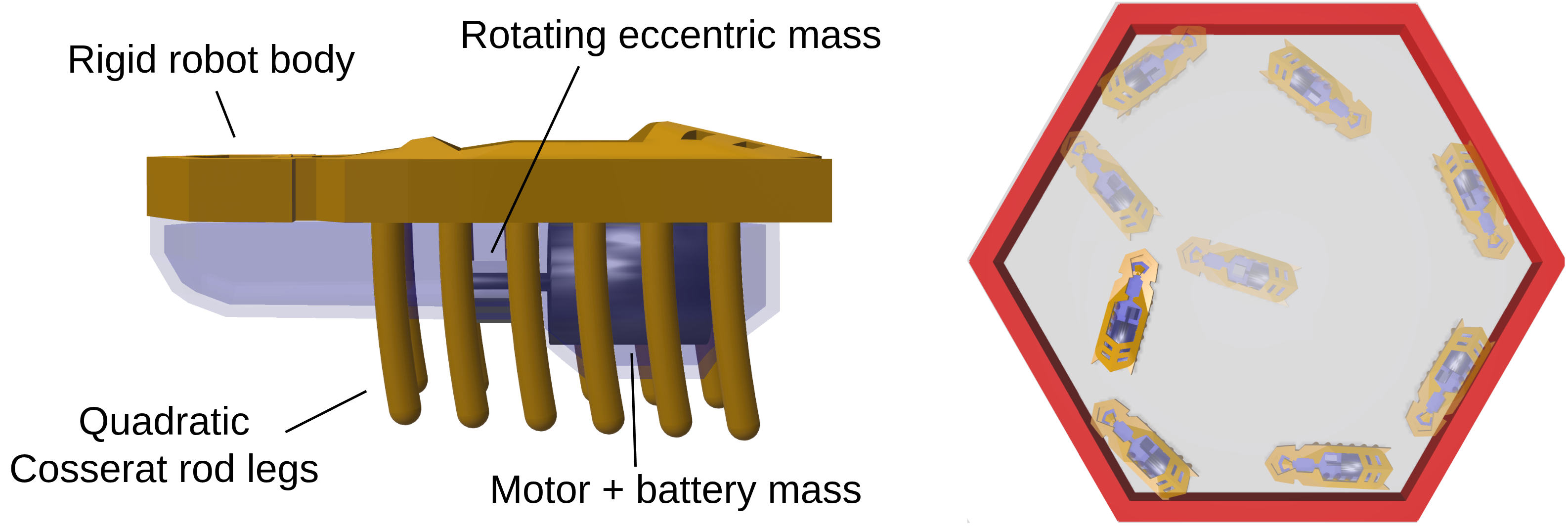}
    \caption{\textbf{Left.} Diagram of the bristlebot construction. \textbf{Right.} The bristlebot colliding with a walled enclosure, visualized every 10 frames.}
    \label{fig:hexbug-closeup}
\end{figure}

The bristlebot shown in Figure \ref{fig:hexbug-closeup} consists of a rigid body mounted on 12 pre-curved compliant legs. A high-speed eccentric rotating mass (>500 Hz) induces periodic vibrations throughout the robot. The compliance and inclination of the legs produce asymmetric stick-slip interactions with the ground, resulting in an anisotropic friction response that converts the oscillatory vibrations into a net forward motion.
The legs are modeled as single-element Cosserat rods with  quadratic basis functions. To limit the effect of numerical damping and to capture the fast, vibrational dynamics of the robot, a time step of $h=1\times10^{-4}$ seconds is used. Remarkably, the scenario in Figure \ref{fig:hexbug-closeup}, right, can be simulated faster than real-time, despite the small time step used.

\begin{table*}[]
    \centering
    \setlength{\tabcolsep}{3.4pt}
    \caption{Timings and parameters for experiments involving Cosserat rods. All quantities reported in SI units unless specified otherwise. A ``-'' indicates the same value as the row above. ``Time/step'' is the average execution time per step, specified in milliseconds.}
    \label{tab:experiment-timings}
    \begin{tabular}{|l|cccccccc|cc|c|}
    \hline
        & \multicolumn{8}{c|}{\textit{Rod params}}
        & \multicolumn{2}{c|}{\textit{Timing}}
        & \\
        & Elem. Type & (\# Rods)\# Elems & Length & Dia. & $E$ & $\nu$ & $\rho$ & $\beta$ &  $h$ [ms] & Time/step [ms] & Collisions \\
    \hline
       \multicolumn{1}{|l|}{\textit{Spring Buckling} (Fig. \ref{fig:spring-buckling})} &  &  & 9.5 & 0.01 & 60e9 & 0.3 & 7650 & 0 & 5 &  & No \\
       \quad Rigid Body & RB & 120 & - & - & - & - & - & - & - & 0.09 & - \\
       \quad Linear     & Lin & 120 & - & - & - & - & - & - & - & 0.10 & - \\
                  & Lin & 250 & - & - & - & - & - & - & - & 0.22 & - \\
       \quad Quadratic  & Quad & 66 & - & - & - & - & - & - & - & 0.22 & - \\
       \quad Cubic      & Cub & 40 & - & - & - & - & - & - & - & 0.28 & - \\
       \hline
       \multicolumn{1}{|l|}{\textit{Plectoneme} (Fig. \ref{fig:plectoneme})} & & & 10 & 0.1 & 5e7 & 0.4 & 1000 & 1e6 & 1 & & Yes \\
       \quad Linear       & Lin & 30 & - & - & - & - & - & - & - & 0.06 & - \\
                    & Lin & 60 & - & - & - & - & - & - & - & 0.11 & - \\
                    & Lin & 120 & - & - & - & - & - & - & - & 0.23 & - \\
       \quad Quadratic    & Quad & 15 & - & - & - & - & - & - & - & 0.11 & - \\
                    & Quad & 30 & - & - & - & - & - & - & - & 0.23 & - \\
                    & Quad & 60 & - & - & - & - & - & - & - & 0.43 & - \\
        \quad Cubic       & Cub & 10 & - & - & - & - & - & - & - & 0.17 & - \\
                    & Cub & 20 & - & - & - & - & - & - & - & 0.35 & - \\
                    & Cub & 40 & - & - & - & - & - & - & - & 0.68 & - \\
        \hline
       \multicolumn{1}{|l|}{\textit{Bristle Bot}} & Quad & (12)1 & 0.015 & 2e-3 & 4.5e6 & 0.4 & 1000 & 0 & 0.1 & & Yes \\
       \quad Single (Fig. \ref{fig:hexbug-closeup})      & - & - & - & - & - & - & - & - & - & 0.07 & - \\
       \quad Single in Enclosure (Fig. \ref{fig:hexbug-closeup}) & - & - & - & - & - & - & - & - & - & 0.08 & - \\
       \quad Multi in Enclosures (Fig. \ref{fig:teaser}) & - & (36)1 & - & - & - & - & - & - & - & 0.23 & - \\
       \hline
       \multicolumn{1}{|l|}{\textit{Bowling} (Fig. \ref{fig:teaser})} & Cub & (10)8 & 1.5 & 0.01 & 1e8 & 0.4 & 1150 & 10 & 2 & 1.18 & Yes \\
      \hline
    \end{tabular}
\end{table*}

\section{Conclusion}
We presented a rigorous framework for incorporating rotations into XPBD using Lie theory, including constraint definitions and their gradients, rotational interpolation, and differentiation. We demonstrated that the choice of constraint definition strongly influences simulation accuracy and convergence: the more linear the constraints, the better their numerical behavior. Guided by this principle, our explicit Lie-theoretic formulations for rigid-body joint constraints achieve primary residuals that are multiple orders of magnitude lower than the implicit constraints of \cite{muller2020detailed}. We further showed that grouping related constraints and solving the resulting XPBD subproblems with block Gauss-Seidel improves convergence and reduces residuals.

Applying this framework to Cosserat rods, Lie theory provides a natural on-manifold interpolation of rotations, enabling arbitrary basis functions for both positions and orientations within a Cosserat finite element. By enforcing constraints at Gauss quadrature points, we establish a direct connection to the geometrically exact Cosserat rod formulation while enabling efficient higher-order rod simulation in XPBD. Compared with the standard chain-of-rigid-bodies approach, linear finite elements achieve higher accuracy at comparable computational cost. We further demonstrated that higher-order basis functions provide smoother solutions and faster convergence across a range of benchmark problems, making them more computationally efficient than linear bases for high-fidelity simulations. Finally, we introduced collision detection and response for higher-order rod elements and demonstrated coupled rigid-body and rod simulations with contact.

Overall, this work demonstrates that Lie theory provides a principled and practical foundation for extending XPBD to manifold-valued state variables. Beyond the applications presented here, the resulting framework offers a general blueprint for incorporating nonlinear geometric structure into position-based simulation and related physics-based methods.

\begin{acks}
Research reported in this publication was supported by the Advanced Research Projects Agency for Health (ARPA-H) under the ALISS project, Award Number D24AC00415-00. The ARPA-H award of up to \$11,935,038 provided 100\% of the financial support for this work. The opinions and findings in this paper is solely the responsibility of the authors and do not necessarily represent the official views of ARPA-H.
\end{acks}

\appendix
\section{Equivalent Definitions of Angular Velocity}
\label{sec:equivalent-definitions-angular-velocity}
The traditional definition of body angular velocity is given as:
\begin{equation}
    \boldsymbol{\omega}(t) = \left( \mathbf{R}(t)^\top \frac{d\mathbf{R}}{dt} \right)^\vee
\end{equation}
with $d\mathbf{R}/dt$ belonging to $T_{\mathbf{R}(t)}SO(3)$. Alternatively, in Section \ref{subsec:differentiation-of-3D-rotations}, we propose that body angular velocity can be defined as
\begin{equation}
    \boldsymbol{\omega}(t) = \lim_{\Delta t \to 0} \frac{\mathbf{R}(t+\Delta t) \boxminus \mathbf{R}(t)}{\Delta t} = \mathcal{D}_t(\mathbf{R}(t))
\end{equation}
To see these are equivalent mathematically, we can use the Taylor expansion $${\mathbf{R}}(t+\Delta t)=\mathbf{R}(t) + \dot{\mathbf{R}}(t) \Delta t + O(\Delta t^2)$$ and the fact that for sufficiently small $\Delta t$ a series expansion of $\log$ yields: $$\log(\mathbf{I} + \Delta t\mathbf{A})=\mathbf{A} + O(\Delta t^2)$$
Starting from the definition of $\mathcal{D}_t(\mathbf{R}(t))$:
\begin{equation}
\begin{aligned}
    \mathcal{D}_t (\mathbf{R}(t)) &= \lim_{\Delta t \to 0} \frac{\left[ \log(\mathbf{R}(t)^\top \mathbf{R}(t + \Delta t)) \right]^\vee}{\Delta t} \\
    &= \lim_{\Delta t \to 0} \frac{\left[ \log( \mathbf{I} + \mathbf{R}(t)^\top\dot{\mathbf{R}}(t) \Delta t + O(\Delta t^2)) \right]^\vee}{\Delta t} \\
    &= \lim_{\Delta t \to 0} \frac{\left[ \mathbf{R}(t)^\top \dot{\mathbf{R}}(t) \Delta t + O(\Delta t^2) \right]^\vee}{\Delta t} \\
    &= \left[ \mathbf{R}(t)^\top \dot{\mathbf{R}}(t) \right]^\vee \\
    &= \boldsymbol{\omega}_b
\end{aligned}
\end{equation}
Note that this logic applies directly to the definition of continuous curvature in Section \ref{sec:continuous-cosserat-rod}:
\begin{equation}
    \mathbf{u}(s) = \left( \mathbf{R}(s)^\top \frac{d\mathbf{R}}{ds} \right)^\vee = \lim_{\Delta s \to 0} \frac{\mathbf{R}(s+\Delta s) \boxminus \mathbf{R}(s)}{\Delta s}
\end{equation}

\section{Forces and Torques Derived from Energy Potential}
\label{sec:foces-torques-derived-from-energy}
By the principle of virtual work, the force derived from an energy potential $U$ is:
\begin{equation}
    \delta W = \mathbf{f} \cdot \delta \mathbf{p}
\end{equation}
Setting $\delta W = -\delta U$:
\begin{equation}
    \delta W = -\frac{\partial U}{\partial \mathbf{p}} \cdot \delta \mathbf{p}
\end{equation}
Therefore, $\mathbf{f} = -\partial_{\mathbf{p}_i} U$. We can apply the exact same argument to torques derived from an energy potential. We express a virtual displacement in rotation as the body-frame rotation vector $\delta \boldsymbol{\theta} \in \mathbb{R}^3$. Then, the virtual work is:
\begin{equation}
    \delta W = \boldsymbol{\tau} \cdot \delta \boldsymbol{\theta}
\end{equation}
where $\boldsymbol{\tau}$ is the body-frame torque associated with the angular perturbation. The variation of $U$ in the direction of $\delta \boldsymbol{\theta}$ is:
\begin{equation}
    \delta U = \frac{d}{d \epsilon} \bigg\rvert_{\epsilon=0} U(\mathbf{R} \exp(\epsilon \widehat{\delta \boldsymbol{\theta}}))
\end{equation}
which can be equivalently written using $\boxplus$ and limits to mirror the definition in \eqref{eq:R-to-SO3-differential}:
\begin{equation}
    \delta U = \lim_{\epsilon \to 0} \frac{U(\mathbf{R} \boxplus \epsilon \delta \boldsymbol{\theta}) - U(\mathbf{R})}{\epsilon} \\
\end{equation}
This can be interpreted as the directional derivative of $U$ along the perturbation $\delta \boldsymbol{\theta}$. Thus,
\begin{equation}
    \delta U = \mathcal{D}_{\mathbf{R}} U \cdot \delta \boldsymbol{\theta},
\end{equation}
and from $\delta W = -\delta U$:
\begin{equation}
    \boldsymbol{\tau} = -\mathcal{D}_{\mathbf{R}}U
\end{equation}

\section{Derivation of Rotational Inertial Update}
\label{sec:derivation-of-rotational-inertial-update}
From \eqref{eq:rotational-dof}, we have the equations of motion for the rotational DOF of a single particle, repeated below:
\begin{equation*}
    \boldsymbol{\mathcal{I}}_i \left( \frac{\boldsymbol{\omega}_i^{n+1} - \boldsymbol{\omega}_i^{n}}{\Delta t}\right) = -\mathcal{D}_{\mathbf{R}_i} U(\boldsymbol{x}^{n+1})^\top + \boldsymbol{\tau}_i^{\mathrm{ext}} - \boldsymbol{\omega}_i^n \times \boldsymbol{\mathcal{I}_i} \boldsymbol{\omega}_i^n
\end{equation*}
We wish to put this in terms of particle rotations, and lump known terms evaluated at the previous time step into an ``inertial prediction'' $\tilde{\mathbf{R}}_i^n$. To start, we substitute in \eqref{eq:omega-n+1},
\begin{equation*}
    \boldsymbol{\omega}_i^{n+1} \approx \frac{\mathbf{R}_i^{n+1} \boxminus \mathbf{R}_i^n}{\Delta t} 
\end{equation*}
and multiply through by $\Delta t^2$ to obtain:
\begin{equation}
    \boldsymbol{\mathcal{I}}_i \left( (\mathbf{R}_i^{n+1} \boxminus \mathbf{R}_i^n) - \Delta t \boldsymbol{\omega}_i^n - \Delta t^2 \boldsymbol{\mathcal{I}}_i^{-1} \boldsymbol{\tau}_i^n \right) = -\Delta t^2 \mathcal{D}_{\mathbf{R}_i} U(\boldsymbol{x}^{n+1})
\end{equation}
where $\boldsymbol{\tau}_i^n = \boldsymbol{\tau}_i^{\mathrm{ext}} - \boldsymbol{\omega}_i^n \times \boldsymbol{\mathcal{I}}_i \boldsymbol{\omega}_i^n$ combines the external applied body torque and the gyroscopic torque on particle $i$. For sufficiently small $\Delta t$, we can assume that the difference $\mathbf{R}_i^{n+1} \boxminus \mathbf{R}_i^n$ is small, and that the sum $\Delta t \boldsymbol{\omega}_i^n + \Delta t^2 \boldsymbol{\mathcal{I}}_i^{-1} \boldsymbol{\tau}_i^n$ is small as well, meaning that we can invoke the identity in \eqref{eq:first-order-expansion-of-relative-rotation-under-right-perturbation} to move the torque term into the inner parentheses:
\begin{equation}
    \boldsymbol{\mathcal{I}}_i \left( \mathbf{R}_i^{n+1} \boxminus (\mathbf{R}_i^n \boxplus (\Delta t \boldsymbol{\omega}_i^n + \Delta t^2 \boldsymbol{\mathcal{I}}_i^{-1} \boldsymbol{\tau}_i^n )) \right) = -\Delta t^2 \mathcal{D}_{\mathbf{R}_i} U(\boldsymbol{x}^{n+1})
\end{equation}
and thus we can define our inertially predicted orientation as:
\begin{equation}
    \tilde{\mathbf{R}}_i^n = \mathbf{R}_i^n \boxplus (\Delta t \boldsymbol{\omega}_i^n + \Delta t^2 \boldsymbol{\mathcal{I}}_i^{-1} \boldsymbol{\tau}_i^n )
\end{equation}
We note that this is equivalent to explicit integration of the orientation over a time step $\Delta t$.

\section{Some Useful Properties and Identities}
\subsection{First-Order Expansion of Relative Rotation Under Right Perturbations}
Define $\boldsymbol{\theta} = \mathbf{R}_1 \boxminus \mathbf{R}_2$ for $\mathbf{R}_1, \mathbf{R}_2 \in SO(3)$ and take some vector $\mathbf{v} \in \mathbb{R}^3$. We will show that for small $\boldsymbol{\theta}$ and small $\mathbf{v}$:
\begin{equation}
\label{eq:first-order-expansion-of-relative-rotation-under-right-perturbation}
    \mathbf{R}_1 \boxminus (\mathbf{R}_2 \boxplus \mathbf{v}) \approx (\mathbf{R}_1 \boxminus \mathbf{R}_2) - \mathbf{v}
\end{equation}
Proof: We start by writing the left hand side in terms of $\log$ and $\exp$:
\begin{equation}
\begin{aligned}
\label{eq:R1-min-R2-plus-v-expansion}
    \mathbf{R}_1 \boxminus (\mathbf{R}_2 \boxplus \mathbf{v}) &= \log \left[ (\mathbf{R}_2 \exp(\widehat{\mathbf{v}}))^\top \mathbf{R}_1 \right]^\vee \\
    &= \log \left[ \exp(-\widehat{\mathbf{v}}) \mathbf{R}_2^\top \mathbf{R}_1\right]^\vee \\
    &= \log \left[ \exp(-\widehat{\mathbf{v}}) \exp(\widehat{\boldsymbol{\theta}}) \right]^\vee
\end{aligned}
\end{equation}
A right perturbation inside the matrix logarithm can be first-order approximated with:
\begin{equation*}
    \log(\mathbf{R} \exp (\widehat{\boldsymbol{\delta \mathbf{R}}}))^\vee \approx \log(\mathbf{R})^\vee + \boldsymbol{\Gamma}^{-1}(\log(\mathbf{R})^\vee) \delta \mathbf{R}
\end{equation*}
where $\boldsymbol{\Gamma}^{-1}$ is the right Jacobian of the $SO(3)$ logarithm map given by \eqref{eq:jacobian-of-log}.
Applying the above expression to \eqref{eq:R1-min-R2-plus-v-expansion}:
\begin{equation}
\begin{aligned}
        \log \left[ \exp(-\widehat{\mathbf{v}}) \exp(\widehat{\boldsymbol{\theta}}) \right]^\vee &\approx \log(\exp(-\widehat{\mathbf{v}}))^\vee + \boldsymbol{\Gamma}^{-1}(\log(\exp(-\widehat{\mathbf{v}}))^\vee) \boldsymbol{\theta} \\
        &= -\mathbf{v} + \boldsymbol{\Gamma}^{-1}(\mathbf{v})  \boldsymbol{\theta}
\end{aligned}
\end{equation}
From the expression in \eqref{eq:jacobian-of-log}, for small $\mathbf{v}$ we have that:
\begin{equation}
    \boldsymbol{\Gamma}^{-1}(\mathbf{v}) \approx \mathbf{I} + \frac{1}{2} \widehat{\mathbf{v}}
\end{equation}
and therefore:
\begin{equation}
    \log \left[ \exp(-\widehat{\mathbf{v}}) \exp(\widehat{\boldsymbol{\theta}}) \right]^\vee = -\mathbf{v} + \boldsymbol{\theta} + \frac{1}{2} \mathbf{v} \times \boldsymbol{\theta}
\end{equation}
When both $\mathbf{v}$ and $\boldsymbol{\theta}$ are small:
\begin{equation}
    \mathbf{R}_1 \boxminus (\mathbf{R}_2 \boxplus \mathbf{v}) \approx (\mathbf{R}_1 \boxminus \mathbf{R}_2) - \mathbf{v}
\end{equation}
\subsection{Transpose of Jacobian of $SO(3)$ Logarithm Map} A useful property of the Jacobian of the logarithm map for $SO(3)$ is:
\begin{equation}
\label{eq:transpose-of-log-jacobian}
    \boldsymbol{\Gamma}^{-1}(\mathbf{R}_1 \boxminus \mathbf{R}_2)^\top = \boldsymbol{\Gamma}^{-1}(\mathbf{R}_1 \boxminus \mathbf{R}_2) \mathbf{R}_1^\top \mathbf{R}_2
\end{equation}
{Proof:}
First, observe that
\begin{equation*}
\begin{aligned}
    \exp(-(\mathbf{R}_1 \boxminus \mathbf{R}_2)^\wedge) &= [\exp((\mathbf{R}_1 \boxminus \mathbf{R}_2)^\wedge)]^{-1}\\ &= [\exp(\log(\mathbf{R}_2^\top \mathbf{R}_1) )]^{-1} \\ &= [\mathbf{R}_2^\top \mathbf{R}_1]^{-1}= \mathbf{R}_1^\top \mathbf{R}_2
\end{aligned}
\end{equation*}
Therefore, with $\boldsymbol{\theta} = \mathbf{R}_1 \boxminus \mathbf{R}_2$, we have that
\begin{equation*}
    \boldsymbol{\Gamma}^{-1}(\boldsymbol{\theta}) \mathbf{R}_1^\top \mathbf{R}_2 = \boldsymbol{\Gamma}^{-1}(\boldsymbol{\theta}) \exp(-\widehat{\boldsymbol{\theta}})
\end{equation*}
Then, substituting the explicit formulas from \eqref{eq:so3-exp} and \eqref{eq:jacobian-of-log} and multiplying through, we find that
\begin{equation*}
\begin{aligned}
    \boldsymbol{\Gamma}^{-1}(\boldsymbol{\theta})  \exp(-\widehat{\boldsymbol{\theta}}) &= \mathbf{I} - \frac{1}{2} \widehat{\boldsymbol{\theta}} + \left(\frac{1}{||\boldsymbol{\theta}||^2}-\frac{1+\cos||\boldsymbol{\theta}||}{2||\boldsymbol{\theta}||\sin||\boldsymbol{\theta}||}\right) \widehat{\boldsymbol{\theta}}^2\\
    &= \boldsymbol{\Gamma}^{-1}(\boldsymbol{\theta})^\top
\end{aligned}
\end{equation*}
and thus \eqref{eq:transpose-of-log-jacobian} is true.

This becomes useful when deriving the Jacobian of the $\boxminus$ operator with respect to its second argument, see \eqref{eq:jacobian-of-boxminus-wrt-R2-derivation}.

\subsection{Jacobian of Exp and Log Post-Multiplied by its Argument}
Another useful property is that
\begin{equation}
\label{eq:jacobian-of-exp-multiplied-by-arg}
\begin{aligned}
    \boldsymbol{\Gamma}(\boldsymbol{\theta}) \boldsymbol{\theta} &= \boldsymbol{\theta} \\ \boldsymbol{\Gamma}^{-1}(\boldsymbol{\theta}) \boldsymbol{\theta} &= \boldsymbol{\theta}
\end{aligned}
\end{equation}
\textbf{Proof:}
This is clearly seen by post-multiplying the formulas \eqref{eq:jacobian-of-exp} and \eqref{eq:jacobian-of-log} by $\boldsymbol{\theta}$. Since $\widehat{\boldsymbol{\theta}} \boldsymbol{\theta} = \boldsymbol{\theta} \times \boldsymbol{\theta} = 0$, all that remains in both cases is $\mathbf{I} \boldsymbol{\theta} = \boldsymbol{\theta}$. Note that \eqref{eq:jacobian-of-exp-multiplied-by-arg} also applies for the transpose of the $\exp$ and $\log$ Jacobians.

\section{Derivation of Common Jacobians}
Below, we derive the Jacobian blocks presented in Section \ref{sec:jacobian-blocks}. These can be chained together via the chain rule, enabling analytical expressions for complicated intrinsic derivatives.

\subsection{Jacobian of Rotation Applied to Vector}:
Here, our function $f: SO(3) \to \mathbb{R}^3$ is $f(\mathbf{R}) = \mathbf{R} \mathbf{p}$ for some arbitrary fixed vector $\mathbf{p} \in \mathbb{R}^3$. Thus, we use the definition of differentiation given in \eqref{eq:SO3-to-Rn-differential}, and examine column $i$ of the derivative:
\begin{equation}
\left[ \mathcal{D}_\mathbf{R} (\mathbf{R}\mathbf{p}) \right]_i = \lim_{\epsilon \to 0} \frac{(\mathbf{R} \boxplus \mathbf{e}_i \epsilon)\mathbf{p} - \mathbf{R}\mathbf{p}}{\epsilon}
\end{equation}
For sufficiently small $\epsilon$, $\exp(\epsilon \widehat{\mathbf{e}}_i) \approx \mathbf{I} + \epsilon \widehat{\mathbf{e}}_i$, therefore:
\begin{equation}
\begin{aligned}
\left[ \mathcal{D}_\mathbf{R} (\mathbf{R}\mathbf{p}) \right]_i &= \lim_{\epsilon \to 0} \frac{\mathbf{R} (\mathbf{I} + \epsilon\widehat{\mathbf{e}}_i) \mathbf{p} - \mathbf{R} \mathbf{p}}{\epsilon} \\
&= \lim_{\epsilon \to 0} \frac{\epsilon \mathbf{R} \widehat{\mathbf{e}}_i \mathbf{p}}{\epsilon} \\
&= \lim_{\epsilon \to 0} \frac{-\epsilon \mathbf{R} \widehat{\mathbf{p}}  \mathbf{e}_i}{\epsilon} \\
&= -\mathbf{R} \widehat{\mathbf{p}} \mathbf{e}_i
\end{aligned}
\end{equation}
Therefore, the full derivative can be written succinctly as:
\begin{equation}
    \mathcal{D}_\mathbf{R} (\mathbf{R} \mathbf{p}) = -\mathbf{R} \widehat{\mathbf{p}}
\end{equation}

\subsection{Jacobian of Inverse}
Here, our function $g: SO(3) \to SO(3)$ is $g(\mathbf{R}) = \mathbf{R}^{-1}$. Thus, we use the definition of differentiation given in \eqref{eq:SO3-to-SO3-differential}, and again examine a single column $i$:
\begin{equation}
\begin{aligned}
\left[ \mathcal{D}_\mathbf{R} (\mathbf{R}^{-1}) \right]_i &= \lim_{\epsilon \to 0} \frac{( \mathbf{R} \boxplus \mathbf{e}_i \epsilon)^{-1} \boxminus \mathbf{R}^{-1}}{\epsilon} \\
&= \lim_{\epsilon \to 0} \frac{\log \left[ \mathbf{R}  \exp(-\widehat{\mathbf{e}}_i \epsilon)  \mathbf{R}^{-1} \right]^\vee}{\epsilon}
\end{aligned}
\end{equation}
Using the identity $\mathbf{R} \exp(\widehat{\boldsymbol{\omega}}) \mathbf{R}^\top = \exp( (\mathbf{R}\boldsymbol{\omega})^\wedge)$:
\begin{equation}
\begin{aligned}
&= \lim_{\epsilon \to 0} \frac{\log \left[\exp(-\epsilon(\mathbf{R}\mathbf{e}_i)^\wedge) \right]^\vee}{\epsilon} \\
&= \lim_{\epsilon \to 0} \frac{-\epsilon \mathbf{R} \mathbf{e}_i }{\epsilon} \\
&= -\mathbf{R}  \mathbf{e}_i
\end{aligned}
\end{equation}
Therefore, the full gradient is:
\begin{equation}
    \mathcal{D}_\mathbf{R} (\mathbf{R}^{-1}) = -\mathbf{R}
\end{equation}

\subsection{Jacobian of Multiplication with respect to First Argument}:
Here, our function $g : SO(3) \to SO(3)$ is $g(\mathbf{R}_1)=\mathbf{R}_1 \mathbf{R}_2$ for some arbitrary fixed $\mathbf{R}_2 \in SO(3)$. Thus, we use the definition of differentiation given in \eqref{eq:SO3-to-SO3-differential}, and examine a single column $i$:
\begin{equation}
\begin{aligned}
\left[ \mathcal{D}_{\mathbf{R}_1} (\mathbf{R}_1 \mathbf{R}_2) \right]_i &= \lim_{\epsilon \to 0} \frac{ \left( (\mathbf{R}_1 \boxplus \mathbf{e}_i \epsilon) \mathbf{R}_2 \right) \boxminus (\mathbf{R}_1 \mathbf{R}_2)}{\epsilon} \\
&= \lim_{\epsilon \to 0} \frac{\log \left[ \mathbf{R}_2^\top \mathbf{R}_1^\top \mathbf{R}_1 \exp(\epsilon \widehat{\mathbf{e}}_i)  \mathbf{R}_2 \right]^\vee}{\epsilon} \\
\end{aligned}
\end{equation}
Again, we use the identity $\mathbf{R} \exp(\widehat{\boldsymbol{\omega}}) \mathbf{R}^\top = \exp( (\mathbf{R}\boldsymbol{\omega})^\wedge)$ to obtain:
\begin{equation}
\begin{aligned}
&= \lim_{\epsilon \to 0} \frac{\log \left[ \exp( \epsilon( \mathbf{R}_2^\top \mathbf{e}_i)^\wedge) \right]^\vee}{\epsilon} \\
&= \lim_{\epsilon \to 0} \frac{\epsilon \mathbf{R}_2^\top \mathbf{e}_i }{\epsilon} \\
&= \mathbf{R}_2^\top \mathbf{e}_i
\end{aligned}
\end{equation}
Therefore, the full gradient is:
\begin{equation}
    \mathcal{D}_{\mathbf{R}_1} (\mathbf{R}_1 \mathbf{R}_2) = \mathbf{R}_2^\top
\end{equation}

\subsection{Jacobian of Multiplication with respect to Second Argument}
Here, our function $g : SO(3) \to SO(3)$ is $g(\mathbf{R}_2)=\mathbf{R}_1 \mathbf{R}_2$ for some arbitrary fixed $\mathbf{R}_1 \in SO(3)$. Thus, we use the definition of differentiation given in \eqref{eq:SO3-to-SO3-differential}, and examine a single column $i$:
\begin{equation}
\begin{aligned}
\left[ \mathcal{D}_{\mathbf{R}_2} (\mathbf{R}_1 \mathbf{R}_2) \right]_i &= \lim_{\epsilon \to 0} \frac{ \left( \mathbf{R}_1   (\mathbf{R}_2 \boxplus \mathbf{e}_i \epsilon) \right) \boxminus (\mathbf{R}_1 \mathbf{R}_2)}{\epsilon} \\
&= \lim_{\epsilon \to 0} \frac{\log \left[ \mathbf{R}_2^\top \mathbf{R}_1^\top \mathbf{R}_1 \mathbf{R}_2 \exp(\epsilon \widehat{\mathbf{e}}_i) \right]^\vee}{\epsilon} \\
&= \lim_{\epsilon \to 0} \frac{\log \left[ \exp(\epsilon \widehat{\mathbf{e}}_i)\right]^\vee}{\epsilon} \\
&= \mathbf{e}_i
\end{aligned}
\end{equation}
Therefore, the full gradient is:
\begin{equation}
    \mathcal{D}_{\mathbf{R}_2} (\mathbf{R}_1 \mathbf{R}_2) = \mathbf{I}
\end{equation}

\subsection{Jacobian of Boxminus Operator}
The gradient of $\mathbf{R}_1 \boxminus \mathbf{R}_2$ with respect to the first argument can be found through the chain rule:
\begin{equation}
    \label{eq:jacobian-of-boxminus-wrt-R1-derivation}
    \begin{aligned}
    \mathcal{D}_{\mathbf{R}_1} (\mathbf{R}_1 \boxminus \mathbf{R}_2) &= \mathcal{D}_{ (\mathbf{R}_2^\top \mathbf{R}_1)}\log(\mathbf{R}_2^\top \mathbf{R}_1)^\vee \cdot\mathcal{D}_{\mathbf{R}_1}(\mathbf{R}_2^\top \mathbf{R}_1) \\
    &= \boldsymbol{\Gamma}^{-1}(\mathbf{R}_1 \boxminus \mathbf{R}_2) \cdot \mathbf{I}_{3\times3} \\
    &= \boldsymbol{\Gamma}^{-1}(\mathbf{R}_1 \boxminus \mathbf{R}_2)
    \end{aligned}
\end{equation}
The gradient with respect to the second argument can be found through the chain rule:
\begin{equation}
    \label{eq:jacobian-of-boxminus-wrt-R2-derivation}
    \begin{aligned}
    \mathcal{D}_{\mathbf{R}_2} (\mathbf{R}_1 \boxminus \mathbf{R}_2) &= \mathcal{D}_{ (\mathbf{R}_2^\top \mathbf{R}_1)}\log(\mathbf{R}_2^\top \mathbf{R}_1)^\vee \cdot \mathcal{D}_{\mathbf{R}_2^\top} (\mathbf{R}_2^\top \mathbf{R}_1)\cdot \mathcal{D}_{\mathbf{R}_2} \mathbf{R}_2^\top \\
    &= \boldsymbol{\Gamma}^{-1}(\mathbf{R}_1 \boxminus \mathbf{R}_2) \cdot \mathbf{R_1}^\top \cdot (-\mathbf{R}_2) \\
    &= -\boldsymbol{\Gamma}^{-1}(\mathbf{R}_1 \boxminus \mathbf{R}_2)^\top
    \end{aligned}
\end{equation}
where the identity \eqref{eq:transpose-of-log-jacobian} is used to simplify the resulting expression.
 
\section{Derivation of Constraint Gradients}
The non-trivial gradients of constraints and other functions used in the paper are derived below, in the order that they appear in the paper.

\subsection{Vector Revolute Joint Constraint}
The rotational part of the vector-valued revolute joint constraint is:
\begin{equation*}
    \mathbf{C}_{\mathrm{rev,rot}} = \left[ \overline{\mathbf{R}}_1 \boxminus \overline{\mathbf{R}}_2 \right]_{1-2}
\end{equation*}
Thus, using \eqref{eq:partial-of-box-minus} and the chain rule, the constraint gradients with respect to rotational DOF are:
\begin{equation}
\begin{aligned}
    \mathcal{D}_{\mathbf{R}_1}(\mathbf{C}_\mathrm{rev,rot}) &= \left[ \mathcal{D}_{\overline{\mathbf{R}}_1} (\overline{\mathbf{R}}_1 \boxminus \overline{\mathbf{R}}_2) \cdot \mathcal{D}_{\mathbf{R}_1} (\overline{\mathbf{R}}_1) \right]_{1-2} \\
    &= \left[ \boldsymbol{\Gamma}^{-1}(\overline{\mathbf{R}}_1 \boxminus \overline{\mathbf{R}}_2) \cdot (\mathbf{R}_1^\mathrm{J})^\top \right]_{1-2}
\end{aligned}
\end{equation}
\begin{equation}
\begin{aligned}
    \mathcal{D}_{\mathbf{R}_2}(\mathbf{C}_\mathrm{rev,rot}) &= \left[ \mathcal{D}_{\overline{\mathbf{R}}_2} (\overline{\mathbf{R}}_1 \boxminus \overline{\mathbf{R}}_2) \cdot \mathcal{D}_{\mathbf{R}_2} (\overline{\mathbf{R}}_2) \right]_{1-2} \\
    &= \left[ -\boldsymbol{\Gamma}^{-1}(\overline{\mathbf{R}}_1 \boxminus \overline{\mathbf{R}}_2)^\top \cdot (\mathbf{R}_2^\mathrm{J})^\top \right]_{1-2}
\end{aligned}
\end{equation}

\subsection{Spherical Joint Swing and Twist Limit Constraints}
For a spherical joint, let us define $\mathbf{R}_\Delta = \overline{\mathbf{R}}_1^\top \overline{\mathbf{R}}_2$, where $\overline{\mathbf{R}}_1$ and $\overline{\mathbf{R}}_2$ are the joint frame orientations as defined in \eqref{eq:joint-frame-orientation}. The swing angle limit constraints constrain the swing angle to be within predefined limits:
\begin{equation}
\begin{aligned}
\label{eq:swing-angle-limit-constraints}
    C_\mathrm{min} &= \theta_\mathrm{swing} - \theta_\mathrm{min} \geq 0 \\
    C_\mathrm{max} &= \theta_\mathrm{max} - \theta_\mathrm{swing} \geq 0
\end{aligned}
\end{equation}
where
\begin{equation}
    \theta_\mathrm{swing} = \arccos(\mathbf{e}_3^\top \mathbf{R}_\Delta \mathbf{e}_3)
\end{equation}
The constraint gradients of \eqref{eq:swing-angle-limit-constraints} follow simply from the derivatives $\mathcal{D}_{\mathbf{R}_1} \theta_\mathrm{swing}$ and $\mathcal{D}_{\mathbf{R}_2} \theta_\mathrm{swing}$, which are evaluated using the chain rule. First, we compute the necessary derivatives of $\mathbf{R}_\Delta$:
\begin{align}
&\begin{aligned}
    \mathcal{D}_{\mathbf{R}_1} \mathbf{R}_\Delta &= \mathcal{D}_{\overline{\mathbf{R}}_1^\top}(\overline{\mathbf{R}}_1^\top \overline{\mathbf{R}}_2) \cdot \mathcal{D}_{\overline{\mathbf{R}}_1}(\overline{\mathbf{R}}_1^\top) \cdot \mathcal{D}_{\mathbf{R}_1} (\overline{\mathbf{R}}_1) \\
    &= \overline{\mathbf{R}}_2^\top \cdot -\overline{\mathbf{R}}_1 \cdot (\mathbf{R}_1^\mathrm{J})^\top \\
    &= -\overline{\mathbf{R}}_2^\top \mathbf{R}_1
\end{aligned}\\
&\begin{aligned}
    \mathcal{D}_{\mathbf{R}_2} \mathbf{R}_\Delta &= \mathcal{D}_{\overline{\mathbf{R}}_2}(\overline{\mathbf{R}}_1^\top \overline{\mathbf{R}}_2) \cdot \mathcal{D}_{\mathbf{R}_2}(\overline{\mathbf{R}}_2) \\
    &= (\mathbf{R}_2^\mathrm{J})^\top
\end{aligned}
\end{align}
Then, the derivatives of $\theta_\mathrm{swing}$ with respect to $\mathbf{R}_1$ and $\mathbf{R}_2$ are:
\begin{align}
&\begin{aligned}
    \mathcal{D}_{\mathbf{R}_1} \theta_\mathrm{swing} &= \frac{1}{\sqrt{1-\theta_\mathrm{swing}^2}} \mathbf{e}_3^\top \mathbf{R}_\Delta \widehat{\mathbf{e}}_3 \cdot \mathcal{D}_{\mathbf{R}_1} \mathbf{R}_\Delta \\
    %
\end{aligned} \\
&\begin{aligned}
    \mathcal{D}_{\mathbf{R}_2} \theta_\mathrm{swing} &= \frac{1}{\sqrt{1-\theta_\mathrm{swing}^2}} \mathbf{e}_3^\top \mathbf{R}_\Delta \widehat{\mathbf{e}}_3 \cdot \mathcal{D}_{\mathbf{R}_2} \mathbf{R}_\Delta 
\end{aligned}
\end{align}

The spherical joint twist angle limit constraints are defined analogously to \eqref{eq:swing-angle-limit-constraints}, with $\theta_\mathrm{twist}$ in place of $\theta_\mathrm{swing}$, where
\begin{equation}
    \theta_\mathrm{twist} = \mathrm{atan2}(\mathbf{R}_\Delta(2,1) - \mathbf{R}_\Delta(1,2), \ \mathbf{R}_\Delta(1,1) + \mathbf{R}_\Delta(2,2))
\end{equation}
Defining $y=\mathbf{R}_\Delta(2,1) - \mathbf{R}_\Delta(1,2)$ and $x=\mathbf{R}_\Delta(1,1) + \mathbf{R}_\Delta(2,2)$, the derivatives of $\theta_\mathrm{twist}$ with respect to $\mathbf{R}_1$ and $\mathbf{R}_2$ are:
\begin{equation}
\begin{aligned}
    \mathcal{D}_{\mathbf{R}_1} \theta_\mathrm{twist} &= \left[ -\frac{y}{x^2+y^2} \mathcal{D}_{\mathbf{R}_\Delta} x  + \frac{x}{x^2+y^2} \mathcal{D}_{\mathbf{R}_\Delta} y  \right] \cdot \mathcal{D}_{\mathbf{R}_1} \mathbf{R}_\Delta \\
    \mathcal{D}_{\mathbf{R}_2} \theta_\mathrm{twist} &= \left[ -\frac{y}{x^2+y^2} \mathcal{D}_{\mathbf{R}_\Delta} x  + \frac{x}{x^2+y^2} \mathcal{D}_{\mathbf{R}_\Delta} y  \right] \cdot \mathcal{D}_{\mathbf{R}_2} \mathbf{R}_\Delta
\end{aligned}
\end{equation}
where
\begin{equation}
\begin{aligned}
    \mathcal{D}_{\mathbf{R}_\Delta} x &= -(\mathbf{e}_1^\top \mathbf{R}_\Delta \widehat{\mathbf{e}}_1 + \mathbf{e}_2^\top \mathbf{R}_\Delta \widehat{\mathbf{e}_2}) \\
    \mathcal{D}_{\mathbf{R}_\Delta} y &= -(\mathbf{e}_2^\top \mathbf{R}_\Delta \widehat{\mathbf{e}}_1 - \mathbf{e}_1^\top \mathbf{R}_\Delta \widehat{\mathbf{e}}_2)
\end{aligned}
\end{equation}
\subsection{Prismatic Joint Constraint}
The positional part of the vector-valued prismatic joint constraint is:
\begin{equation}
    \mathbf{C}_\mathrm{pr,pos} = \left[ \overline{\mathbf{R}}_1^\top \mathbf{p}_2 \right]_{1-2}
\end{equation}
The derivation of the constraint gradients can be done using the chain rule:
\begin{equation}
\begin{aligned}
    \mathcal{D}_{\mathbf{R}_1} \mathbf{C}_\mathrm{pr,pos} &= \left[ \mathcal{D}_{\overline{\mathbf{R}}_1^\top}(\overline{\mathbf{R}}_1^\top \mathbf{p}_2) \cdot \mathcal{D}_{\overline{\mathbf{R}}_1} (\overline{\mathbf{R}}_1^\top) \cdot \mathcal{D}_{\mathbf{R}_1} \overline{\mathbf{R}}_1 \right]_{1-2} \\
    &= \left[ -\overline{\mathbf{R}}_1^\top \widehat{\mathbf{p}}_2 \cdot -\overline{\mathbf{R}}_1 \cdot (\mathbf{R}_1^\mathrm{J})^\top \right]_{1-2} \\
    &= \left[ \overline{\mathbf{R}}_1^\top \widehat{\mathbf{p}}_2 \mathbf{R}_1 \right]_{1-2}
\end{aligned}
\end{equation}

\subsection{Cosserat Rod Shear Constraint}
\begin{equation}
    \mathbf{C}_j^\mathrm{v} = \frac{1}{2l} (\mathbf{R}_{j}^\top + \mathbf{R}_{j+1}^\top ) \left( \mathbf{p}_{j+1} - \mathbf{p}_{j}\right) - \mathbf{e}_3
\end{equation}
The gradients with respect to the position are trivial. The gradients with respect to the rotational DOF are found using the chain rule:
\begin{equation}
\begin{aligned}
\label{eq:grad-cv-orj}
\mathcal{D}_{\mathbf{R}_j} \mathbf{C}_{j}^\mathrm{v} &= \frac{1}{2l} \cdot \mathcal{D}_{\mathbf{R}_j^\top} \left( \mathbf{R}_j^\top (\mathbf{p}_{j+1}-\mathbf{p}_j) \right) \cdot \mathcal{D}_{\mathbf{R}_j} \mathbf{R}_j^\top \\
&= \frac{1}{2 l} \cdot -\mathbf{R}_j^\top  (\mathbf{p}_{j+1} - \mathbf{p}_j)^\wedge \cdot -\mathbf{R}_j \\
&= \frac{1}{2 l} \left( \mathbf{R}_j^\top(\mathbf{p}_{j+1} - \mathbf{p}_j)\right)^\wedge
\end{aligned}
\end{equation}
\begin{equation}
\begin{aligned}
\label{eq:grad-cv-orjplus1}
\mathcal{D}_{\mathbf{R}_{j+1}} \mathbf{C}_{j}^\mathrm{v} &= \frac{1}{2l} \cdot \mathcal{D}_{\mathbf{R}_{j+1}} \left( \mathbf{R}_{j+1}^\top(\mathbf{p}_{j+1}-\mathbf{p}_j) \right) \cdot \mathcal{D}_{\mathbf{R}_{j+1}} \mathbf{R}_{j+1}^\top \\
&= \frac{1}{2 l} \cdot  -\mathbf{R}_{j+1}^\top (\mathbf{p}_{j+1} - \mathbf{p}_j)^\wedge \cdot -\mathbf{R}_{j+1} \\
&= \frac{1}{2 l} \left( \mathbf{R}_{j+1}^\top(\mathbf{p}_{j+1} - \mathbf{p}_j)\right)^\wedge
\end{aligned}
\end{equation}

\subsection{Cosserat Rod Bending Constraint}
\begin{equation}
    \mathbf{C}_j^\mathrm{u} = \frac{1}{l} \left( \mathbf{R}_{j+1} \boxminus \mathbf{R}_j \right)
\end{equation}
The gradients with respect to the positional DOF are 0, as they are not involved in the constraint. The gradients with respect to the rotational DOF are found using \eqref{eq:partial-of-box-minus}:
\begin{equation}
\begin{aligned}
\mathcal{D}_{\mathbf{R}_j} \mathbf{C}_{j}^\mathrm{u} &= -\frac{1}{l}  \boldsymbol{\Gamma}^{-1}(\mathbf{R}_{j+1} \boxminus \mathbf{R}_j)^\top \\
\mathcal{D}_{\mathbf{R}_{j+1}} \mathbf{C}_{j}^\mathrm{u} &= \frac{1}{l}  \boldsymbol{\Gamma}^{-1}(\mathbf{R}_{j+1} \boxminus \mathbf{R}_j)
\end{aligned}
\end{equation}

\section{Derivation of FEM Cosserat Rod Strain Measures and Gradients}
Deriving the expression for $\mathbf{v}(s)$ is a simple application of the chain rule:
\begin{equation*}
\begin{aligned}
\mathbf{p}(s) &= \sum_{i=1}^{N_{en}} \phi_i(s) \mathbf{p}_i \\
\frac{\partial}{\partial s} \mathbf{p}(s) &= \frac{1}{l_i} \sum_{i=1}^{N_{en}} \phi'(s) \mathbf{p}_i \\
\mathbf{v}(s) &= \mathbf{R}(s)^\top \mathbf{p}'(s) = \frac{1}{l_i} \mathbf{R}(s)^\top \sum_{i=1}^{N_{en}} \phi'(s) \mathbf{p}_i 
\end{aligned}
\end{equation*}
Deriving the expression for $\mathbf{u}(s)$ is a bit more involved, requiring us to use the limit definition of differentiation.
\begin{equation*}
\begin{aligned}
	\mathbf{u}(s) &= \mathcal{D}_s (\mathbf{R}(s)) \\
    &= \lim_{\Delta s \to 0} \frac{\mathbf{R}(s + \Delta s) \boxminus \mathbf{R}(s)}{\Delta s} \\
	&= \lim_{\Delta s \to 0} \frac{ (\mathbf{R}_1 \exp(\widehat{\boldsymbol{\theta}}(s + \Delta s))) \boxminus (\mathbf{R}_1 \exp(\widehat{\boldsymbol{\theta}}(s)))}{\Delta s} \\
	&= \lim_{\Delta s \to 0} \frac{\log\left[ \exp(\widehat{\boldsymbol{\theta}}(s))^\top \exp(\widehat{\boldsymbol{\theta}}(s + \Delta s)) \right]^\vee}{\Delta s}
\end{aligned}
\end{equation*}
To simplify this expression, we use two Taylor approximations.
The first is a standard Taylor expansion about $\boldsymbol{\theta}(s)$:
\begin{equation*}
\boldsymbol{\theta}(s + \Delta s) = \boldsymbol{\theta}(s) + \Delta s \boldsymbol{\theta}'(s) + O(\Delta s^2)
\end{equation*}
and from \eqref{eq:first-order-approx-exponential-map}:
\begin{equation*}
    \exp ((\boldsymbol{\theta} + t\Delta \boldsymbol{\theta})^\wedge) = \exp(\widehat{\boldsymbol{\theta}}) \exp \left( (t \boldsymbol{\Gamma}(\boldsymbol{\theta}) \Delta \boldsymbol{\theta} + O(t^2) )^\wedge \right)
\end{equation*}
Combining these, we find that for sufficiently small $\Delta s$:
\begin{equation*}
\begin{aligned}
\exp(\widehat{\boldsymbol{\theta}}(s + \Delta s)) &\approx \exp( (\boldsymbol{\theta}(s) + \Delta s \boldsymbol{\theta}'(s))^\wedge) \\
&\approx \exp(\boldsymbol{\theta}(s)) \exp \left( (\Delta s \boldsymbol{\Gamma}(\boldsymbol{\theta}(s)) \boldsymbol{\theta}'(s))^\wedge \right)
\end{aligned}
\end{equation*}
Substituting this expression in and simplifying, we have that
\begin{equation*}
\begin{aligned}
\mathbf{u}(s) &= \lim_{\Delta s \to 0} \frac{\Delta s \boldsymbol{\Gamma}(\boldsymbol{\theta}(s)) \boldsymbol{\theta}'(s)}{\Delta s} \\
&=  \boldsymbol{\Gamma}(\boldsymbol{\theta}(s)) \boldsymbol{\theta}'(s)
\end{aligned}
\end{equation*}

The gradients of $\mathbf{v}(s)$ and $\mathbf{u}(s)$ with respect to the element degrees of freedom ($\mathbf{p}_i$, $\mathbf{R}_i$, with $i=1,...,N_{en}$) can be found via the chain rule. The gradients of $\mathbf{v}(s)$ with respect to the positional DOF are simply:
\begin{equation*}
\frac{\partial \mathbf{v}(s)}{\partial \mathbf{p}_i} = \mathbf{R}(s)^\top \phi_i'(s)
\end{equation*}
The gradients of $\mathbf{v}(s)$ with respect to the rotational DOF are:
\begin{equation*}
\begin{aligned}
	\mathcal{D}_{\mathbf{R}_i} (\mathbf{v}(s)) &= \mathcal{D}_{\mathbf{R}(s)} \left( \mathbf{R}(s)^\top \mathbf{p}'(s) \right) \cdot \mathcal{D}_{\mathbf{R}(s)} \mathbf{R}(s)^\top \cdot \mathcal{D}_{\mathbf{R}_i} \mathbf{R}(s) \\
	&= -\mathbf{R}(s)^\top (\mathbf{p}'(s))^\wedge \cdot -\mathbf{R}(s) \cdot \mathcal{D}_{\mathbf{R}_i} \mathbf{R}(s) \\
	&= \left( \mathbf{v}(s) \right)^\wedge \cdot \mathcal{D}_{\mathbf{R}_i} \mathbf{R}(s)
\end{aligned}
\end{equation*}
where
\begin{equation*}
\mathcal{D}_{\mathbf{R}_1} \mathbf{R}(s) = \mathrm{Exp}(\boldsymbol{\theta}(s))^\top + \boldsymbol{\Gamma}(\boldsymbol{\theta}(s)) \cdot \mathcal{D}_{\mathbf{R}_1} \boldsymbol{\theta}(s)
\end{equation*} 
and for $i=2,...,N_{en}$:
\begin{equation*}
\mathcal{D}_{\mathbf{R}_i} \mathbf{R}(s) = \boldsymbol{\Gamma}(\boldsymbol{\theta}(s)) \cdot \mathcal{D}_{\mathbf{R}_i} \boldsymbol{\theta}(s)
\end{equation*}
The final quantities needed are $\mathcal{D}_{\mathbf{R}_i} \boldsymbol{\theta}(s)$, given by:
\begin{equation*}
\mathcal{D}_{\mathbf{R}_1} \boldsymbol{\theta}(s) = -\sum_{i=2}^{N_{en}} \phi_i (s) \boldsymbol{\Gamma}^{-1}(\mathbf{R}_i \boxminus \mathbf{R}_1)^\top
\end{equation*}
and for $i=2,...,N_{en}$:
\begin{equation*}
\mathcal{D}_{\mathbf{R}_i} \boldsymbol{\theta}(s) = \phi_i (s) \boldsymbol{\Gamma}^{-1}(\mathbf{R}_i \boxminus \mathbf{R}_1)
\end{equation*}
The gradients of $\mathbf{u}(s)$ with respect to the positional DOF are 0, so that only leaves the gradients with respect to the rotational DOF:
\begin{equation*}
\mathcal{D}_{\mathbf{R}_i} \mathbf{u}(s) = D\boldsymbol{\Gamma}(\boldsymbol{\theta}(s)) \cdot \mathcal{D}_{\mathbf{R}_i} (\boldsymbol{\theta}(s)) \boldsymbol{\theta}'(s) + \boldsymbol{\Gamma}(\boldsymbol{\theta}(s)) \mathcal{D}_{\mathbf{R}_i} (\boldsymbol{\theta}'(s))
\end{equation*}
where
\begin{equation*}
\mathcal{D}_{\mathbf{R}_1} (\boldsymbol{\theta}'(s)) = -\sum_{i=2}^{N_{en}} \phi_i'(s) \boldsymbol{\Gamma}^{-1}(\mathbf{R}_i \boxminus \mathbf{R}_1)^\top
\end{equation*}
and for $i=2,...,N_{en}$:
\begin{equation*}
\mathcal{D}_{\mathbf{R}_i} (\boldsymbol{\theta}'(s)) = \phi_i'(s) \boldsymbol{\Gamma}^{-1}(\mathbf{R}_i \boxminus \mathbf{R}_1) 
\end{equation*}
The first term is tricky because $\partial \boldsymbol{\Gamma} (\boldsymbol{\theta}) / \partial \boldsymbol{\theta}$ is a 3rd order tensor. Just looking at the first term in tensor notation, a change in $\mathbf{R}_i$ results in a change in $u(s)$ through:
\begin{equation*}
\delta \mathbf{u} = \left( \frac{\partial \boldsymbol{\Gamma}(\boldsymbol{\theta})}{\partial \boldsymbol{\theta}} \right)_{abc} \cdot \left( \mathcal{D}_{\mathbf{R}_i} \boldsymbol{\theta} \right)_{cd} \cdot (\delta \mathbf{R}_i)_d \cdot \boldsymbol{\theta}'_b 
\end{equation*}
where 
\begin{equation*}
\begin{aligned}
\frac{(\partial \boldsymbol{\Gamma}(\boldsymbol{\theta}))_{ab}}{\partial \boldsymbol{\theta}_c} = &-\frac{\partial \alpha}{\partial ||\boldsymbol{\theta}||} \frac{\boldsymbol{\theta}_c}{||\boldsymbol{\theta}||} \widehat{\boldsymbol{\theta}}_{ab} - \alpha  \frac{\partial \widehat{\boldsymbol{\theta}}_{ab}}{\partial \boldsymbol{\theta}_c} + \\
&\frac{\partial \beta}{\partial ||\boldsymbol{\theta}||} \frac{\boldsymbol{\theta}_c}{||\boldsymbol{\theta}||} \widehat{\boldsymbol{\theta}}_{ab}^2 + \beta \left( \frac{\partial \widehat{\boldsymbol{\theta}}}{\partial \boldsymbol{\theta}_c} \widehat{\boldsymbol{\theta}} + \widehat{\boldsymbol{\theta}} \frac{\partial \widehat{\boldsymbol{\theta}}}{\partial \boldsymbol{\theta}_c} \right)_{ab}
\end{aligned}
\end{equation*}
with
\begin{equation*}
\begin{aligned}
\alpha(||\boldsymbol{\theta}||) &= \frac{1 - \cos ||\boldsymbol{\theta}||}{||\boldsymbol{\theta}||^2} \\
\beta(||\boldsymbol{\theta}||) &= \frac{||\boldsymbol{\theta}|| - \sin ||\boldsymbol{\theta}|| }{||\boldsymbol{\theta}||^3} \\
\frac{\partial \alpha}{\partial ||\boldsymbol{\theta}|| } &= -2 \frac{1-\cos ||\boldsymbol{\theta}||}{||\boldsymbol{\theta}||^3} + \frac{\sin ||\boldsymbol{\theta}||}{||\boldsymbol{\theta}||^2} \\
\frac{\partial \beta}{\partial ||\boldsymbol{\theta}|| } &= \frac{1 - \cos ||\boldsymbol{\theta}||}{||\boldsymbol{\theta}||^3} - 3 \frac{||\boldsymbol{\theta}|| - \sin ||\boldsymbol{\theta}||}{||\boldsymbol{\theta}||^4}
\end{aligned}
\end{equation*}
and
\begin{equation*}
\frac{\partial \widehat{\boldsymbol{\theta}}}{\boldsymbol{\theta}_1} = 
\begin{bmatrix}
	0 & 0 & 0 \\
	0 & 0 & -1 \\
	0 & 1 & 0
\end{bmatrix} \ \ \ \ 
\frac{\partial \widehat{\boldsymbol{\theta}}}{\boldsymbol{\theta}_2} = 
\begin{bmatrix}
	0 & 0 & 1 \\
	0 & 0 & 0 \\
	-1 & 0 & 0
\end{bmatrix} \ \ \ \
\frac{\partial \widehat{\boldsymbol{\theta}}}{\boldsymbol{\theta}_3} = 
\begin{bmatrix}
	0 & -1& 0 \\
	1 & 0 & 0 \\
	0 & 0 & 0
\end{bmatrix}
\end{equation*}
Ultimately, we want a matrix expression for the gradient $\partial \mathbf{u} / \partial \mathbf{R}_i$ such that we map perturbations $\delta \mathbf{R}_i$ to changes in $\mathbf{u}$. Thus, we rearrange the first term to be:
\begin{equation*}
\delta \mathbf{u} = \left( \frac{\partial \boldsymbol{\Gamma}(\boldsymbol{\theta})}{\partial \boldsymbol{\theta}} \right)_{abc} \cdot \boldsymbol{\theta}'_b \cdot \left( \mathcal{D}_{\mathbf{R}_i} \boldsymbol{\theta} \right)_{cd} \cdot (\delta \mathbf{R}_i)_d
\end{equation*}
and compute a contraction of the derivative of $\boldsymbol{\Gamma}$ over the second index (i.e. over tensor index $b$) as:
\begin{equation*}
\begin{aligned}
\left( \frac{\partial \boldsymbol{\Gamma}(\boldsymbol{\theta})}{\partial \boldsymbol{\theta}} \right)_{abc} \cdot \boldsymbol{\theta}'_b = &-\frac{\partial \alpha}{\partial ||\boldsymbol{\theta}||} \frac{\boldsymbol{\theta}_c}{||\boldsymbol{\theta}||} \widehat{\boldsymbol{\theta}}_{ab} \boldsymbol{\theta}'_b - \alpha  \frac{\partial \widehat{\boldsymbol{\theta}}_{ab}}{\partial \boldsymbol{\theta}_c} \boldsymbol{\theta}'_b + \\
&\frac{\partial \beta}{\partial ||\boldsymbol{\theta}||} \frac{\boldsymbol{\theta}_c}{||\boldsymbol{\theta}||} \widehat{\boldsymbol{\theta}}_{ab}^2 \boldsymbol{\theta}'_b + \beta \left( \frac{\partial \widehat{\boldsymbol{\theta}}}{\partial \boldsymbol{\theta}_c} \widehat{\boldsymbol{\theta}} + \widehat{\boldsymbol{\theta}} \frac{\partial \widehat{\boldsymbol{\theta}}}{\partial \boldsymbol{\theta}_c} \right)_{ab} \boldsymbol{\theta}'_b
\end{aligned}
\end{equation*}
The above equation gives the expression for the $c$th column of the matrix.

\typeout{=== BIBLIOGRAPHY TEST ===}
\bibliographystyle{ACM-Reference-Format}
\bibliography{root}

\end{document}


\maketitle

In this supplement, we make detailed comparisons between constraints and gradients derived with our Lie-theoretic approach and the formulas presented in \cite{muller2020detailed}. Though \cite{muller2020detailed} do not present explicit constraint definitions, we can extract the effective constraint definition and constraint gradient from their provided XPBD update equations. From there, we compute the constraint gradient with our Lie theoretic approach and compare the gradient expressions. We rewrite the XPBD update formulas for XPBD with orientation derived in the main document:
\begin{equation}
    \Delta\lambda_j=\frac{-C_j(\boldsymbol{x}^k) - \tilde\alpha_j\lambda_{j}^k}{\nabla C_j(\boldsymbol{x}^k)\boldsymbol{\mathcal{M}}^{-1}\nabla C_j(\boldsymbol{x}^k)^\top + \tilde\alpha_j}
\end{equation}
\begin{equation}
    \Delta \boldsymbol{x}^k = \boldsymbol{\mathcal{M}}^{-1} \nabla \mathbf{C}(\boldsymbol{x}^k)^\top \Delta \boldsymbol{\lambda}^k
\end{equation}

We use our joint frame notation from Section 5.2.1, namely that the joint frame on body 1 is defined by its position $\overline{\mathbf{p}}_1= \mathbf{p}_1 + \mathbf{R}_1 \mathbf{r}_1^\mathrm{J}$ and its orientation $\overline{\mathbf{R}}_1 = \mathbf{R}_1 \mathbf{R}_1^\mathrm{J}$, with analogous definitions for body 2.

\section{Positional and Angular Constraint Comparison}
\cite{muller2020detailed} define core projection operations that they use for all other rigid body constraints. These are (1) applying a positional correction between two points on separate rigid bodies and (2) applying an angular correction between two rigid bodies. For these operations, the update formulas that \cite{muller2020detailed} provide are equivalent to explicitly defining constraints. This is detailed below.  


\subsection{Positional Correction}
\label{sec:positional-constraint-comparison}
\subsubsection{Impulse-Based Formulas}
\label{sec:muller2020-positional-constraints}
To constrain the positions of two joint frames to be co-located, we wish that the positional difference between the two joint frames $$\Delta \overline{\mathbf{p}} = \overline{\mathbf{p}}_1 - \overline{\mathbf{p}}_2 =\mathbf{p}_1 + \mathbf{R}_1 \mathbf{r}_1^\mathrm{J} - \mathbf{p}_2 - \mathbf{R}_2 \mathbf{r}_2^\mathrm{J}$$ is 0. 
The generalized inverse inertia at each joint frame are
\begin{equation}
    \label{eq:muller2020-positional-inverse-inertia}
    \begin{aligned}
        w_1 &= \frac{1}{m_1} + \left(\mathbf{r}_1^\mathrm{J} \times (\mathbf{R}_1^\top \mathbf{n})\right)^\top \boldsymbol{\mathcal{I}}_1^{-1} \left(\mathbf{r}_1^\mathrm{J} \times (\mathbf{R}_1^\top \mathbf{n})\right) \\
        w_2 &= \frac{1}{m_2} + \left(\mathbf{r}_2^\mathrm{J} \times (\mathbf{R}_2^\top \mathbf{n})\right)^\top \boldsymbol{\mathcal{I}}_2^{-1} \left(\mathbf{r}_2^\mathrm{J} \times (\mathbf{R}_2^\top \mathbf{n})\right)
    \end{aligned}
\end{equation}
where $\mathbf{n} = \Delta \overline{\mathbf{p}} / ||\Delta \overline{\mathbf{p}}||$ is the direction of the positional difference, expressed in the global frame, and $\boldsymbol{\mathcal{I}}_1$ and $\boldsymbol{\mathcal{I}}_2$ are the body-frame rotational inertia tensors for body $1$ and body $2$, respectively. Then, the scalar XPBD Lagrange multiplier update for this constraint is
\begin{equation}
    \label{eq:muller2020-positional-xpbd-lambda-update}
    \begin{aligned}
    \Delta \lambda &= \frac{-||\Delta \overline{\mathbf{p}}|| - \tilde{\alpha} \lambda}{w_1 + w_2 + \tilde{\alpha}} \\
    \lambda &= \lambda + \Delta \lambda
    \end{aligned}
\end{equation}
Then, the position and orientation update is:
\begin{equation}
    \label{eq:muller2020-positional-xpbd-position-update}
    \begin{aligned}
        \mathbf{p}_1 &= \mathbf{p}_1 + \Delta\lambda \mathbf{n} / m_1 \\
        \mathbf{p}_2 &= \mathbf{p}_2 - \Delta\lambda \mathbf{n} / m_2 \\
        \mathbf{R}_1 &= \mathbf{R}_1 \boxplus \left[\boldsymbol{\mathcal{I}}_1^{-1}\left(\mathbf{r}_1^\mathrm{J} \times (\mathbf{R}_1^\top \mathbf{n}) \right) \Delta \lambda \right] \\
        \mathbf{R}_2 &= \mathbf{R}_2 \boxplus \left[-\boldsymbol{\mathcal{I}}_2^{-1}\left(\mathbf{r}_2^\mathrm{J} \times (\mathbf{R}_2^\top \mathbf{n}) \right) \Delta \lambda \right]
    \end{aligned}
\end{equation}
For notational consistency, we have replaced the linearized quaternion update formulas used in \cite{muller2020detailed} with $SO(3)$ updates using the $\boxplus$ operator. Also, since $\boldsymbol{\mathcal{I}}_1$ and $\boldsymbol{\mathcal{I}}_2$ are expressed in their respective body frames, $\mathbf{n}$ must also be transformed into the appropriate body frame. 

By inspection of \eqref{eq:muller2020-positional-xpbd-lambda-update}, we can see that the effective constraint function is:
\begin{equation}
\label{eq:muller2020-positional-constraint}
    C_\mathrm{pos} = ||\Delta \overline{\mathbf{p}}||
\end{equation}
and from \eqref{eq:muller2020-positional-xpbd-position-update} we can see that the effective constraint gradients are:
\begin{equation}
\label{eq:muller2020-positional-constraint-gradients}
\begin{aligned}
    \frac{\partial C_\mathrm{pos}}{\partial\mathbf{p}_1} &= \mathbf{n}^\top, \ \  && \mathcal{D}_{\mathbf{R}_1} C_\mathrm{pos} = \left( \mathbf{r}_1^\mathrm{J} \times (\mathbf{R}_1^\top \mathbf{n}) \right)^\top \\
    \frac{\partial C_\mathrm{pos}}{\partial\mathbf{p}_2} &= -\mathbf{n}^\top, \ \ && \mathcal{D}_{\mathbf{R}_2} C_\mathrm{pos} = \left(\mathbf{r}_2^\mathrm{J} \times (\mathbf{R}_2^\top \mathbf{n}) \right)^\top
\end{aligned}
\end{equation}

\subsubsection{Explicit Constraint-Based Formulas}
We can differentiate \eqref{eq:muller2020-positional-constraint} using our Lie-theoretic approach, obtaining:
\begin{equation}
\label{eq:1D-positional-constraint-gradients}
\begin{aligned}
    \frac{\partial C_\mathrm{pos}}{\partial\mathbf{p}_1} &= \mathbf{n}^\top, \ \  && \mathcal{D}_{\mathbf{R}_1} C_\mathrm{pos} = -\mathbf{n}^\top \mathbf{R}_1  \widehat{\mathbf{r}}_1 \\
    \frac{\partial C_\mathrm{pos}}{\partial\mathbf{p}_2} &= -\mathbf{n}^\top, \ \ && \mathcal{D}_{\mathbf{R}_2} C_\mathrm{pos} =\mathbf{n}^\top \mathbf{R}_2 \widehat{\mathbf{r}}_2
\end{aligned}
\end{equation}
which exactly match the gradients in \eqref{eq:muller2020-positional-constraint-gradients}. Therefore, the XPBD updates in \cite{muller2020detailed} are correct.

\subsection{Angular Correction}
\label{sec:angular-constraint-comparison}
\subsubsection{Impulse-Based Formulas}
\label{sec:muller2020-angular-constraints}
Now, we wish to constrain the relative orientation between two bodies to be 0. The angular difference is the rotation vector defined by $(\mathbf{R}_1 \boxminus \mathbf{R}_2) = \Delta \boldsymbol{\theta} \in \mathbb{R}^3$. The magnitude $||\Delta \boldsymbol{\theta}||$ is the rotation angle and the direction $\mathbf{n}=\Delta \boldsymbol{\theta} / ||\Delta \boldsymbol{\theta}||$ is the rotation axis, expressed in the local frame\footnote{Note that in \cite{muller2020detailed}, the rotation vector is represented in the \textit{global} frame, and computed with the global analog of the $\boxminus$ operator: $\log(\mathbf{R}_1 \mathbf{R}_2^\top) ^\vee$. The global direction is transformed to the body frame by pre-multiplying by the appropriate body rotation matrix transposed, which is equivalent to the direction of $\mathbf{R}_1 \boxminus \mathbf{R}_2$. We have opted to use the body rotation vector when presenting the equations of \cite{muller2020detailed} because it aligns better with the conventions used in this paper.}. The generalized inverse inertias for the rotational update are:
\begin{equation}
    \label{eq:muller2020-angular-inverse-inertia}
    \begin{aligned}
    w_1 &= \mathbf{n}^\top \boldsymbol{\mathcal{I}}_1^{-1} \mathbf{n} \\
    w_2 &=\mathbf{n}^\top \boldsymbol{\mathcal{I}}_2^{-1} \mathbf{n}
    \end{aligned}
\end{equation}
The scalar XPBD Lagrange multiplier update is:
\begin{equation}
    \label{eq:muller2020-angular-xpbd-lambda-update}
    \begin{aligned}
    \Delta \lambda &= \frac{-||\Delta \boldsymbol{\theta}|| - \tilde{\alpha} \lambda}{w_1 + w_2 + \tilde{\alpha}} \\
    \lambda &= \lambda + \Delta \lambda
    \end{aligned}
\end{equation}
and the orientation update is:
\begin{equation}
    \label{eq:muller2020-angular-xpbd-orientation-update}
    \begin{aligned}
        \mathbf{R}_1 &= \mathbf{R}_1 \boxplus \left[\Delta \lambda \cdot \boldsymbol{\mathcal{I}}_1^{-1}\mathbf{n}  \right] \\
        \mathbf{R}_2 &= \mathbf{R}_2 \boxplus \left[-\Delta \lambda \cdot \boldsymbol{\mathcal{I}}_2^{-1}\mathbf{n} \right]
    \end{aligned}
\end{equation}
where again $SO(3)$ updates using the $\boxplus$ operator are used in place of the linearized quaternion updates used in \cite{muller2020detailed}.

By inspection of \eqref{eq:muller2020-angular-xpbd-lambda-update}, we see that the constraint function is the rotation angle:
\begin{equation}
\label{eq:muller2020-angular-constraint}
    C_\mathrm{ang} = || \Delta \boldsymbol{\theta} ||
\end{equation}, and from \eqref{eq:muller2020-angular-xpbd-orientation-update} the constraint gradients are:
\begin{equation}
\label{eq:muller2020-angular-constraint-gradients}
 \mathcal{D}_{\mathbf{R}_1} C_\mathrm{ang} = \mathbf{n}^\top \ \ \ \ \ \ \ \ \  \mathcal{D}_{\mathbf{R}_2} C_\mathrm{ang} = -\mathbf{n}^\top 
\end{equation}

\subsubsection{Explicit Constraint-Based Formulas}
Differentiating \eqref{eq:muller2020-angular-constraint} using our Lie-theoretic approach we obtain:
\begin{equation}
\begin{aligned}
    \mathcal{D}_{\mathbf{R}_1} C_\mathrm{ang} &= \frac{\partial ||\boldsymbol{\theta} ||}{\partial \boldsymbol{\theta}} \cdot \mathcal{D}_{\mathbf{R}_1} \boldsymbol{\theta} \\
    &= \mathbf{n}^\top \cdot \boldsymbol{\Gamma}^{-1}( \boldsymbol{\theta}) \\
    &= \mathbf{n}^\top \\
    \mathcal{D}_{\mathbf{R}_2} C_\mathrm{ang} &= \frac{\partial ||\boldsymbol{\theta} ||}{\partial \boldsymbol{\theta}} \cdot \mathcal{D}_{\mathbf{R}_2} \boldsymbol{\theta} \\
    &= \mathbf{n}^\top \cdot -\boldsymbol{\Gamma}^{-1}( \boldsymbol{\theta})^\top \\
    &= -\mathbf{n}^\top
\end{aligned}
\end{equation}
where the identity $\boldsymbol{\Gamma}^{-1}(\boldsymbol{\theta}) \boldsymbol{\theta} = \boldsymbol{\theta}$ was used to simplify. The result is exactly that of \eqref{eq:muller2020-angular-constraint-gradients}.

\section{Joint Constraint Comparison}
Here, we compare the two approaches for rigid body joint constraints. As before, we extract the effective constraint definition from the XPBD update provided in \cite{muller2020detailed} and use our Lie theory approach to derive its gradient. To align with \cite{muller2020detailed}, we split each joint constraint into a positional projection and rotational projection.

\subsection{Fixed Joint}
A fixed joint constrains both the position and orientation of the two joint frames to be exactly aligned. The positional part of the constraint is given by \eqref{eq:muller2020-positional-constraint}, and as shown in Section \ref{sec:positional-constraint-comparison}, the impulse-based gradients and constraint-based gradients agree. The rotational part of the constraint is slightly different than \eqref{eq:muller2020-angular-constraint}, since a constant rotational joint frame offset needs to be included. Therefore, are angular correction vector is now $\Delta \overline{\boldsymbol{\theta}} = \overline{\mathbf{R}}_1 \boxminus \overline{\mathbf{R}}_2$, so the constraint is:
\begin{equation}
\label{eq:muller2020-fixed-angular-constraint}
    C_\mathrm{fix,ang} = || \overline{\mathbf{R}}_1 \boxminus \overline{\mathbf{R}}_2 ||
\end{equation}
where $\overline{\mathbf{R}}_1 = \mathbf{R}_1 \mathbf{R}_1^\mathrm{J}$ and $\overline{\mathbf{R}}_2 = \mathbf{R}_2 \mathbf{R}_2^\mathrm{J}$. If we use the angular projection formulas from Section \ref{sec:angular-constraint-comparison}, we get:
\begin{equation}
\label{eq:muller2020-fixed-constraint-gradients}
\mathcal{D}_{\mathbf{R}_1} C_\mathrm{fix,ang} = \overline{\mathbf{n}}^\top (\mathbf{R}_1^\mathrm{J})^\top \ \ \ \ \ \ \ \ \  \mathcal{D}_{\mathbf{R}_2} C_\mathrm{fix,ang} = -\overline{\mathbf{n}}^\top (\mathbf{R}_2^\mathrm{J})^\top
\end{equation}
where $\overline{\mathbf{n}}= \Delta \overline{\boldsymbol{\theta}} / ||\Delta \overline{\boldsymbol{\theta}}||$, and $\mathbf{R}_1^\mathrm{J}$ and $\mathbf{R}_2^\mathrm{J}$ are used to transform $\overline{\mathbf{n}}$ from the joint frame into the local frame of each body, per \cite{muller2020detailed}. However, differentiating \eqref{eq:muller2020-fixed-angular-constraint} with a Lie-theoretic approach yields:
\begin{equation}
\begin{aligned}
    \mathcal{D}_{\mathbf{R}_1} C_\mathrm{fix,ang} &= \overline{\mathbf{n}}^\top \mathcal{D}_{\overline{\mathbf{R}}_1}(\overline{\mathbf{R}}_1 \boxminus \overline{\mathbf{R}}_2) \cdot \mathcal{D}_{\mathbf{R}_1}(\overline{\mathbf{R}}_1) \\
    &= \overline{\mathbf{n}}^\top \boldsymbol{\Gamma}^{-1}(\overline{\mathbf{R}}_1 \boxminus \overline{\mathbf{R}}_2) (\mathbf{R}_1^\mathrm{J})^\top \\
    &= \overline{\mathbf{n}}^\top (\mathbf{R}_1^\mathrm{J})^\top \\
    \mathcal{D}_{\mathbf{R}_2} C_\mathrm{fix,ang} &= \overline{\mathbf{n}}^\top \mathcal{D}_{\overline{\mathbf{R}}_2}(\overline{\mathbf{R}}_1 \boxminus \overline{\mathbf{R}}_2) \cdot \mathcal{D}_{\mathbf{R}_2}(\overline{\mathbf{R}}_2) \\
    &= -\overline{\mathbf{n}}^\top \boldsymbol{\Gamma}^{-1}(\overline{\mathbf{R}}_1 \boxminus \overline{\mathbf{R}}_2)^\top (\mathbf{R}_1^\mathrm{J})^\top \\
    &= -\overline{\mathbf{n}}^\top (\mathbf{R}_2^\mathrm{J})^\top \\
\end{aligned}
\end{equation}
The gradients between the two approaches match.

\subsection{Revolute Joint}
\cite{muller2020detailed} compute the global angular correction for a revolute joint as $\Delta \boldsymbol{\theta}_g~=~(\overline{\mathbf{R}}_1 \mathbf{e}_3) \times (\overline{\mathbf{R}}_2 \mathbf{e}_3)$, and then use angular constraint projection formulas from Section \ref{sec:muller2020-angular-constraints} to compute the Lagrange multiplier update and orientation updates. The effective constraint function is therefore:
\begin{equation}
\label{eq:muller2020-revolute-angular-constraint}
    C_\mathrm{rev,ang} = ||\Delta \boldsymbol{\theta}_g||
\end{equation} with effective constraint gradients:
\begin{equation}
\label{eq:muller2020-revolute-angular-constraint-gradient}
\mathcal{D}_{\mathbf{R}_1} C_\mathrm{rev,ang} = \mathbf{n}_g^\top \mathbf{R}_1 \ \ \ \ \ \ \ \ \  \mathcal{D}_{\mathbf{R}_2} C_\mathrm{fix,ang} = -\mathbf{n}_g^\top \mathbf{R}_2
\end{equation}
where $\mathbf{n}_g=\Delta \boldsymbol{\theta}_g / ||\Delta \boldsymbol{\theta}_g||$ and the body rotation matrices are used to transform the global normal into each body frame.

We can derive the constraint gradient by differentiating \eqref{eq:muller2020-revolute-angular-constraint}. Let $\mathbf{v} = (\overline{\mathbf{R}}_1 \mathbf{e}_3) \times (\overline{\mathbf{R}}_2 \mathbf{e}_3)$ and denote the two joint axes $\mathbf{a}_1 = \overline{\mathbf{R}}_1 \mathbf{e}_3$ and $\mathbf{a}_2 = \overline{\mathbf{R}}_2 \mathbf{e}_3$. By the chain rule:
\begin{equation}
\begin{aligned}
\mathcal{D}_{\mathbf{R}_1} C_\mathrm{rev} &= \frac{\partial ||\mathbf{v}||}{\partial \mathbf{v}} \cdot \frac{\partial \mathbf{v}}{\partial \mathbf{a}_1} \cdot \mathcal{D}_{\mathbf{R}_1} \mathbf{a}_1 \\
    &= \frac{\mathbf{v}^\top}{||\mathbf{v}||} \cdot -\widehat{\mathbf{a}}_2 \cdot -\mathbf{R}_1 (\mathbf{R}_1^\mathrm{J} \mathbf{e}_3)^\wedge
\end{aligned}
\end{equation}
Using the identity $\mathbf{R} \widehat{\mathbf{v}} = (\mathbf{R} \mathbf{v})^\wedge \mathbf{R}$:
\begin{equation}
    = \frac{\mathbf{v}^\top}{||\mathbf{v}||} \cdot \widehat{\mathbf{a}}_2 \cdot \widehat{\mathbf{a}}_1 \cdot \mathbf{R}_1
\end{equation}
Then, using the identity $\mathbf{a}_2 \times (\mathbf{a}_1 \times \mathbf{p}) = \mathbf{a}_1(\mathbf{a}_2^\top \mathbf{p}) - \mathbf{p}(\mathbf{a}_1^\top \mathbf{a}_2)$ for any vector $\mathbf{p} \in \mathbb{R}^3$, we see that:
\begin{equation}
    \frac{\mathbf{v}^\top}{||\mathbf{v}||} \widehat{\mathbf{a}}_2 \widehat{\mathbf{a}}_1 = \frac{\mathbf{v}^\top}{||\mathbf{v}||}(\mathbf{a}_1 \mathbf{a}_2^\top) - (\mathbf{a}_1^\top \mathbf{a}_2) \frac{\mathbf{v}^\top}{||\mathbf{v}||}
\end{equation}
and since $\mathbf{v} = \mathbf{a}_1 \times \mathbf{a}_2$, by definition of the cross product $\mathbf{v}^\top \mathbf{a}_1=0$, meaning:
\begin{equation}
    \frac{\mathbf{v}^\top}{||\mathbf{v}||} \widehat{\mathbf{a}}_2 \widehat{\mathbf{a}}_1 = -(\mathbf{a}_1^\top \mathbf{a}_2) \frac{\mathbf{v}^\top}{||\mathbf{v}||}
\end{equation}
Therefore we have that:
\begin{equation}
    \mathcal{D}_{\mathbf{R}_1} C_\mathrm{rev} = -\left( \mathbf{a}_1^\top \mathbf{a}_2 \right) \mathbf{n}_g^\top \mathbf{R}_1
\end{equation}
The process can be repeated straightforwardly with respect to $\mathbf{R}_2$ to derive:
\begin{equation}
    \mathcal{D}_{\mathbf{R}_2} C_\mathrm{rev} = \left( \mathbf{a}_1^\top \mathbf{a}_2 \right) \mathbf{n}_g^\top \mathbf{R}_2
\end{equation}
The extra factor $(\mathbf{a}_1^\top \mathbf{a}_2)$ becomes significant as $||\Delta \boldsymbol{\theta}_g||$ grows, that is, as the joint axes become increasingly misaligned. Practically speaking, this discrepancy will make little difference, as due to the XPBD algorithm and small time steps, the axes will almost always be well-aligned. However, note where the negative signs are; in the typical case where the angle between the two joint axes is in $[-\frac{\pi}{2}, \frac{\pi}{2}]$, the dot product $\mathbf{a}_1^\top \mathbf{a}_2 \geq 0$, and the gradient is in the direction exactly opposite what is proposed in \cite{muller2020detailed}.


\subsection{Spherical Joint}
A spherical joint constrains the positions of the two joint frames to be aligned, so from the equivalence shown in Section \ref{sec:positional-constraint-comparison}, the two approaches are equivalent for spherical joints.

\subsection{Prismatic Joint}
In \cite{muller2020detailed}, the global positional correction for a prismatic joint is found by subtracting the components of the positional difference\footnote{In \cite{muller2020detailed}, they write $\Delta\overline{\mathbf{p}}=(\overline{\mathbf{p}}_2 - \overline{\mathbf{p}}_1)$ for prismatic joints, but this would create a sign error, so we flip the order for consistency.} $\Delta \overline{\mathbf{p}}=(\overline{\mathbf{p}}_1 - \overline{\mathbf{p}}_2)$ that are in the direction of the joint axis (which we take as the local z-axis of body 1):
\begin{equation}
\label{eq:prismatic-global-position-correction}
\begin{aligned}
    \Delta \mathbf{x}_g &= \Delta \mathbf{p} - \overline{\mathbf{R}}_1 \mathbf{e}_3 \left( (\overline{\mathbf{p}}_2 - \overline{\mathbf{p}}_1)^\top (\overline{\mathbf{R}}_1 \mathbf{e}_3)\right) \\
    &= \overline{\mathbf{R}}_1 \mathbf{e}_1 \left( \Delta \mathbf{p}^\top (\overline{\mathbf{R}}_1 \mathbf{e}_1)\right) + \overline{\mathbf{R}}_1 \mathbf{e}_2 \left( \Delta \mathbf{p}^\top (\overline{\mathbf{R}}_1 \mathbf{e}_2)\right) \\
    &= \overline{\mathbf{R}}_1 \mathbf{S}_{12} \overline{\mathbf{R}}_1^\top \Delta \mathbf{p}
\end{aligned}
\end{equation}
where $\mathbf{S}_{12}=\left[ \begin{smallmatrix}
    1 & 0 & 0 \\ 0 & 1 & 0 \\ 0 & 0 & 0
\end{smallmatrix} \right]$ is a $3\times3$ selection matrix that selects the first two rows of a vector or matrix, and zeroes the third.
Therefore, from \eqref{eq:muller2020-positional-xpbd-lambda-update}, the effective constraint is:
\begin{equation}
\label{eq:muller2020-prismatic-positional-constraint}
C_\mathrm{pr,pos} = ||\Delta \mathbf{x}_g|| = ||\mathbf{S}_{12} \overline{\mathbf{R}}^\top \Delta \mathbf{p} ||
\end{equation}
Note that because rotations are length-preserving, we removed the $\overline{\mathbf{R}}_1$ from the left of $\Delta \mathbf{x}_g$ inside the norm.
Substituting this positional correction into \eqref{eq:muller2020-positional-xpbd-position-update}, we can find the effective constraint gradient expressions used by \cite{muller2020detailed}:
\begin{equation}
\begin{aligned}
    \frac{\partial C_\mathrm{pr,pos}}{\partial \mathbf{p}_1}  &= \frac{\Delta \mathbf{x}_g}{|| \Delta \mathbf{x}_g ||} \\
    &= \mathbf{n}^\top \overline{\mathbf{R}}_1^\top \\
    \frac{\partial C_\mathrm{pr,pos}}{\partial \mathbf{p}_2}  &= -\frac{\Delta \mathbf{x}_g}{|| \Delta \mathbf{x}_g ||} \\
    &= -\mathbf{n}^\top \overline{\mathbf{R}}_1^\top \\
\end{aligned}
\end{equation}
\begin{equation}
\begin{aligned}
\label{eq:muller2020-prismatic-positional-constraint-gradient-wrt-rotation}
    \mathcal{D}_{\mathbf{R}_1} C_\mathrm{pr,pos} &= \left( \mathbf{r}_1^\mathrm{J} \times (\mathbf{R}_1^\top \frac{\Delta \mathbf{x}_g}{|| \Delta \mathbf{x}_g ||}) \right)^\top \\
    &= \frac{1}{|| \Delta \mathbf{x}_g ||} \left( (\mathbf{r}_1^\mathrm{J})^\wedge \mathbf{R}_1^\top \overline{\mathbf{R}}_1 \mathbf{S}_{12} \overline{\mathbf{R}}_1^\top \Delta \overline{\mathbf{p}} \right)^\top \\
    &= -\mathbf{n}^\top (\mathbf{R}_1^\mathrm{J})^\top (\mathbf{r}_1^\mathrm{J})^\wedge \\
    \mathcal{D}_{\mathbf{R}_2} C_\mathrm{pr,pos} &= \left( \mathbf{r}_2^\mathrm{J} \times (\mathbf{R}_2^\top \frac{\Delta \mathbf{x}_g}{|| \Delta \mathbf{x}_g ||}) \right)^\top \\
    &= \frac{1}{|| \Delta \mathbf{x}_g ||} \left( (\mathbf{r}_2^\mathrm{J})^\wedge \mathbf{R}_2^\top \overline{\mathbf{R}}_1 \mathbf{S}_{12} \overline{\mathbf{R}}_1^\top \Delta \overline{\mathbf{p}} \right)^\top \\
    &= \mathbf{n}^\top \overline{\mathbf{R}}_1^\top \mathbf{R}_2 (\mathbf{r}_2^\mathrm{J})^\wedge
\end{aligned}
\end{equation}
where $\mathbf{n} = (\mathbf{S}_{12} \overline{\mathbf{R}}_1^\top \Delta \overline{\mathbf{p}}) / ||\mathbf{S}_{12} \overline{\mathbf{R}}_1^\top \Delta \overline{\mathbf{p}}||$

Differentiating \eqref{eq:muller2020-prismatic-positional-constraint} using the Lie-theoretic approach yields:
\begin{equation}
\begin{aligned}
    \frac{\partial C_\mathrm{pr,pos}}{\partial \mathbf{p}_1}  &= \mathbf{n}^\top \mathbf{S}_{12} \overline{\mathbf{R}}_1^\top \frac{\partial \Delta \overline{\mathbf{p}}}{\partial\mathbf{p}_1} \\
    &= \mathbf{n}^\top \overline{\mathbf{R}}_1^\top \\
    \frac{\partial C_\mathrm{pr,pos}}{\partial \mathbf{p}_2}  &= \mathbf{n}^\top \mathbf{S}_{12} \overline{\mathbf{R}}_1^\top \frac{\partial \Delta \overline{\mathbf{p}}}{\partial\mathbf{p}_2} \\
    &= -\mathbf{n}^\top \overline{\mathbf{R}}_1^\top \\
\end{aligned}
\end{equation}
\begin{equation}
\begin{aligned}
    \mathcal{D}_{\mathbf{R}_1} C_\mathrm{pr} &= \mathbf{n}^\top \cdot \mathbf{S}_{12} \left[ \mathcal{D}_{\overline{\mathbf{R}}_1^\top}(\overline{\mathbf{R}}_1^\top \Delta \overline{\mathbf{p}}) \cdot \mathcal{D}_{\overline{\mathbf{R}}_1}(\overline{\mathbf{R}}_1^\top) \cdot \mathcal{D}_{\mathbf{R}_1}(\overline{\mathbf{R}}_1) + \overline{\mathbf{R}}_1^\top \mathcal{D}_{\mathbf{R}_1} (\Delta \overline{\mathbf{p}}) \right] \\
    &= \mathbf{n}^\top \left[ (-\overline{\mathbf{R}}_1^\top (\Delta \overline{\mathbf{p}})^\wedge) \cdot (-\overline{\mathbf{R}}_1 (\mathbf{R}_1^\mathrm{J})^\top) + \overline{\mathbf{R}}_1^\top(-\mathbf{R}_1 (\mathbf{r}_1^\mathrm{J})^\wedge) \right] \\ 
    &= \mathbf{n}^\top \left( \overline{\mathbf{R}}_1^\top (\Delta \overline{\mathbf{p}})^\wedge \mathbf{R}_1 - (\mathbf{R}_1^\mathrm{J})^\top (\mathbf{r}_1^\mathrm{J})^\wedge \right) \\
    \mathcal{D}_{\mathbf{R}_2} C_\mathrm{pr} &= \mathbf{n}^\top \cdot \mathcal{D}_{\mathbf{R}_2} (\mathbf{S}_{12} \overline{\mathbf{R}}_1^\top \Delta \overline{\mathbf{p}}) \\
    &= \mathbf{n}^\top\overline{\mathbf{R}}_1^\top \mathbf{R}_2 (\mathbf{r}_2^\mathrm{J})^\wedge 
\end{aligned}
\end{equation}
where $\mathbf{n}^\top \mathbf{S}_{12} = \mathbf{n}^\top$ was used to simplify. The two approaches mostly agree, except for the $\mathbf{R}_1$ gradient, where the term in the parentheses $\overline{\mathbf{R}}_1^\top (\Delta \mathbf{p})^\wedge \mathbf{R}_1$ is missing from the corresponding gradient expression in \eqref{eq:muller2020-prismatic-positional-constraint-gradient-wrt-rotation}. This term arises from the dependence of the joint translation axis $\overline{\mathbf{R}}_1 \mathbf{e}_3$ on the body $1$ rotation $\mathbf{R}_1$. This is not accounted for in \cite{muller2020detailed}, and would be difficult to account for without explicitly defining the constraint function and taking proper derivatives.

\bibliographystyle{ACM-Reference-Format}
\bibliography{root}